%% file: main_arxiv.tex
\documentclass[10pt,journal,twocolumn]{IEEEtran}
\newif\ifCLASSOPTIONromanappendices \CLASSOPTIONromanappendicestrue
\usepackage[utf8x]{inputenc}
\usepackage[T1]{fontenc}
\usepackage{url}
\usepackage{xfrac}
\usepackage{multicol}
\usepackage{bigints}
\usepackage{xcolor}
\usepackage{ifthen}
\usepackage{amsfonts}
\usepackage{amssymb}
\usepackage{relsize}
\usepackage{amsmath}
 
\usepackage{bm}
\usepackage{sansmath}
\usepackage{siunitx}
\usepackage{mathtools}

\usepackage{cite}
\usepackage{systeme}
\usepackage{spalign}
\usepackage{subfiles}

\usepackage{algpseudocode}
\usepackage{algorithm}
\algnewcommand\algorithmicforeach{\textbf{for each}}
\algdef{S}[FOR]{ForEach}[1]{\algorithmicforeach\ #1\ \algorithmicdo}

\usepackage{hyperref}
\usepackage{eqparbox}
\usepackage{caption}

\usepackage[labelformat=simple]{subcaption}

\newdimen{\algindent}
\algnewcommand\LeftComment[2]{%
\hspace{#1\algindent}$\triangleright$ \eqparbox{COMMENT}{#2} \hfill %
}
\algnewcommand\LeftCommentNoTriangle[2]{%
\hspace{#1\algindent} \eqparbox{COMMENT}{#2} \hfill %
}
\algnewcommand\LeftCommentNoIntent[1]{%
$\triangleright$ \eqparbox{COMMENT}{#1} \hfill %
}

\newcommand*{\myDots}{\ifmmode\mathellipsis\else.\kern1.40em.\kern 1.40em.\fi}
\newcommand*{\yourtightDots}{\ifmmode\mathellipsis\else.\kern0.1em.\kern 0.18em.\kern 0.1em.\fi}
\newcommand*{\mytightDots}{\ifmmode\mathellipsis\else.\kern0.05em.\kern 0.05em.\kern 0.05em\fi}

\newcommand\Tstrut{\rule{0pt}{2.6ex}}       
\newcommand\Bstrut{\rule[-1.7ex]{0ex}{0ex}} 

\usepackage{verbatim}
\usepackage{tikz}
\usetikzlibrary{positioning,quotes,calc,fit,shapes,shadows,arrows}
\tikzset{block/.style={draw,very thick,text width=2cm,minimum height=4cm,align=center},
         line/.style={-latex}}
\tikzset{blockV/.style={draw,very thick,text width=2cm,minimum height=2cm, minimum width=4cm,align=center},
         line/.style={-latex}}
\tikzset{blockExt/.style={draw,very thick,minimum height=0.6cm, minimum width=0.6cm,align=center},
         line/.style={-latex}}
\tikzset{blockTranspose/.style={draw,very thick, minimum height=0.1cm, minimum width=0.1cm,align=center},
         line/.style={-latex}}
\usepgflibrary{arrows}
\usetikzlibrary{bayesnet}

\newcommand{\overbar}[1]{\mkern 1.5mu\overline{\mkern-1.5mu#1\mkern-1.5mu}\mkern 1.5mu}

\usepackage{enumitem,kantlipsum}

\definecolor{light-gray}{HTML}{E0E0E0}
\definecolor{carnelian}{rgb}{0.7, 0.11, 0.11}
\definecolor{github-link}{RGB}{0,0,139}
\newcommand*{\rv}{\fontfamily{cmss}\selectfont}

\usepackage{booktabs}
\usepackage{multirow}
\usepackage{dblfloatfix}

\usepackage{pifont}

\makeatletter
\newcommand\notsotiny{\@setfontsize\notsotiny{6.6}{6.6}}
\makeatother

\definecolor{darkpastelgreen}{rgb}{0.01, 0.75, 0.24}
\newcommand{\StateGreen}[1]{\algrenewcommand{\alglinenumber}[1]{\footnotesize\textcolor{darkpastelgreen}{##1}:}\State #1}

\definecolor{blue-violet}{rgb}{0.54, 0.17, 0.89}
\newcommand{\StateBlue}[1]{\algrenewcommand{\alglinenumber}[1]{\footnotesize\textcolor{blue-violet}{##1}:}\State #1}

\newcommand{\StateRed}[1]{\algrenewcommand{\alglinenumber}[1]{\footnotesize\textcolor{carnelian}{##1}:}\State #1}
\newcommand{\StateBlack}[1]{\algrenewcommand{\alglinenumber}[1]{\footnotesize\textcolor{black}{##1}:}\State #1}

\usepackage{balance}

\newcommand{\vast}{\bBigg@{4}}

\DeclareUnicodeCharacter{2212}{-}

\begin{document}
\title{Unlabeled Compressed Sensing\\from Multiple Measurement Vectors}

\author{Mohamed Akrout, \IEEEmembership{Member, IEEE},   Amine Mezghani, \IEEEmembership{Member, IEEE}, and Faouzi  Bellili, \IEEEmembership{Member, IEEE}
 \vspace{0.3cm}
\\\small E2-390 E.I.T.C, 75 Chancellor's Circle  Winnipeg, MB, Canada, R3T 5V6.
  \vspace{0.1cm}
  \\\small Emails:  akroutm@myumanitoba.ca, \{amine.mezghani,\,faouzi.bellili\}@umanitoba.ca.
  \vspace{0.3cm}
\thanks{The authors are with the Department of Electrical and Computer Engineering at the University of Manitoba, Winnipeg, MB, Canada.  This work was supported by the Discovery Grants Program of the Natural Sciences and Engineering Research Council of Canada (NSERC).}}

\maketitle
\begin{abstract}
This paper introduces an algorithmic solution to a broader class of unlabeled sensing problems with multiple measurement vectors (MMV). The goal is to recover an unknown structured signal matrix, $\bm{X}$, from its noisy linear observation matrix, $\bm{Y}$, whose rows are further randomly shuffled by an unknown permutation matrix $\bm{U}$. A new Bayes-optimal unlabeled compressed sensing  (UCS) recovery algorithm is developed from the bilinear approximate message passing (Bi-VAMP) framework using non-separable and coupled priors on the rows and columns of the permutation matrix $\bm{U}$. In particular, standard unlabeled sensing is a special case of the proposed framework, and UCS further generalizes it by neither assuming a partially shuffled signal matrix $\bm{X}$ nor a small-sized permutation matrix $\bm{U}$. For the sake of theoretical performance prediction, we also conduct a state evolution (SE) analysis of the proposed algorithm and show its consistency with the asymptotic empirical mean-squared error (MSE). Numerical results demonstrate the effectiveness of the proposed UCS algorithm and its advantage over state-of-the-art baseline approaches in various applications. We also numerically examine the phase transition diagrams of UCS, thereby characterizing the detectability region as a function of the signal-to-noise ratio (SNR).

\end{abstract}


\section{Introduction}
\subsection{Background and motivation}
\IEEEPARstart{C}onsider the following unlabeled compressed sensing problem with multiple measurement vectors (MMV). Given a known sensing matrix $\boldsymbol{A} \in \mathbb{R}^{N \times R}$, reconstruct a signal matrix $\boldsymbol{X} \in \mathbb{R}^{R \times M}$ from a noisy observation matrix $\boldsymbol{Y} \in \mathbb{R}^{N \times M}$:
\begin{equation}
\label{eq:unlabeled-sensing}
\boldsymbol{Y} = \boldsymbol{U} \boldsymbol{A}\,\boldsymbol{X}+ \boldsymbol{W}.
\end{equation}
Here, $\bm{U}\in \mathbb{R}^{N\times N}$ is an unknown permutation matrix and $\boldsymbol{W}$ $\in$ $\mathbb{R}^{N \times M}$ is a noise matrix whose entries, $w_{ij}$, are modelled by independent and identically distributed (i.i.d.) zero-mean Gaussian random variables with variance $\gamma_w^{-1}$, i.e., $w_{ij} \sim$ $\mathcal{N}({w};0, \gamma_w^{-1})$. The noiseless case (i.e., $\boldsymbol{W}=\bm{0}$) with $M=1$ in (\ref{eq:unlabeled-sensing}) corresponds to the single measurement vector (SMV) ``\textit{unlabeled sensing}'' problem recently formulated in \cite{unnikrishnan2018unlabeled}. For convenient reasons which will become apparent later on, we define the matrix $\bm{V} \triangleq (\bm{AX})^{\mathsf{T}}$ and we denote the noiseless signal part as $\boldsymbol{Z} \triangleq \boldsymbol{U} \boldsymbol{V}^\top$. Moreover, by following a Bayesian approach, we treat $\boldsymbol{X}= [\bm{x}_1, \dots, \bm{x}_M]$ as a realization of a random matrix $\textrm{\textbf{{\rv{X}}}}=[\bm{\mathsf{x}}_1, \dots, \bm{\mathsf{x}}_M]$ with a known separable prior:
\begin{equation}
\label{eq:pdf-x}
p_\textrm{\textbf{{\rv{X}}}}(\boldsymbol{X})=\prod_{i=1}^{M} \prod_{j=1}^{R} p_{\textrm{{\rv{x}}}_{ij}}(x_{ij}).
\end{equation}
The distribution $p_{\mathsf{x}_{ij}}(x_{ij})$ can be used to capture any prior knowledge about the entries of $\bm{X}$. 
In particular, if the $x_{ij}$'s are sparse, then they can be modeled by the following prior distribution, widely used in compressed sensing (CS) recovery:
\begin{equation}
    p_{\mathsf{x}_{ij}}(x_{ij})~=~\rho\,\delta(x_{ij})~+~(1-\rho)\,q_{\mathsf{x}_{ij}}(x_{ij}),
\end{equation}
where  $\delta(.)$ is the Dirac distribution, $0<\rho<1$ is the sparsity ratio, and $q_{\mathsf{x}_{ij}}(x_{ij})$ is another distribution that captures any prior knowledge about the non-zero elements (e.g., discreteness).
The unknown permutation matrix $\boldsymbol{U} = [\bm{u}_1, \dots, \bm{u}_N]^\top$ is also a realization of a random matrix $\textrm{\textbf{{\rv{U}}}} = [\bm{\mathsf{u}}_1, \dots, \bm{\mathsf{u}}_N]^\top$, which can in principle be modeled by the following noninformative prior over the set, $\mathcal{P}_{N}$, of all $(N \times N)$ permutation matrices:
\begin{equation}
\label{eq:untractable-permutation-prior}
p_\textrm{\textbf{{\rv{U}}}}(\boldsymbol{U})= \frac{1}{N!}.
\end{equation}
This leads, however, to computationally prohibitive inference for moderately large values of $N$ and we shall describe in Section \ref{sec:unlabeled-as-problem-bilinear} a novel approach that tractably approximates the exact noninformative permutation prior in (\ref{eq:untractable-permutation-prior}). The UCS-MMV problem studied in this paper is encountered in many applications as discussed in the sequel:

\begin{itemize}[leftmargin=*]
    \item \textit{Communication systems}: time interleaving is used between parallel analog-to-digital converters (ADCs) to aggregate all their samples so as to construct a high-data-rate representation of the input signal. Indeed, the ADCs introduce various analog impairments, e.g., the so-called ``residual spurs'', whose suppression can be achieved by randomly shuffling the order in which each ADC samples the input. Once the residual spurs are removed, it is important to cancel the shuffling delay which boils down to recovering the associated permutation matrix $\bm{U}$.
    \item \textit{Robotics and tracking}: the well-known problem of ``simultaneous localization and mapping'' (SLAM) \cite{durrant2006simultaneous} consists in constructing or updating a map of the unknown environment during the robot navigation, which requires finding the relative permutation between the measurements. Similarly, the multi-target tracking (MTT) problem consists in estimating the target states (e.g., locations, velocities) from randomly shuffled measurements such as in radar applications \cite{blackman1986multiple}, wherein knowing the order of the measurements is critical to correctly assign them to targets.
    
    \item \textit{Genomics}: under the finite chemical bases alphabet $\{A, T, G,C\}$ constraint, genome assembly \cite{huang1999cap3} is a reconstruction problem of DNA sequences $\boldsymbol{X}=(\boldsymbol{x}_1,\dots,\boldsymbol{x}_M )$ from their assembled and permuted sub-vector measurements $\boldsymbol{Y}=(\boldsymbol{y}_1,\dots,\boldsymbol{y}_M )$, called ``shotgun reads''. In this case, the matrix $\boldsymbol{A}$ selects sub-vectors of every DNA sequence $\boldsymbol{x}_i$, and as such it is composed of block identity matrices. The unknown permutation matrix $\boldsymbol{U}$ permutes the underlying sub-vectors. Formulating the problem in the MMV setting makes it possible to take advantage of the priors on the DNA structure between the different sequences $\boldsymbol{x}_i$, instead of the i.i.d. assumption on the occurrence of the elements of the chemical bases alphabet.
    
    \item \textit{Data de-anonymization}: when the order of measurements is lost or intentionally deleted to preserve privacy, the problem of data de-anonymization \cite{narayanan2008robust} is equivalent to recovering the unknown permutation from the randomly shuffled observations. This problem arises naturally in many applications in natural science and engineering, such as $i)$ internet-of-things (IoT) networks where each sensor node can communicate without identity information \cite{song2018permuted} while the server is tasked with recovering sensor identities, $ii)$ archaeological deposits where the measurement samples inherently lose their chronological order, $iii)$ electronic health record linkage \cite{lahiri2005regression} which consists of joining together two files that contain information on the same individual whose unique personal identification is missing.
    \item \textit{Biology}: consider the problem of detecting and measuring the physical and chemical characteristics of a population of cells using the flow cytometry technique \cite{shapiro2005practical}. In this case, the order of the cells as they pass through the light is unknown, resulting in permuted cell measurements (e.g., granularity, affinity to a particular target, etc.). Inferring the permutation matrix allows biologists to identify the measurement order of the cells, which can potentially remove the need for the cell sorting step to understand cellular properties that are obscured or non-trivial within populations of cells. A similar permutation ambiguity arises in the molecular communication context when indistinguishable molecules are released at different times to jointly express a message. In this case, the receiving cell detects these molecules in the arrival order instead of the sending order.
\end{itemize}
\subsection{Related work}

\subsubsection{SMV setup (i.e., $M=1$ \textrm{and} $\bm{x}\in\mathbb{R}^R$)}
Since the original formulation of the unlabeled sensing problem in \cite{unnikrishnan2018unlabeled}, several theoretical advances have been made toward understanding its different variations in greater generality. Without bothering with computational feasibility, the authors of \cite{unnikrishnan2018unlabeled} studied the fundamental limits of the problem by assuming that an exhaustive search over $N!$ permutation matrices is possible. Specifically, they showed that for the noiseless SMV case (i.e., $\bm{y}=\bm{UA\,x}$), exact signal recovery is possible\footnote{Under the assumption that the entries of the sensing matrix $\bm{A}$ are drawn from a continuous
distribution over $\mathbb{R}$.} as long as the number of measurements, $N$, is at least twice the number of unknowns (i.e., $N \geq 2R$).

\noindent In \cite{pananjady2016linear}, the authors showed that finding the maximum likelihood (ML) estimate of $\boldsymbol{U}$ is a NP-hard problem, and established the statistical limits of both the exact and approximate recovery of $\boldsymbol{U}$ as a function of the SNR and the number of measurements $N$. A polynomial-complexity approximate ML solution that is not amenable to practical implementation\footnote{As recognized by the authors in \cite{hsu2017linear}.} was proposed in \cite{hsu2017linear}, wherein the minimum required SNR for approximate signal recovery was characterized as a function of $N$ and $R$. When a limited number of measurements are shuffled, both the signal $\boldsymbol{x}$ and the permutation $\boldsymbol{U}$ can be recovered by treating the small fraction of the permuted measurements as outliers, thereby motivating the use of regression methods in \cite{hsu2017linear,slawski2020sparse}. Another line of work has considered the dual problem wherein the roles of $\boldsymbol{y}$ and $\boldsymbol{x}$ are exchanged and $\boldsymbol{U}$ selects a fraction of the measurements in a preserving order \cite{haghighatshoar2017signal}. More recently, an algebraic abstraction of the SMV unlabeled sensing problem called ``homomorphic sensing'' was proposed in \cite{tsakiris2019homomorphic} to devise a branch-and-bound scheme whose computational complexity is exponential and is applicable when $R$ is very small only.
\begin{table*}
\caption{Summary of existing algorithms for unlabeled sensing}
\centering
\begin{tabular}{ *7c }
\cmidrule[1pt]{2-7}
    \multirow{2}{*}{} & \multirow{2}{*}{\textbf{Paper}}
    &\multirow{2}{*}{\textbf{Algorithm}}
    & \multirow{2}{*}{\textbf{Limitation}}
    & \multicolumn{2}{c}{\textbf{Unknown}}
    & \multirow{2}{*}{\textbf{Noisy}}\\
     \cmidrule(l{2pt}r{2pt}){5-6}
     & & & & $\boldsymbol{U}$ & $\boldsymbol{X}$\\
    \midrule 
   \multirow{3}{*}{\shortstack{Single measurement \\ (i.e., $M=1$)}}   
   &\cite{abid2018stochastic} & Stochastic EM & rows of $\boldsymbol{A}$ partially shuffled  &  & \ding{51} & \ding{51} \\
 &\cite{pananjady2016linear} & ML estimation & $R=1$ & \ding{51} &  & \ding{51} \\
 &\cite{tsakiris2019homomorphic} & Dynamic programming & $R$ must be small & \ding{51} & \ding{51} & \ding{51}\\
 \midrule
 
 \multirow{3}{*}{\shortstack{Multiple measurements \\ (i.e., $M \geq 2$)}} & \cite{slawski2020sparse} & ADMM & rows of $\boldsymbol{A}$ partially shuffled  & \ding{51} & \ding{51} & \ding{51}\\
   &\cite{emiya2014compressed} & Branch and Bound & $N$ must be small  & \ding{51} & \ding{51} \\
   &\cite{zhang2019permutation} & ADMM & $R$ must be small & \ding{51} & \ding{51} & \ding{51}\\
   &\cite{slawski2020two,zhang2020optimal} & Mismatch estimation + Coordinate descent & rows of $\boldsymbol{A}$ partially shuffled  & \ding{51} & \ding{51} & \ding{51} \\
   &\cite{abbasi2021r} &  Proximal Alternating minimization  &  $\boldsymbol{U}$ is block diagonal&\ding{51} & \ding{51} & \ding{51} \\
   &\cite{pananjady2017denoising} &  Spectral computation + matching step  &  $\textrm{rank}(\boldsymbol{A})\leq \textrm{rank}(\boldsymbol{X})$ &\ding{51} & \ding{51} &  \ding{51}\\
   &{\bf This work}  &  {\bf Bilinear Vector Approximate  Message Passing}  &  $N$ and $M$ must be large enough  & \ding{51} & \ding{51} & \ding{51} \\
   \bottomrule
\end{tabular}
\vspace{-0.3cm}
\label{tab:algorithms-review}
\end{table*}
\subsubsection{MMV setup (i.e., $M>1$ \textit{and} $\bm{X}\in\mathbb{R}^{R\times M}$)}
Since ML estimation of an arbitrary permutation matrix   $\bm{U}$ is NP-hard \cite{pananjady2017denoising}, regression methods based on outlier detection were investigated in \cite{slawski2020two, slawski2020sparse}, assuming a small number of permuted observations. The recovery algorithm in \cite{slawski2020two} alternates between the block coordinate descent update of $\boldsymbol{X}$ and the least-squares reconstruction of the partially unknown permutation matrix $\boldsymbol{U}$. Similarly, the optimization method used in \cite{slawski2020sparse} alternates between permutation reconstruction via successive outlier removal and signal estimation via ordinary least squares. The need for a small number of permuted observations as assumed in \cite{slawski2020two,slawski2020sparse} has been relaxed to a certain extent in \cite{zhang2019permutation} by reformulating the problem as a bi-convex optimization which can be solved by the alternating direction method of multipliers (ADMM).

Despite the interest that the unlabeled sensing problem has spurred in the recent literature, very few practical algorithms have been proposed to solve it for either SMV or MMV setups. These are classified into three major categories: $i)$ those which recover $\boldsymbol{X}$ only, $ii)$ those which recover $\boldsymbol{U}$ only, and $iii)$ those which jointly reconstruct $\boldsymbol{X}$ and $\boldsymbol{U}$. These three variants have been studied under different (and possibly combined) settings, namely SMV vs. MMV and/or noiseless vs. noisy.

\noindent Table~\ref{tab:algorithms-review} on the top of the next page summarizes the aforementioned categories and provides an overview of the existing algorithmic solutions to the unlabeled sensing problem. At a glance, the reader can notice that each of the available algorithms suffers from one of the following defects:
\begin{itemize}[leftmargin=*]
    \item a scalar unknown signal, i.e., $M=1$ and $R=1$ in (\ref{eq:unlabeled-sensing})
    which is the only tractable case for ML estimation \cite{pananjady2016linear},
    \item a small-size permutation matrix $\boldsymbol{U}$ due to the factorial complexity induced by the exhaustive search over the set, $\mathcal{P}_N$, of all $N\times N$ permutation matrices.
    \item a small fraction of the rows of $\boldsymbol{A}$ is permuted to recover $\bm{U}$, e.g., a sufficiently large superset  of $\mathcal{P}_N$ of size $\Omega(N)$ should be correctly matched \cite{slawski2020two}.
\end{itemize}

\noindent While the unlabeled sensing problem, and generally speaking, regression problems with arbitrarily permuted data are notoriously challenging, no practical algorithm which overcomes the three aforementioned limitations has so far been introduced in the open literature, even for small values of $N$ (e.g., $1<N\leqslant10$). In fact, existing algorithms in Table \ref{tab:algorithms-review} can only handle either: $i)$ small values of $N$, typically $N \in \{2, 3, 4\}$, or $ii)$ limited partial shuffling of the rows of $\bm{A}$.

\subsection{Contributions}
We propose a novel approximate message passing (AMP)-based algorithm that overcomes all the aforementioned limitations to jointly recover both the signal and permutation matrices, $\boldsymbol{X}$ and $\boldsymbol{U}$, respectively. To sidestep the problem of using the exact but untractable prior in (\ref{eq:untractable-permutation-prior}), the permutation matrix is reconstructed via two denoisers that operate side by side on the rows and columns of $\boldsymbol{U}$ under two coupled assignment priors. The flexibility of the AMP paradigm promotes such a divide-and-conquer approach by alternating between the estimation of the rows and columns of $\boldsymbol{U}$, and the reconstruction of $\boldsymbol{X}$, while exchanging extrinsic messages between the various denoising modules. Unlike \cite{slawski2020sparse,slawski2020two}, the algorithm that we propose does not alternate between outlier detection and regression methods, but rather relies entirely on message passing. It builds upon the Bi-VAMP framework recently introduced in \cite{akrout2020bilinear} which is restricted, however, to separable priors on both unknown matrices. In summary, this paper embodies the following main contributions toward the development of a practical recovery algorithm for the broader unlabeled compressed sensing problem: 
 \begin{itemize}[leftmargin=*]
     \item We reformulate the unlabeled compressive sensing problem as a bilinear signal recovery problem under a general i.i.d. prior on the signal matrix $\bm{X}$,
     \item We develop a new AMP-based algorithm (referred to hereafter as UCS)  which extends Bi-VAMP to the case of non separable priors on the unknown permutation matrix $\boldsymbol{U}$,
     \item We provide the theoretical performance guarantees of UCS by deriving its state evolution (SE) update equations that predict the empirical MSE in the large system limit,
     \item We demonstrate numerically the effectiveness of the proposed UCS algorithm in solving the unlabeled compressed sensing problem and the validity of our SE analysis. The code of the UCS algorithm is available at {\scriptsize{\href{https://github.com/makrout/Unlabeled-Compressed-Sensing}{\textcolor{github-link}{\texttt{https://github.com/makrout/Unlabeled-Compressed-Sensing}}}}}.
 \end{itemize}

\subsection*{Notation}
We use Sans Serif fonts (e.g., {\rv{x}}) for random variables and Serif fonts (e.g., $x$) for their realizations. We use boldface lowercase letters for random vectors and their realizations (e.g., {\textbf{\rv{x}}} and $\boldsymbol{x}$) and boldface uppercase letters for random matrices and their realizations (e.g., {\textbf{\rv{X}}} and $\boldsymbol{X}$). Vectors are in column-wise orientation by default. Given any matrix $\boldsymbol{X}$, we use $\boldsymbol{x}_i$  and $x_{ij}$ to denote its $i$th column and $ij$th entry, respectively. We also denote the $i$th component of a vector $\bm{x}$ as $[\bm{x}]_i$ or $x_i$.  We denote  the $k$th canonical basis vector in $\mathbb{R}^N$ as $\bm{e}_k=[0,\cdots,0,1,0,\cdots,0]^{\mathsf{T}}$, which has a single 1 at position $k$. The operator diag($\boldsymbol{X}$) stacks the diagonal elements of $\boldsymbol{X}$ in a vector while Diag($\bm{X}$) returns a diagonal matrix by keeping the diagonal elements only. We use $\mathbf{I}_N$ and $\boldsymbol{J}_N$ to denote the $N \times N$ identity and all-ones matrix, respectively. We also use $p_{\textrm{{\rv{x}}}} (x;\bm{\theta})$, $p_{\bm{\mathsf{x}}} (\bm{x};\bm{\theta})$, and $p_{\bm{\mathsf{X}}} (\bm{X};\bm{\theta})$ to denote the probability density function (pdf) of random variables/vectors/matrices; as being parameterized by a parameter vector $\bm{\theta}$. Moreover,  $\mathcal{N}(\boldsymbol{x}; \widehat{\bm{x}}, \boldsymbol{R})$ stands for the multivariate Gaussian pdf of a random  vector $\bm{\mathsf{x}}$ with mean $\widehat{\bm{x}}$ and covariance matrix $\bm{R}$. The Bernoulli distribution with probability $p$ is denoted as $\mathcal{B}(p)$. We use  $\sim$ and  $\propto$ as short-hand notations for ``distributed according to'' and ``proportional to'', respectively.  We also use $\mathbb{E}[{\textbf{\rv{x}}}|d(\boldsymbol{x})]$ and $\text{Cov}[{\textbf{\rv{x}}}|d(\boldsymbol{x})]$ to denote the expectation and the covariance matrix of $\textrm{\textbf{{\rv{x}}}} \sim d(\boldsymbol{x})$, respectively, and $\delta(\boldsymbol{x})$ denotes the Dirac delta distribution. Moreover, $\langle \boldsymbol{x}\rangle$ and $\langle \boldsymbol{X}\rangle$ return the (empirical) mean of vectors and matrices, i.e., $\langle \boldsymbol{x}\rangle \triangleq$ $\frac{1}{N} \sum_{i=1}^{N} x_{i}$ for $\boldsymbol{x} \in \mathbb{R}^{N}$ and $\langle \boldsymbol{X}\rangle \triangleq$ $\frac{1}{N M} \sum_{i=1}^{N} \sum_{j=1}^{M} x_{ij}$ for $\boldsymbol{X} \in \mathbb{R}^{N \times M}$, where $\triangleq$ is used for definitions. The symbol $\odot$ denotes the Hadamard (i.e., elementwise) product between any two vectors/matrices and we refer to the Frobenius norm of $\bm{X}$ by $\|\bm{X}\|_{\rm{F}}$.

\subsection{Outline}
We structure the rest of this paper as follows. In Section \ref{sec:unlabeled-as-problem-bilinear}, we recast unlabeled compressed sensing as a bilinear recovery problem. In Section \ref{sec:u-big-vamp}, we introduce the UCS algorithm and derive the required belief propagation (BP) messages. In Section \ref{sec:SE}, we establish the theoretical performance guarantees of UCS by deriving its SE update equations. We then assess the performance of the proposed UCS algorithm in Section~\ref{sec:simulation-results} before drawing out some concluding remarks in Section \ref{sec:conclusion}.

\section{Unlabeled compressed sensing as a bilinear signal recovery problem}\label{sec:unlabeled-as-problem-bilinear}
In this section, we start by highlighting the existing bilinear recovery algorithms and explain how Bi-VAMP \cite{akrout2020bilinear} sets the appropriate framework for solving the UCS problem. We then extend Bi-VAMP to account for the presence of a known sensing matrix $\bm{A}$, thereby leading to a parametric version of Bi-VAMP (referred to as P-Bi-VAMP). We finally recast the UCS problem as a bilinear recovery and devise a tractable method for accommodating the non-separable prior enforced by the unknown permutation matrix $\boldsymbol{U}$.

\subsection{Bi-VAMP: the adequate framework for UCS}

Unlike all existing competitors for bilinear recovery (e.g., BAd-VAMP \cite{sarkar2019bilinear}, Bi-GAMP \cite{parker2014bilinear} and its parametric version P-BiG-AMP \cite{parker2016parametric}), Bi-VAMP \cite{akrout2020bilinear} is the only bilinear algorithm that is based entirely on a factor graph consisting of vector-valued variable nodes in both $\boldsymbol{U}$ and $\boldsymbol{V}$. This unique feature of BiG-VAMP makes it possible to use two \textit{inseparable} priors on the columns and rows of $\boldsymbol{U}$ so as to enforce the overall permutation prior. In other words, the factor graphs of the existing bilinear algorithms consist of scalar-valued variable nodes and hence do not capture the correlation inside the rows/columns of $\boldsymbol{U}$. Moreover, when $\boldsymbol{U}$ and $\boldsymbol{V}$ are i.i.d. Gaussian matrices, Bi-GAMP, BAd-VAMP, and Bi-VAMP exhibit almost the same reconstruction performance. However, in the presence of discrete priors on either matrices (e.g., binary i.i.d., permutations in UCS, etc.), the blind reconstruction capabilities of Bi-VAMP \cite{akrout2020bilinear} stands out among its competitors. In this case, BAd-VAMP cannot recover $\boldsymbol{U}$ due to the intractability of the associated E-step in the expectation maximization (EM) procedure.

\subsection{Background on Bi-VAMP for bilinear signal recovery}\label{subsec:frombilinear-to-unlabeled}
In bilinear recovery, the goal is to reconstruct two unknown matrices $\boldsymbol{U}=[\bm{u}_1,\cdots,\bm{u}_N]^{\top} \in \mathbb{R}^{N \times R}$ and $\boldsymbol{V}=[\bm{v}_1,\cdots,\bm{v}_M]^{\top} \in \mathbb{R}^{M \times R}$ from an observation matrix $\boldsymbol{Y} \in$ $\mathbb{R}^{N \times M}$ given by:
\begin{equation}\label{eq:bilinear-recovery}
    \boldsymbol{Y}~ =~ \boldsymbol{U}\,\boldsymbol{V}^{\top} +\, \boldsymbol{W},
\end{equation}
wherein $\boldsymbol{W}$ $\in$ $\mathbb{R}^{N \times M}$ is an additive white Gaussian noise matrix whose entries are assumed to be mutually independent with mean zero and precision $\gamma_w$, i.e., $w_{ij}~\,\sim\,\mathcal{N}(w_{ij}; 0, \gamma_w^{-1})$. The model in (\ref{eq:bilinear-recovery}) assumes separable priors on $\boldsymbol{U}$ and $\boldsymbol{V}$, i.e.:
\begin{subequations}\label{eq:iid-priors-U-V}
    \begin{align}
    p_{\bm{\mathsf{U}}}(\boldsymbol{U}) &=\prod_{i=1}^{N} \prod_{j=1}^{R} p_{\mathsf{u}_{ij}}\left(u_{ij}\right), \label{eq:iid-prior-U}\\
    p_{\bm{\mathsf{V}}}(\boldsymbol{V}) &=\prod_{k=1}^{M} \prod_{l=1}^{R} p_{\mathsf{v}_{kl}}\left(v_{kl}\right). \label{eq:iid-prior-V}
    \end{align}
\end{subequations}

\begin{figure}[h!]
\centering
\input{CLT_approx_factor_graph.tex}
\begin{center}
\begin{tabular}{ |c|c| } 
 \hline
 \protect\tikz[inner sep=.25ex,baseline=-.75ex] \protect\node[circle,draw] {{$1$}}; & $\mathcal{N}\left(\boldsymbol{u}_i;\mathbf{\Lambda}_{\bm{U}}^{-1}\,\bm{b}_{\textrm{\textbf{{\rv{u}}}}_i}, \bm{\Lambda}^{-1}_{\bm{U}}\right)$ \Tstrut\\
 \protect\tikz[inner sep=.25ex,baseline=-.75ex] \protect\node[circle,draw] {{$2$}}; & $\mathcal{N}\left(\boldsymbol{v}_j;\mathbf{\Lambda}_{\bm{V}}^{-1}\,\bm{b}_{\textrm{\textbf{{\rv{v}}}}_j}, \bm{\Lambda}^{-1}_{\bm{V}}\right)$\Tstrut \Bstrut\\
 \hline
\end{tabular}
\end{center}
\caption{Factor graph of Bi-VAMP and its outgoing messages that are used by the denoisers of $\bm{u}_i$ and $\bm{v}_j$ (taken from \cite{akrout2020bilinearArxiv}).}
\vspace{-0.3cm}
\label{fig:factor-graph-bivamp}
\end{figure}
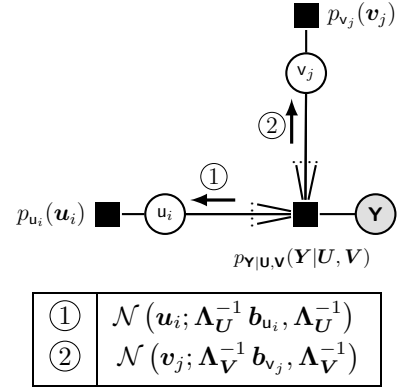

\noindent Unfortunately, discrete priors of practical interest are known to be of combinatorial nature \cite{van1981another}, under which the bilinear recovery of the unknown matrices $\boldsymbol{U}$ and $\boldsymbol{V}$ becomes analytically intractable. To sidestep this problem, Bi-VAMP relies on the central limit theorem (CLT) in the large system limit due to the linear mix in $\boldsymbol{Y}$ that is induced on $\boldsymbol{V}$ (resp. $\boldsymbol{U}$) given an estimate of $\boldsymbol{U}$ (resp. $\boldsymbol{V}$). Fig~\ref{fig:factor-graph-bivamp} depicts the factor graph associated with the exchanged messages during the bilinear LMMSE step of Bi-VAMP. There, it is seen that the factor node $p_{\bm{\mathsf{Y}}|\bm{\mathsf{U}}, \bm{\mathsf{V}}}(\bm{Y}|\bm{U},\bm{V})$ approximates the messages to each random column $\bm{\mathsf{v}}_j$ in $\bm{\mathsf{V}}^\top$ by a Gaussian density denoted by {\scriptsize \protect\tikz[inner sep=0.4ex,baseline=.4ex] \protect\node[circle,draw,yshift=0.15cm] {{$2$}};} with mean vector and covariance matrix\footnote{Bi-VAMP does not track a covariance matrix estimate ${\bm{\Sigma}}_{\bm{v}_j}$ for each random column $\bm{\mathsf{v}}_j$, but rather a single covariance matrix for all $\bm{\mathsf{v}}_j$'s by neglecting vanishing order terms.}:
\begin{subequations}
    \begin{align}
    \widehat{\bm{v}}_j &= \mathbb{E}\big[\bm{\mathsf{v}}_j\big|p_{\bm{\mathsf{v}}_{j}}(\bm{v}_{j})\big] = \mathbf{\Lambda}_{\bm{V}}^{-1}\,\bm{b}_{\bm{v}_j},\\
    {\bm{\Sigma}}_{\bm{v}_j} &= \text{Cov}\big[\bm{\mathsf{v}}_j\big|p_{\bm{\mathsf{v}}_{j}}(\bm{v}_{j})\big] = \bm{\Lambda}^{-1}_{\bm{V}}.
    \end{align}
\end{subequations}

\noindent A similar CLT approximation is also applied to the message {\scriptsize \protect\tikz[inner sep=0.4ex,baseline=.4ex] \protect\node[circle,draw,yshift=0.15cm] {{$1$}};} sent by the factor node $p_{\bm{\mathsf{Y}}|\bm{\mathsf{U}}, \bm{\mathsf{V}}}(\bm{Y}|\bm{U},\bm{V})$ to each $\bm{\mathsf{u}}_i$ due to the inherent symmetry among the variable nodes $\bm{\mathsf{u}}_i$ and $\bm{\mathsf{v}}_j$. The variables $\bm{b}_{\bm{u}_i}$, $\bm{b}_{\bm{v}_j}$, $\mathbf{\Lambda}_{\bm{U}}$, and $\mathbf{\Lambda}_{\bm{V}}$ arise from the CLT approximations in the bilinear LMMSE (Bi-LMMSE) step of Bi-VAMP and we refer the reader to \cite{akrout2020bilinear} for further details.\vspace{-0.3cm}

\subsection{From Bi-VAMP to Parametric Bi-VAMP}

To solve the UCS problem in (\ref{eq:unlabeled-sensing}), Bi-VAMP must be carefully tailored to the bilinear observation model in (\ref{eq:bilinear-recovery}) in which $\boldsymbol{\mathsf{V}} \triangleq \boldsymbol{\mathsf{X}}^{\top}\boldsymbol{A}^{\top}$ with $\boldsymbol{A}$ being some known sensing matrix and $\bm{\mathsf{X}}$ is the unknown random matrix. The presence of a known sensing matrix in the bilinear structure is encountered in various applications such as blind deconvolution and low-rank-plus-sparse recovery models \cite{parker2016parametric}. We call the associated version of Bi-VAMP ``Parametric Bi-VAMP'' (P-Bi-VAMP).

\noindent Given any separable prior $p_{\bm{\mathsf{X}}}(\boldsymbol{X})$ on $\boldsymbol{X}$, the message associated to each random column $\bm{\mathsf{v}}_j=\bm{A}\,\bm{\mathsf{x}}_j$ of $\bm{\mathsf{V}}^\top$ is approximated by a Gaussian density with mean and covariance matrix:
\begin{subequations}\label{eq:change-variable-V-AX}
    \begin{align}
    \mathbb{E}\big[\bm{\mathsf{v}}_j\big|p_{\bm{\mathsf{v}}_{j}}(\bm{v}_{j})\big] &= \boldsymbol{A} ~ \mathbb{E}\big[\bm{\mathsf{x}}_j\big|p_{\bm{\mathsf{x}}_{j}}(\bm{x}_{j})\big],\\
    \text{Cov}\big[\bm{\mathsf{v}}_j\big|p_{\bm{\mathsf{v}}_{j}}(\bm{v}_{j})\big] &= \boldsymbol{A}~ \,\text{Cov}\big[\bm{\mathsf{x}}_j\big|p_{\bm{\mathsf{x}}_{j}}(\bm{x}_{j})\big]\,\boldsymbol{A}^{\top}.
    \end{align}
\end{subequations}

\noindent Finding the second order statistics of $\bm{\mathsf{x}}_j$ from those of $\bm{\mathsf{v}}_j$, however, requires revisiting the Bi-LMMSE step of Bi-VAMP \cite{akrout2020bilinear}, already depicted in Fig. \ref{fig:factor-graph-bivamp}. We start by rewriting  message {\scriptsize \protect\tikz[inner sep=0.4ex,baseline=.4ex] \protect\node[circle,draw,yshift=0.15cm] {{$2$}};} which represents the prior $p_{\bm{\mathsf{v}}_{j}}(\bm{v}_{j})$ as

\begin{equation}\label{eq:prior-V-rewritten}
    \begin{aligned}[b]
        p_{\bm{\mathsf{v}}_{j}}(\bm{v}_{j}) &= \mathcal{N}\left(\boldsymbol{v}_j;\mathbf{\Lambda}_{\bm{V}}^{-1}\,\bm{b}_{\bm{v}_j}, \bm{\Lambda}^{-1}_{\bm{V}}\right)\\
        &\propto\exp\left( -\frac{1}{2} \bm{v}_j^\top\, \bm{\Lambda}_{\bm{V}} \, \bm{v}_j + \bm{b}_{\bm{v}_j}^\top \,\bm{v}_j\right).
    \end{aligned}
\end{equation}

\noindent By setting $\bm{v}_j = \bm{A}\,\bm{x}_j$ in (\ref{eq:prior-V-rewritten}) then completing the square, we obtain:
\begin{equation}\label{eq:prior-V-rewritten-Ax}
    \begin{aligned}[b]
        p_{\bm{\mathsf{v}}_{j}}(\bm{A}\,\bm{x}_{j}) &\propto\exp\left( -\frac{1}{2} \bm{x}_j^\top\,\bm{A}^\top\, \bm{\Lambda}_{\bm{V}} \, \bm{A}\,\bm{x}_j +  \left(\bm{A}^\top\,\bm{b}_{\bm{v}_j}\right)^\top\bm{x}_j\right)\\
        &\propto \mathcal{N}\left(\boldsymbol{x}_j;\mathbf{\Lambda}_{\bm{X}}^{-1}\,\bm{b}_{\bm{x}_j},\mathbf{\Lambda}_{\bm{X}}^{-1}\right) \\
        &\triangleq p_{\bm{\mathsf{x}}_{j}}(\bm{x}_{j}),
    \end{aligned}
\end{equation}
where
\begin{subequations}\label{eq:change-variable-X-to-V-appendix}
    \begin{align}    \bm{b}_{\bm{x}_j}&=\mathbb{E}\big[\bm{\mathsf{x}}_j\big|p_{\bm{\mathsf{x}}_{j}}(\bm{x}_{j})\big] = \bm{A}^{\top}\,\bm{b}_{\bm{v}_j},\\
    \mathbf{\Lambda}_{\bm{X}} &= \text{Cov}\big[\bm{\mathsf{x}}_j\big|p_{\bm{\mathsf{x}}_{j}}(\bm{x}_{j})\big]^{-1} = \boldsymbol{A}^\top \,\mathbf{\Lambda}_{\bm{V}}\,\boldsymbol{A},
    \end{align}
\end{subequations}
or equivalently:
\begin{subequations}\label{eq:change-variable-X-to-V}
    \begin{align}
    \mathbb{E}\big[\bm{\mathsf{x}}_j\big|p_{\bm{\mathsf{x}}_{j}}(\bm{x}_{j})\big] &= \bm{A}^{\top}\,\mathbb{E}\big[\bm{\mathsf{v}}_j\big|p_{\bm{\mathsf{v}}_{j}}(\bm{v}_{j})\big],\\
     \text{Cov}\big[\bm{\mathsf{x}}_j\big|p_{\bm{\mathsf{x}}_{j}}(\bm{x}_{j})\big]^{-1} &= \boldsymbol{A}^\top \,\text{Cov}\big[\bm{\mathsf{v}}_j\big|p_{\bm{\mathsf{v}}_{j}}(\bm{v}_{j})\big]^{-1}\,\boldsymbol{A}.
    \end{align}
\end{subequations}

\noindent In summary, eqs. (\ref{eq:change-variable-V-AX}) and (\ref{eq:change-variable-X-to-V}) allow P-Bi-VAMP to retrieve the first- and second-order statistics of $\bm{\mathsf{v}}_j$ from $\bm{\mathsf{x}}_j$ and vice-versa.

\subsection{From Parametric Bi-VAMP to UCS}
\subsubsection{Addressing the assignment prior on $\bm{U}$}
The i.i.d. assumption on the entries of $\boldsymbol{U}$ in (\ref{eq:iid-prior-U}) which is required by all existing bilinear recovery algorithms is incompatible with the fact that  $\boldsymbol{U}$ is a permutation matrix in the UCS observation model (\ref{eq:unlabeled-sensing}). Indeed, the columns/rows of any ($N\times N$) permutation matrix form a canonical basis for $\mathbb{R}^{N}$ and hence must be captured by a non-separable prior.

\noindent While the exact inference of permutation matrices from bilinear  models as in (\ref{eq:unlabeled-sensing}) is known to be NP-hard \cite{pananjady2016linear}, one of the key ideas embodied by this paper is to design two coupled denoisers for the columns and rows of $\boldsymbol{U}$ so as to overcome the intractability associated with the noninformative permutation prior in (\ref{eq:untractable-permutation-prior}). To that end, we split the variable $\bm{{U}}$ into two auxiliary variables $\bm{{U}}^+$ and $\bm{{U}}^-$ with equality constraint in between. To better define the row-wise and column-wise denoisers in the next section, we define $\bm{{U}}^-$ and $\bm{{U}}^+$ in terms of their rows and columns, respectively. That is to say:
\begin{subequations}
    \begin{align}
        \bm{U}^-~&=~[\bm{u}_1^-,\bm{u}_2^-,\cdots, \bm{u}_N^-]^\top,\\
        \bm{U}^+~&=~[\bm{u}_1^+, \bm{u}_2^+,\dots, \bm{u}_N^+].
    \end{align}
\end{subequations}
\noindent Then, we design its two-step denoiser  by simultaneously enforcing a \textit{non-uniform} assignment prior, $p_{\bm{\mathsf{u}}_i^-}(.)$, on each $i$th row $(\boldsymbol{u}_i^-)^\top$:
\begin{equation} \label{eq:assignment-prior}
p_{\bm{\mathsf{u}}_i^-}(\boldsymbol{u}_i^-) = \sum_{n=1}^{N} \omega_{i,n}\,\delta\left(\boldsymbol{u}_i^- - \boldsymbol{e}_n\right),\quad \, i = 1,2, \dots, N,  
\end{equation}
and a \textit{uniform} assignment prior, $p_{\bm{\mathsf{u}}_i^+}(.)$, on each $i$th column $\bm{u}_i^+$:
\begin{equation} \label{eq:assignment-prior-column}
p_{\bm{\mathsf{u}}_i^+}(\bm{u}_i^+) = \frac{1}{N}\sum_{n=1}^{N} \delta\left(\boldsymbol{u}_i^+ - \boldsymbol{e}_n\right),    \quad \, i = 1,2, \dots, N.
\end{equation}
Note here that the coefficients $\{\omega_{i,n}\}_{i,n}$ in (\ref{eq:assignment-prior}) which must satisfy $\omega_{i,n}\geq 0,$ and $~\sum_{n=1}^N\omega_{i,n}=1~\forall i$  will be updated iteratively by the UCS algorithm as will be described later in Section \ref{sec:u-big-vamp}. From this perspective, customizing P-Bi-VAMP to UCS recovery requires further fundamental changes to the Bi-VAMP formulation itself so as to accommodate the non-separable permutation prior $p_{\bm{\mathsf{U}}}(\boldsymbol{U})$.

\subsubsection{Bi-VAMP with row-wise assignment priors}\label{subsubsec:bivamp-with-one-assignment}
Although Bi-VAMP was originally devised assuming entry-wise separable denoising functions $\mathbf{g}_{\mathsf{u}}(\cdot)$, its modular structure makes it amenable to possible extensions to a number of practical non-separable priors such as the one in (\ref{eq:assignment-prior}). Interestingly, in the latter case, the required posterior mean $\boldsymbol{\widehat{U}}$ of $\bm{\mathsf{U}}$ and the associated covariance matrix $\boldsymbol{{\Sigma}}_{{\boldsymbol{U}}}$ can be computed in closed form, i.e., without the need for the Gaussian approximations used in the original Bi-VAMP framework. To see this, observe that the posterior mean and covariance matrix of the $i$th row, $\boldsymbol{u}_i^\top$, of $\boldsymbol{U}$ can in general be obtained as follows \cite{matsushita2013low,lesieur2016phase}:
\begin{equation}
    \label{eq:denoising-assignment-prior-integral_1}
    \begin{aligned}
        \boldsymbol{\widehat{u}}_i &\,=\, \mathbf{g}_{\mathsf{u}}(\boldsymbol{b}_{\bm{{u}}_i}, \mathbf{\Lambda}_{\bm{U}};p_{\bm{\mathsf{u}}_i})~~~\textrm{and}~~~ \boldsymbol{{\Sigma}}_{\bm{U}} \,=\, \frac{\partial  \mathbf{g}_{\mathsf{u}}(\boldsymbol{b}_{\bm{{u}}_i}, \mathbf{\Lambda}_{\bm{U}};p_{\bm{\mathsf{u}}_i})}{\partial \boldsymbol{b}_{\bm{\mathsf{u}}_i}}, 
    \end{aligned}
\end{equation}
wherein the denoising function $ \mathbf{g}_{\mathsf{u}}(.,.;p_{\bm{\mathsf{u}}_i})$, parameterized by the prior $p_{\bm{\mathsf{u}}_i}(.)$, is given by:
\begin{equation}
    \label{eq:denoising-assignment-prior-integral_2}
   \begin{aligned}[b]
       \!\!\!\!  \mathbf{g}_{\mathsf{u}}(\bm{b}_{\bm{u}_i}, \mathbf{\Lambda}_{\bm{U}};p_{\bm{\mathsf{u}}_i})&\\ &\!\!\!\!\!\!\!\!\!\!\!\!\!\!\!\!\!\!\!\!\!\!\!\!=\frac{\int \boldsymbol{u}_i\, p_{\bm{\mathsf{u}}_i}(\boldsymbol{u}_i)\,\exp\left(-\frac{1}{2} \,\boldsymbol{u}_i^{\top} \mathbf{\Lambda}_{\bm{U}}\, \boldsymbol{u}_i+\boldsymbol{u}_i^{\top}\boldsymbol{b}_{\bm{u}_i} \right) \,\text{d}\boldsymbol{u}_i}{\int p_{\bm{\mathsf{u}}_i}(\boldsymbol{u}_i)\,\exp\left(-\frac{1}{2}\, \boldsymbol{u}_i^{\top} \mathbf{\Lambda}_{\bm{U}}\, \boldsymbol{u}_i+\boldsymbol{u}_i^{\top}\boldsymbol{b}_{\bm{u}_i} \right) \,\text{d}\boldsymbol{u}_i}.
    \end{aligned}
\end{equation}
Here, $\bm{b}_{\bm{{u}}_i}$ and $\mathbf{\Lambda}_{\bm{U}}$ stand, respectively, for the mean vector and precision matrix of the Gaussian density that approximates all the incoming messages to the variable node $\bm{\mathsf{u}}_i$ in the factor graph associated to (\ref{eq:bilinear-recovery}). We refer the reader to \cite{akrout2020bilinear} and the supplementary materials in \cite{matsushita2013low} for further details.\\
Now, by substituting the non-uniform assignment prior given in (\ref{eq:assignment-prior}) into (\ref{eq:denoising-assignment-prior-integral_1}), one obtains the posterior mean  of $\boldsymbol{u}_i$,  as follows:
\begin{equation}
\label{eq:denoising-assignment-prior-mean}
\boldsymbol{\widehat{u}}_{i}\,=\,\mathbf{g}_{\mathsf{u}}(\boldsymbol{b}_{\bm{{u}}_i}, \mathbf{\Lambda}_{\bm{U}})\,=\,\frac{\bm{\omega}_i\odot\exp  \left(\boldsymbol{b}_{\bm{{u}}_i}-\frac{1}{2}\,\textrm{diag}(\bm{\Lambda}_{\bm{U}})\right)}{\sum\limits_{n=1}^{N}\omega_{i,n} \exp\left([\boldsymbol{b}_{\bm{{u}}_i}]_n-\frac{1}{2}\,[\bm{\Lambda}_{\bm{U}}]_{nn}\right)},
\end{equation}
where $\bm{\omega}_i\triangleq[\omega_{i,1}, \omega_{i,2},\cdots,\omega_{i,N} ]^{\mathsf{T}}$.
The corresponding posterior variance for each $k$th component, $\left[\bm{u}_{i}\right]_k$,  of $\boldsymbol{u}_i$ is also obtained as follows:
\begin{equation}\label{eq:denoising-assignment-prior-variance}
    \sigma^{2}_{\left[\boldsymbol{\widehat{u}}_{i}\right]_k} \,=\, \left[\boldsymbol{\widehat{u}}_{i}\right]_k - \left[\boldsymbol{\widehat{u}}_{i}\right]_k^2.
\end{equation}
Since Bi-VAMP can be adapted to handle the assignment prior on the rows of $\boldsymbol{U}$, we now describe the fundamental novelties brought by the proposed UCS algorithm, namely its ability to handle the permutation prior on $\boldsymbol{U}$ based on two coupled denoisers for its rows and columns.

\section{The UCS algorithm}\label{sec:u-big-vamp}
Before delving into the derivation details of UCS, we first present its algorithmic steps in Algorithm \ref{algo:big-vamp} (displayed on the top of the next page). There, we use the subscript \textit{t} to denote the iteration index and the hat symbol ``$~\,\widehat{}~\,$'' for mean values. We also distinguish between ``posterior'' and ``extrinsic'' variables through subscripts $ \mathsf{p}$ and $ \mathsf{e}$, respectively. For the sake of computational efficiency, most of the updated variables in Algorithm \ref{algo:big-vamp} are matrix-valued. For instance, at each iteration $t$, the entire matrix $\bm{B}_{\bm{U},t} ~\triangleq~ [\boldsymbol{b}_{\bm{u}_1,t}, \boldsymbol{b}_{\bm{u}_2,t},\ldots,\boldsymbol{b}_{\bm{u}_N,t}]^{\mathsf{T}}$ is updated with $\{\boldsymbol{b}_{\bm{u}_i,t}\}_{i=1}^{N}$ being the $t^{th}$ update for the mean of the Gaussian message incoming to $\{\textrm{\textbf{{\rv{u}}}}_i\}_{i=1}^{N}$. To avoid numerical overflow problems, we update log-likelihood ratios (LLRs) for the involved binary random variables. The LLR for a Bernoulli-distributed random variable $\mathsf{x} \sim \mathcal{B}\left(\widehat{x}\right)$, is defined as:
    \begin{equation}\label{eq:LR}
         \textrm{LLR}_{{x}} ~\triangleq~\ln\left(\frac{\textrm{Pr}[\mathsf{x}=1]}{\textrm{Pr}[\mathsf{x}=0]}\right) ~=~ \ln\left(\frac{\widehat{x}}{1-\widehat{x}}\right),
    \end{equation}
\noindent from which it follows that:
\begin{equation}\label{eq:mean-function-LLR}
    \widehat{x} ~=~ \frac{\exp\left(\text{LLR}_{{x}}\right)}{1+\exp\left(\text{LLR}_{{x}}\right)}.
\end{equation}

\noindent We also use $\textrm{LLR}_{\bm{{X}}}$ to denote the matrix whose $nm$th entry is $\textrm{LLR}_{{{x}}_{nm}}$.

\noindent For better illustration, Algorithm~\ref{algo:big-vamp} is also pictorially depicted using the block diagram representation shown in Fig. \ref{fig:block-diagram}. There, the $+$ and $-$ labels are used to distinguish the auxiliary variable nodes $\bm{X}^+$ and $\bm{X}^-$ for $\bm{X}$, as well as, $\bm{U}^+$ and $\bm{U}^-$ for $\bm{U}$. The reader is referred to the general Bi-VAMP framework in \cite{akrout2020bilinear} for more details on the role of auxiliary variables. In particular, these are subject to equality constraints enforced by delta factor nodes, i.e., $\delta(\bm{X}^+-\bm{X}^-)$ and $\delta(\bm{U}^+-\,\bm{U}^-)$ as dictated by their joint posterior density:
\begin{eqnarray}
\label{eq:p-uv-given-y-with-equality-constraints}
\!\!\!\!\!\!\!\!\!\!\!\!p\left(\boldsymbol{U}^+, \boldsymbol{U}^-, \boldsymbol{X}^+, \boldsymbol{X}^- \,|\, \bm{Y}\right)&\!\!\!\!\!\!\!\!&\nonumber\\
 \!\!\!\!\!\!\!\!\!\!\!\!&&\!\!\!\!\!\!\!\!\!\!\!\!\!\!\!\!\!\!\!\!\!\!\!\!\!\!\!\!\!\!\!\!\!\!\!\!\!\!\!\!\!\!\!\!\!\!\!\!\!\!\!\!\!\!\!\!\!\!\!\!\propto p\left(\bm{Y} \,|\, \boldsymbol{U}^-, \boldsymbol{X}^-\right)\delta(\boldsymbol{U}^-\!-\boldsymbol{U}^+)\,
 p_{\textrm{\textbf{{\rv{U}}}}}(\boldsymbol{U}^+)\nonumber\\
 \!\!\!\!\!\!\!\!\!\!\!\!&& \!\!\!\!\times\,\delta(\boldsymbol{X}^-\!-\boldsymbol{X}^+)  
 \, p_{\textrm{\textbf{{\rv{X}}}}}(\boldsymbol{X}^+).
\end{eqnarray}
The Bi-LMMSE module in Fig. \ref{fig:block-diagram} performs the following three key steps:
\begin{itemize}
    \item [$i)$] Bilinear approximate message passing between the variable nodes $\bm{U}^-$ and $\bm{X}^-$ and the factor node $p(\bm{Y}|\bm{U}^-,\bm{X}^-)$\footnote{Note that this step already encompasses the row-wise denoiser of $\bm{U}$ given in (\ref{eq:denoising-assignment-prior-integral_1}) and (\ref{eq:denoising-assignment-prior-integral_2}) and a column-wise denoiser of $\mathbf{X}$ under a Gaussian prior.}.
    \item [$ii)$] Denoising the rows of $\bm{U}^-$  according to (\ref{eq:denoising-assignment-prior-mean}) and (\ref{eq:denoising-assignment-prior-variance}) under non-uniform assignment priors whose parameters $\omega_{i,n}$ are computed from the extrinsic LLRs provided by the ``ext-BP'' block.
    \item [$iii)$] Denoising the columns $\bm{v}_j=\bm{A}\,\bm{x}_j^-$ of $\bm{V} = \bm{A}\,\bm{X}^-$ under Gaussian priors whose means and precisions are computed from the output of the ``ext-EP'' block.
\end{itemize}
In the sequel, we describe how the Bi-LMMSE module interacts with another column-wise denoiser of $\boldsymbol{U}^+$ so as to enforce the permutation structure on $\bm{U}$. The involved BP messages will be established in Section \ref{subsec:BP-denoiser-U}.
\subfile{algorithm2-new}
%
\subfile{block-diagram-with-transpose}
\subsection{Extrinsic information calculation through ext-BP and ext-EP blocks}

In Fig.~\ref{fig:block-diagram}, the computation of the outgoing extrinsic information messages from the Bi-LMMSE module to the outer denoisers of $\bm{X}^+$ and $\bm{U}^+$ relies on the following two message propagation strategies:
\begin{itemize}[leftmargin=*]
    \item Expectation propagation (EP) employed by the ``ext-EP'' block to find the extrinsic message \big($\boldsymbol{\widehat{X}}_{ \textsf{e}}^{+}$,$\gamma_{\boldsymbol{X}^{+}_{ \textsf{e}}}$\big) from the posterior message \big($\boldsymbol{\widehat{X}}_{ \textsf{p}}^{-}$,$\gamma_{\boldsymbol{X}^{-}_{ \textsf{p}}}$\big).
    \item Belief propagation (BP) employed by the ``ext-BP'' block to find the extrinsic LLR message $\text{LLR}_{\bm{{U}}^+,{\textsf{e}}}$ from the posterior LLR message $\text{LLR}_{\bm{{U}}^-,{\textsf{p}}}$.
\end{itemize}

\subsubsection{Extrinsic information update through the ext-EP block}
    The EP principle approximates the posterior pdf of $\boldsymbol{X}$ with a Gaussian distribution, thereby requiring the characterization of its second-order statistics only. To better explain this mechanism, consider the Gaussian belief $\mathcal{N}(\bm{x}_j^+; \widehat{\bm{x}}^+_{j,\mathsf{e}},\gamma_{\bm{\bm{X}}^+_{\mathsf{e}}}^{-1}\mathbf{I})$ on $\bm{x}_j^+$ that is sent by the Bi-LMMSE module to the denoiser of $\boldsymbol{X}^+$. Given some prior, $p_{\bm{\mathsf{x}}}(\bm{x}_j^+)$, the posterior mean or MMSE estimate, $ \widehat{\bm{x}}^{+}_{j,\mathsf{p}}$, of $\boldsymbol{x}_j^+$ and the associated  element-wise posterior precision, $\gamma_{\bm{\bm{X}}^{+}_{\mathsf{p}}}^{-1}$, are updated as follows:
    \begin{subequations}\label{posterior_x}
        \begin{align}
            \widehat{\bm{x}}^{+}_{j,\mathsf{p}} &=\mathbf{g}_{\mathsf{x}}(\widehat{\boldsymbol{x}}^{+}_{j,\mathsf{e}}, \gamma^{-1}_{\boldsymbol{X}^{+}_{ \mathsf{e}}})\triangleq\frac{\int \bm{x}_j\,p_{\bm{\mathsf{x}}}(\bm{x}_j)\,\mathcal{N}(\bm{x}_j; \widehat{\bm{x}}^+_{j,\mathsf{e}},\gamma_{\bm{\bm{X}}^+_{\mathsf{e}}}^{-1}\mathbf{I})\,\text{d}\bm{x}_j}{\int p_{\bm{\mathsf{x}}}(\bm{x}_j)\,\mathcal{N}(\bm{x}_j; \widehat{\bm{x}}^+_{j,\mathsf{e}},\gamma_{\bm{\bm{X}}^+_{\mathsf{e}}}^{-1}\mathbf{I})\,\text{d}\bm{x}_j}, \label{eq:turbo-mean}\\
            \gamma_{\bm{X}^{+}_{\mathsf{p}}}^{-1} &= \gamma_{\bm{X}^+_{\mathsf{e}}}\left(\frac{1}{M}\sum_{j=1}^M\big\langle \mathbf{g}^{\prime}_{\mathsf{x}}(\widehat{\boldsymbol{x}}^{+}_{j,\mathsf{e}}, \gamma^{-1}_{\bm{X}^{+}_{ \mathsf{e}}})\big\rangle\right)^{-1},\label{eq:turbo-precision}
        \end{align}
    \end{subequations}
    \vspace{-0.2cm}in which the divergence, $\mathbf{g_{\mathsf{x}}^{\prime}}(\cdot,\cdot)$, is given by:
    \begin{equation}\label{eq:element-wise-derivative}
      \big[\mathbf{g}_{\mathsf{x}}^{\prime}(\widehat{\bm{x}} , \gamma^{-1}_{\boldsymbol{X}})\big]_{\ell} ~ =~ \frac{\partial\big[\mathbf{g}_{\mathsf{x}}( \widehat{\bm{x}}, \gamma^{-1}_{\boldsymbol{X}})\big]_{\ell}}{\partial\big[ \widehat{\bm{x}}\big]_{\ell}},~\ell=1,\ldots,R.
    \end{equation}
    Note here that for ease of notation the iteration index, $t$, was dropped in (\ref{posterior_x}). The adequate scheduling of messages can be found in Algorithm \ref{algo:big-vamp}. The extrinsic mean and precision of the Gaussian belief $\mathcal{N}(\bm{x}_j^-; \widehat{\bm{x}}^-_{j,\mathsf{e}},\gamma_{\bm{\bm{X}}^-_{\mathsf{e}}}^{-1}\mathbf{I})$ that is fed back to the Bi-LMMSE module are then computed as follows:
    \begin{subequations}\label{eq:extrinsic-computation}
        \begin{align}
        \gamma_{\bm{X}^{-}_{\mathsf{e}}} &\,\,= \,\, \gamma_{\bm{X}^{+}_{\mathsf{p}}} - \gamma_{\bm{X}^+_{\mathsf{e}}},\label{eq:extrinsic-precision}\\
        \widehat{\bm{x}}^{-}_{j,\mathsf{e}}  &\,\,=\,\, \gamma^{-1}_{\bm{X}^{-}_{\mathsf{e}}}\left(\gamma_{\bm{X}^{+}_{\mathsf{p}}} \,\widehat{\bm{x}}^{+}_{j,\mathsf{p}}- \gamma_{\bm{X}^+_{\mathsf{e}}}\,\widehat{\bm{x}}^{+}_{j,\mathsf{e}}\right). \label{eq:extrinsic-mean}
        \end{align}
    \end{subequations}
    
    \noindent Such extrinsic information calculation, updated in lines \ref{eq:algo-diag-sigma-v-e}--\ref{eq:algo-diag-mean-v} and \ref{algo:eq-extrinsic-precision}--\ref{eq:algo-mean-Ve+} of Algorithm~\ref{algo:big-vamp}, is performed by the two ext-EP blocks between the Bi-LMMSE module and the denoiser of $\boldsymbol{X}$ in Fig.~\ref{fig:block-diagram}.
    \subsubsection{Extrinsic information update through the ext-BP block}
    Unlike EP which approximates the posterior distribution by  Gaussian beliefs, BP uses the posterior mean of $\boldsymbol{U}^-$ in closed-form without any approximations. In the binary case, this boils down to finding the element-wise posterior and extrinsic LLR messages $\text{LLR}_{\bm{{U}}^{-}\hspace{-0.06cm},{\textsf{p}}}$, $\text{LLR}_{\bm{{U}}^{+}\hspace{-0.06cm},{\textsf{p}}}$, $\text{LLR}_{\bm{{U}}^{-}\hspace{-0.06cm},{\textsf{e}}}$ and $\text{LLR}_{\bm{{U}}^{+}\hspace{-0.06cm},{\textsf{e}}}$ that are exchanged between the column-wise denoiser of $\boldsymbol{U}^+$ and the Bi-LMMSE module as depicted in Fig.~\ref{fig:block-diagram}. To find the extrinsic LLR message $\text{LLR}_{\bm{{U}}^{-}\hspace{-0.06cm},{\textsf{e}}}$ from the posterior LLR message $\text{LLR}_{\bm{{U}}^{+}\hspace{-0.06cm},{\textsf{p}}}$, we first compute the posterior mean $\widehat{u}^+_{ik,\mathsf{p}}$ of $\mathsf{u}^+_{ik}\,\forall i,k$ as follows:
    
    \begin{equation}\label{eq:posterior-BP}
    \begin{aligned}[b]
    \widehat{u}^+_{ik,\mathsf{p}} &= \frac{ \sum\limits_{u_{ik} \in \{0,1\}} u_{ik}\,\, p_{\mathsf{u}^+_{ik},\mathsf{e}}(u_{ik}) \,\,p_{\textsf{u}_{ik}^{-},\mathsf{e}}(u_{ik})}{ \sum\limits_{u_{ik} \in \{0,1\}}  p_{\mathsf{u}^+_{ik},\mathsf{e}}(u_{ik}) \,\,p_{\textsf{u}_{ik}^{-},\mathsf{e}}(u_{ik})},\\
    &=  \frac{ p_{\mathsf{u}^+_{ik},\mathsf{e}}(1) \,\,p_{\textsf{u}_{ik}^{-},\mathsf{e}}(1)}{p_{\mathsf{u}^+_{ik},\mathsf{e}}(1) \,\,p_{\textsf{u}_{ik}^{-},\mathsf{e}}(1) + p_{\mathsf{u}^+_{ik},\mathsf{e}}(0) \,\,p_{\textsf{u}_{ik}^{-},\mathsf{e}}(0)},\\
    & = \frac{ \widehat{u}^+_{ik, \mathsf{e}} \, \widehat{u}^-_{ik, \mathsf{e}}}{ \widehat{u}^+_{ik, \mathsf{e}} \, \widehat{u}^-_{{ik},\mathsf{e}} + (1-\widehat{u}^+_{ik, \mathsf{e}}) \, (1-\widehat{u}^-_{ik, \mathsf{e}})}.
    \end{aligned}
    \end{equation}
\noindent Solving for $\widehat{u}_{ik,\textsf{e}}^{-}$ in (\ref{eq:posterior-BP}) yields the following expression of the extrinsic mean of $\mathsf{u}^-_{ik}$:
\begin{equation}\label{eq:extrinsic-BP}
\begin{aligned}[b]
     \widehat{u}_{ik,\textsf{e}}^{-} &= \frac{\widehat{u}^+_{ik,\mathsf{p}}\,\,(1-\widehat{u}^+_{ik, \mathsf{e}})}{\widehat{u}_{ik, \textsf{p}}^{+}\,(1-\widehat{u}_{ik, \textsf{e}}^{+})+\widehat{u}_{ik, \textsf{e}}^{+}\,(1-\widehat{u}_{ik, \textsf{p}}^{+})}.
\end{aligned}
\end{equation}
Using (\ref{eq:LR}), it follows that (\ref{eq:extrinsic-BP}) can be rewritten as:
\begin{equation}\label{eq:LR-extrinsic-plus-division}
    \textrm{LLR}_{{u}_{ik}^-, \textsf{e}} = \textrm{LLR}_{{u}^{+}_{ik}, \textsf{p}} -\,\, \text{LLR}_{{u}^+_{ik},\textsf{e}}.
\end{equation}

\noindent The extrinsic LLR value, $\textrm{LLR}_{{u}^-_{ik},\textsf{e}}$ in (\ref{eq:LR-extrinsic-plus-division}), is calculated by the ext-BP block connected to the Bi-LMMSE module shown in Fig.~\ref{fig:block-diagram} and is updated in matrix-form in line  \ref{algo:eq-LRe-from-LRp-minus} of Algorithm~\ref{algo:big-vamp}.

\subsection{BP messages of the column-wise denoiser of $\bm{U}$}\label{subsec:BP-denoiser-U}
 
\noindent Recall that the posterior matrix $\widehat{\bm{U}}^-_{\textsf{p}}$ is reconstructed (inside the Bi-LMMSE module)  using the denoising function $\mathbf{g}_{\mathsf{u}}(\cdot,\cdot,\cdot)$ in (\ref{eq:denoising-assignment-prior-mean}) under a row-wise assignment prior and as such is not necessarily a permutation matrix. To enforce a permutation prior, we couple $\mathbf{g}_{\mathsf{u}}(\cdot,\cdot,\cdot)$ with an outer denoiser, $\mathbf{f}_{\mathsf{u}}(\cdot,\cdot,\cdot)$, that is applied on the columns of $\boldsymbol{U}^+$  using the  uniform assignment prior given in (\ref{eq:assignment-prior-column}). The column-wise denoiser, $\mathbf{f}_{\mathsf{u}}(\cdot,\cdot,\cdot)$, depicted by the green module in Fig.~\ref{fig:block-diagram} provides a second belief about the probability that the $k$th column $\bm{\mathsf{u}}_k^+$ in $\bm{\mathsf{U}}^+$
is the $k$th canonical basis vector $\boldsymbol{e}_{k}\in \mathbb{R}^{N}$. This belief is fed back in the form of LLR values ($\text{LLR}_{\boldsymbol{{U}}^{+}\hspace{-0.06cm},{\textsf{e}}}$ in Fig.~\ref{fig:block-diagram}) to the Bi-LMMSE module. Each extrinsic message,  $\text{LLR}_{{u}^+_{ik},\textsf{e}}$, sets an independent prior on the  $i$th component of the column vector,  $\bm{\mathsf{u}}^+_k$, in $\bm{\mathsf{U}}^+$ being equal  to $0$ or $1$:
\begin{equation}\label{elementary_extrinsic_prior}
    \mathbb{P}\big({{\mathsf{u}}}_{ik}^+=1\big) \,=\,\widehat{u}_{ik, \mathsf{e}}^+~~~\textrm{and} ~~~~\mathbb{P}\big({{\mathsf{u}}}_{ik}^+=0\big) \,=\,1-\widehat{u}_{ik, \mathsf{e}}^+,
\end{equation}
where the extrinsic mean, $\widehat{u}^{+}_{ik,\textsf{e}}$, is computed from $\text{LLR}_{{u}^+_{ik},\textsf{e}}$, using (\ref{eq:mean-function-LLR}). The result in (\ref{elementary_extrinsic_prior}) sets an overall probability measure on the entire vector $\bm{\mathsf{u}}^+_k$, being equal to each binary vector of size $N$. More specifically, we let ${\mathcal{C}}$ be the set of all binary vectors of size $N$. We also denote the support of  each $j$th vector $\mathbf{c}_j\in \mathcal{C}$ by the subset $E_j\subseteq E=\{1,2,\ldots,N\}$  (i.e., the set of positions of all nonzero entries in $\mathbf{c}_j$):
\begin{equation}
   E_j=\textrm{Supp}(\mathbf{c}_j)=\{i:[\mathbf{c}_j]_i = 1\}, ~ j=1,2,\ldots,2^N. 
\end{equation}
It then follows that the Bi-LMMSE module provides independent extrinsic information about the likelihood of each $\bm{\mathsf{u}}^+_k$ being equal to any of the binary vectors in $\mathcal{C}$: 
\begin{equation}\label{eq:message-prior-biAMP}
\begin{aligned}[b] 
P_{\bm{\mathsf{u}}_k^+,\mathsf{e}}(\mathbf{c}_j)
&= \prod_{i\in E_j}\mathbb{P}\big({{\mathsf{u}}}_{ik}^+=1\big) \, \prod_{i\in E_j^c}\mathbb{P}\big({{\mathsf{u}}}_{ik}^+=0\big)\\
    &= \prod_{i\in E_j}\widehat{u}_{ik, \mathsf{e}}^+\,\prod_{i\in E_j^c} \big(1-\widehat{u}^+_{ik, \mathsf{e}}\big).
    \end{aligned} 
\end{equation}
\noindent Here, $E_j^c$ is the complement of the subset $E_j$ inside the set $E = \{1,2,\ldots,N\}$. To enforce the assignment prior on, $\bm{\mathsf{u}}_k^+$, the column-wise denoiser,   $\mathbf{f}_{\mathsf{u}}(\cdot,\cdot,\cdot)$, of $\bm{\mathsf{U}}^+$ combines the extrinsic prior in (\ref{eq:message-prior-biAMP}) with\footnote{Recall here that $\mathcal{B}$ is the set of canonical basis vectors in $\mathbb{R}^N$.}: 
\begin{equation} \label{eq:assignment-prior-U-transpose-element-wise}
P_{\bm{\mathsf{u}}_k^+}(\boldsymbol{c}_j) \,=\,
\begin{cases}
    \frac{1}{N}, &~~\textrm{for}~\mathbf{c}_j\in\mathcal{B},\\
    0,&~~\textrm{for}~\mathbf{c}_j\in\mathcal{C}\backslash\mathcal{B},  
\end{cases}
\end{equation}
 to compute the posterior mean, $\widehat{\bm{u}}_{k, \textsf{p}}^{+}$, as follows:
 \begin{equation}\label{eq:posterior-U-transpose-intermediate}
\begin{aligned}[b]    \widehat{\bm{u}}^{+}_{k,\textsf{p}}&~=~\frac{\sum\limits_{j=1}^{2^N} \mathbf{c}_j\, P_{\bm{\mathsf{u}}_k^+}(\mathbf{c}_j) \,P_{\bm{\mathsf{u}}_{k}^{+},\mathsf{e}}(\mathbf{c}_j)}{\sum\limits_{j=1}^{2^N} P_{\bm{\mathsf{u}}_k^+}(\mathbf{c}_j) \,P_{\bm{\mathsf{u}}_{k}^{+},\mathsf{e}}(\mathbf{c}_j)}
   ~=~\frac{\sum\limits_{i=1}^{N} \mathbf{e}_{i}\, \,P_{\bm{\mathsf{u}}_{k}^{+},\mathsf{e}}(\mathbf{e}_{i})}{\sum\limits_{i=1}^{N} \,P_{\bm{\mathsf{u}}_{k}^{+},\mathsf{e}}(\mathbf{e}_{i})}.
    \end{aligned}
\end{equation}
Owing to (\ref{eq:message-prior-biAMP}), we have:
\begin{equation}\label{eq:message-prior-biAMP-final}
\begin{aligned}[b] 
P_{\bm{\mathsf{u}}_k^+,\mathsf{e}}(\mathbf{e}_i)
    &~=~ \widehat{u}_{i k, \mathsf{e}}^+\,\prod_{\substack{i'=1\\i' \neq i}}^{N} \big(1-\widehat{u}^+_{i'k, \mathsf{e}}\big),
    \end{aligned} 
\end{equation} 
from which we obtain the $i$th component of (\ref{eq:posterior-U-transpose-intermediate}) as
\begin{equation}\label{eq:posterior-U-transpose-intermediate2}
\begin{aligned}[b]    \widehat{u}^{+}_{ik,\textsf{p}}&= \frac{\widehat{u}_{ik, \mathsf{e}}^+\,\prod_{\substack{i'=1\\i' \neq i}}^{N} \big(1-\widehat{u}^+_{i'k, \mathsf{e}}\big)}{\sum\limits_{i = 1}^N\, \widehat{u}_{i k, \mathsf{e}}^+\,\prod_{\substack{i'=1\\i' \neq i}}^{N} \big(1-\widehat{u}^+_{i' k, \mathsf{e}}\big)}.
    \end{aligned}
\end{equation}

\noindent After dividing both the numerator and the denominator of  (\ref{eq:posterior-U-transpose-intermediate2}) by $\prod\limits_{i=1}^{N}(1-\widehat{u}^+_{ik, \mathsf{e}})$, it can be shown that:\vspace{-0.2cm}
\begin{equation}\label{eq:posterior-U-transpose-final}
    \begin{aligned}
    \widehat{u}^{+}_{ik,\textsf{p}} =\frac{1}{1+\sum\limits_{\substack{k'=1\\k' \neq k}}^{N}\exp\big(\textrm{LLR}_{{u}^{+}_{ik'},\textsf{e}}-\,\textrm{LLR}_{{u}^{+}_{ik},\textsf{e}}\big)}
    \end{aligned}
\end{equation}

\noindent Now, injecting (\ref{eq:posterior-U-transpose-final}) back into the computation performed by the ext-BP block in (\ref{eq:extrinsic-BP}) yields the extrinsic LLR value:

\begin{equation}\label{eq:LR-extrinsic-minus-division}    \textrm{LLR}_{{u}^-_{ik},\textsf{e}} = \textrm{LLR}_{{u}^{+}_{ik},\textsf{p}}-\,\,\textrm{LLR}_{{{u}^+_{ik}},\textsf{e}}.
\end{equation}

\noindent The extrinsic LLR value, $\textrm{LLR}_{{u}^-_{ik},\textsf{e}}$ in  (\ref{eq:LR-extrinsic-minus-division}) is calculated by the ext-BP block connected to the prior module of $\bm{U}^+$ shown in Fig.~\ref{fig:block-diagram} and is updated in matrix-form in line  \ref{algo:eq-LRe-from-LRp-plus} of Algorithm~\ref{algo:big-vamp}. Note how Eq. (\ref{eq:LR-extrinsic-minus-division}) relies on the posterior LLR value $\textrm{LLR}_{{u}^{+}_{ik}, \textsf{p}}$ to establish the relationship between the extrinsic LLR values $\textrm{LLR}_{{u}_{ik}^-, \textsf{e}}$ and $\text{LLR}_{{u}^+_{ik},\textsf{e}}$. However, it is also possible to obtain the explicit relationship between the extrinsic LLRs by using (\ref{eq:LR-extrinsic-minus-division}) and injecting (\ref{eq:posterior-U-transpose-final}) back into (\ref{eq:extrinsic-BP}) to get:
\begin{equation}\label{eq:LR-extrinsic-plus}
    \textrm{LLR}_{{u}^{-}_{ik},\textsf{e}} = -\ln\sum\limits_{\substack{k'=1\\k' \neq k}}^{N} \exp\left(\textrm{LLR}_{{u}^{+}_{ik'},\textsf{e}}\right).
\end{equation}

\noindent We have described the BP computation involved in the column-wise denoiser of $\bm{U}^+$ from the incoming extrinsic message $\text{LLR}_{\bm{{U}}^{-}\hspace{-0.06cm},{\textsf{e}}}$ to produce the outgoing extrinsic messages $\text{LLR}_{\bm{{U}}^{+}\hspace{-0.06cm},{\textsf{e}}}$. Now, we show how the latter is incorporated into the Bi-LMMSE module.

\subsection{Integrating the $\text{LLR}_{\bm{{U}}^{+}\hspace{-0.06cm},{\textsf{e}}}$ values into the Bi-LMMSE module}\label{subsec:EP-on-U}

We reviewed in Section \hyperref[subsubsec:bivamp-with-one-assignment]{II-B.2} the closed-form denoising step (\ref{eq:denoising-assignment-prior-mean}) and (\ref{eq:denoising-assignment-prior-variance}) of $\bm{U}$ in the BiVAMP's Bi-LMMSE module based on the row-wise assignment prior in (\ref{eq:assignment-prior}) only. In UCS, however, the Bi-LMMSE module combines the message $\text{LLR}_{\bm{{U}}^{-}\hspace{-0.06cm},{\textsf{e}}}$ established in (\ref{eq:LR-extrinsic-plus}) with the row-wise prior given in (\ref{eq:assignment-prior}). Each extrinsic LLR message,  $\text{LLR}_{{u}^-_{ik},\textsf{e}}$, provides an independent prior on the  $k$th component of the row vector,  $(\bm{\mathsf{u}}^+_i)^\top$. By adopting the equivalent argument about $u^+_{ik}$ from (\ref{elementary_extrinsic_prior}) to (\ref{eq:message-prior-biAMP-final}) and applying it to $u^-_{ik}$, it follows that:
\begin{equation}\label{eq:message-prior-biAMP-row-wise}
\begin{aligned}[b] 
P_{\bm{\mathsf{u}}_k^-,\mathsf{e}}(\mathbf{e}_i)
    &~=~ \widehat{u}_{i k, \mathsf{e}}^-\,\prod_{\substack{i'=1\\i' \neq i}}^{N} \big(1-\widehat{u}^-_{i'k, \mathsf{e}}\big),
    \end{aligned} 
\end{equation} 

\noindent The row-wise denoiser of $\boldsymbol{U}^-$ in the Bi-LMMSE module combines the incoming prior $P_{\bm{\mathsf{u}}_k^-,\mathsf{e}}(\cdot)$ from the column-wise module given in (\ref{eq:message-prior-biAMP-row-wise}) with its row-wise prior $p_{\bm{\mathsf{u}}_i^-}(\cdot)$ given in (\ref{eq:assignment-prior}). Injecting the two independent row-wise priors (\ref{eq:message-prior-biAMP-row-wise}) and (\ref{eq:assignment-prior}) back into (\ref{eq:denoising-assignment-prior-integral_2}) yields the expression of the $ik$th entry of the posterior mean of $\boldsymbol{U}^-$:
\begin{equation}
\label{eq:posterior-mean-U-intermediate1}
\normalsize{
\begin{aligned}
    \widehat{u}^{-}_{ik,\textsf{p}} = \frac{\widehat{u}^{-}_{ik,\textsf{e}}\,\prod\limits_{\substack{i'=1\\i' \neq i}}^{N}  \left(1-\widehat{u}^{-}_{i'k,\textsf{e}}\right)\,\exp \left([\bm{b}_{\bm{u}_i}]_k-\frac{1}{2}\,[\mathbf{\Lambda}_{\bm{U}}]_{kk}\right)}{\sum\limits_{k=1}^{N}\widehat{u}^{-}_{ik,\textsf{e}}\,\prod\limits_{\substack{i'=1\\i \neq i}}^{N}  \left(1-\widehat{u}^{-}_{i'k,\textsf{e}}\right)\, \exp\left([\bm{b}_{\bm{u}_i}]_k-\frac{1}{2}\,[\mathbf{\Lambda}_{\bm{U}}]_{kk}\right)}.
\end{aligned}
}
\end{equation}

\noindent Upon division of the numerator and denominator in  (\ref{eq:posterior-mean-U-intermediate1}) by $\prod\limits_{i'=1}^{N} \left(1-\widehat{u}^-_{i'k, \mathsf{e}}\right)$, the posterior mean and variance of $\mathsf{u}^-_{ik}$ becomes:
\begin{subequations} \label{eq:posterior-mean-variance-U-final}
    \begin{align}
        \widehat{u}^-_{ik, \mathsf{p}} &=\frac{\exp \left([\bm{b}_{\bm{u}_i}]_k-\frac{1}{2}\,\mathbf{[\Lambda}_{\bm{U}}]_{kk} + \textrm{LLR}_{{u}^{-}_{ik},\textsf{e}}\right)}{\sum\limits_{k=1}^{N}\exp \left([\boldsymbol{b}_{\bm{u}_i}]_k-\frac{1}{2}\,[\mathbf{\Lambda}_{\bm{U}}]_{kk} + \textrm{LLR}_{{u}^{-}_{ik},\textsf{e}}\right)}\nonumber\\
        &\triangleq\ h_{\mathsf{u}_{ik}}\left(\boldsymbol{b}_{\bm{{u}}_i}, \mathbf{\Lambda}_{\bm{U}},\text{LLR}_{\bm{{U}}^{-}\hspace{-0.06cm},{\textsf{e}}}\right), \label{eq:posterior-mean-U-final}\\
\sigma^{2}_{\mathsf{u}_{ik, \textsf{p}}^{-}} &= \widehat{u}^-_{ik, \mathsf{p}} - \left(\widehat{u}^-_{ik, \mathsf{p}}\right)^2\nonumber\\
&= \frac{\partial h_{{\mathsf{u}_{ik}}}\left(\boldsymbol{b}_{\bm{{u}}_i},\mathbf{\Lambda}_{\bm{U}},\text{LLR}_{\bm{{U}}^{-}\hspace{-0.06cm},{\textsf{e}}}\right)}{\partial [\boldsymbol{b}_{\bm{{u}}_i}]_k}\nonumber\\
&\triangleq\ h_{\mathsf{u}_{ik}}^\prime\left(\boldsymbol{b}_{\bm{{u}}_i}, \mathbf{\Lambda}_{\bm{U}},\text{LLR}_{\bm{{U}}^{-}\hspace{-0.06cm},{\textsf{e}}}\right).\label{eq:posterior-variance-U-final}
    \end{align}
\end{subequations}

\noindent Clearly, the posterior mean in (\ref{eq:posterior-mean-U-final}) is a tractable estimate of the $ik$th entry of $\bm{U}^-$ that overcomes the use of the untractable prior in (\ref{eq:untractable-permutation-prior}). By shifting each term $\left([\bm{b}_{\bm{u}_i}]_k -\frac{1}{2}\,[\mathbf{\Lambda}_{\bm{U}}]_{kk}\right)$ by a value of $\textrm{LLR}_{{u}^{-}_{il},\textsf{e}}$ stemming from the column-wise denoiser of $\bm{U}^+$, both column- and row-wise assignment priors on $\bm{U}$ are enforced. The update expressions (\ref{eq:posterior-mean-U-final}) and (\ref{eq:posterior-variance-U-final}) correspond to the updates in lines \ref{algo:eq-Up-minus-with-LRs} and \ref{algo:eq-varUp-minus-with-LRs} of Algorithm \ref{algo:big-vamp}, respectively. It is worth noting the relationship between our two-stage permutation denoiser $h_{\mathsf{u}_{ik}}\left(\boldsymbol{b}_{\bm{{u}}_i}, \mathbf{\Lambda}_{\bm{U}},\text{LLR}_{\bm{{U}}^{-}\hspace{-0.06cm},{\textsf{e}}}\right)$ obtained in (\ref{eq:posterior-mean-U-final}) and the row-wise assignment denoiser $\left[\mathbf{g}_{\mathsf{u}}(\boldsymbol{b}_{\bm{{u}}_i}, \mathbf{\Lambda}_{\bm{U}})\right]_k$ in (\ref{eq:denoising-assignment-prior-mean}). By setting each weight $\omega_{i,k} \propto \exp\left(\textrm{LLR}_{{u}^{-}_{ik},\textsf{e}}\right)$ in (\ref{eq:denoising-assignment-prior-mean}), one recovers (\ref{eq:posterior-mean-U-final}). In other words, the two-stage permutation denoiser employs the LLR values ensuing from the BP algorithm as weighting factors instead of the uniform weights (i.e., $\omega_{i,k} = 1/N$) used in the row-wise assignment denoiser.

Owing to the computation performed by the ext-BP block to get $\widehat{u}^+_{ik, \mathsf{e}}$ (instead of $\widehat{u}^-_{ik, \mathsf{e}}$ as in (\ref{eq:extrinsic-BP})), we write the extrinsic message of $\mathsf{u}_{ik}^+$ as follows:
\begin{equation}
\label{eq:extrinsic-U-final}
\normalsize{
\begin{aligned}
\widehat{u}^{+}_{ik,\textsf{e}} = \frac{ \widehat{u}^{-}_{ik,\textsf{p}}\,\left(1-\widehat{u}^{-}_{ik,\textsf{e}}\right)}{
        \widehat{u}^{-}_{ik,\textsf{e}}\left(1-\widehat{u}^{-}_{ik,\textsf{p}}\right) + \widehat{u}^{-}_{ik,\textsf{p}}\left(1-\widehat{u}^{-}_{ik,\textsf{e}}\right)},
\end{aligned}
}
\end{equation}
which can be written using (\ref{eq:LR}) as:
\begin{equation}\label{eq:LR-extrinsic-minus-division-new}    \textrm{LLR}_{{u}^+_{ik},\textsf{e}} = \textrm{LLR}_{{u}^{-}_{ik},\textsf{p}}-\,\,\textrm{LLR}_{{{u}^-_{ik}},\textsf{e}}.
\end{equation}
\noindent The extrinsic LLR value, $\textrm{LLR}_{{u}^+_{ik},\textsf{e}}$, in (\ref{eq:LR-extrinsic-minus-division-new}) is provided by the Bi-LMMSE module to the column-wise denoiser of $\boldsymbol{U}^+$ and corresponds to the matrix-form update in line  \ref{algo:eq-LRe-from-LRp-plus} of Algorithm~\ref{algo:big-vamp}.\\
\noindent Finally, it is also possible to obtain the relationship between extrinsic LLRs without the use of posterior LLRs. By injecting (\ref{eq:posterior-mean-U-final}) back into (\ref{eq:extrinsic-U-final}) and using (\ref{eq:LR-extrinsic-minus-division-new}), we get:
\begin{equation}\label{eq:LR-extrinsic-minus}
    \begin{aligned}[b]
    \textrm{LLR}_{{u}^{+}_{ik},\textsf{e}} &= [\bm{b}_{\bm{u}_i}]_k-\frac{1}{2}\,[\mathbf{\Lambda}_{\bm{U}}]_{kk}\\
    &~~~-\,\ln \sum\limits_{\substack{i'=1\\i' \neq i}}^{N} \exp\left([\bm{b}_{\bm{u}_i}]_k-\frac{1}{2}\,[\mathbf{\Lambda}_{\bm{U}}]_{kk}+\textrm{LLR}_{{u}^{-}_{i'k},\textsf{e}}\right).
    \end{aligned}
\end{equation}
\vspace{-0.4cm}
\section{State Evolution}\label{sec:SE}
An appealing property of AMP-like algorithms such as UCS is that its large system limit behavior can be characterized through the so-called state evolution (SE) analysis. In this section, we first highlight the key differences between the SE equations of UCS and Bi-VAMP to put our contribution in a proper perspective. We then derive the SE equations of UCS and provide their associated pseudo-code.

\subsection{Difference with the state evolution of Bi-VAMP}\label{subsec:diff-SE-UCS-BiVAMP}
The SE equations of Bi-VAMP have been fully characterized in \cite{akrout2020bilinear} under the assumption that both $\bm{U}$ and $\bm{V}$ have entry-wise separable priors in the observation model (\ref{eq:bilinear-recovery}). As depicted in Fig. \ref{fig:Bi-VAMP-SE-bloc-diagram}, this characterization boils down to the derivation of the following four MSE functions:
\begin{itemize}
    \item $\mathcal{E}_{u^-}(\cdot)$ and $\mathcal{E}_{v^-}(\cdot)$ pertaining to the LMMSE estimators of the matrices $\bm{U}^-$ and $\bm{V}^-$,
    \item $\mathcal{E}_{u^+}(\cdot)$ and $\mathcal{E}_{v^+}(\cdot)$ pertaining to the MMMSE estimators of the matrices $\bm{U}^+$ and $\bm{V}^+$.
\end{itemize}

\begin{figure*}[!t]
    \centering
    \subfloat[Bi-VAMP SE with i.i.d. priors on $\bm{U}$ and $\bm{V}$.]{
    \input{block-diagram-SE-bivamp.tex}\label{fig:Bi-VAMP-SE-bloc-diagram}}%
    \qquad
    \subfloat[UCS SE with row-wise and column-wise assignment priors on $\bm{U}$ and an i.i.d. prior on $\bm{X}$.]{
    \input{block-diagram-SE-permutation.tex}\label{fig:UCS-SE-bloc-diagram}}%
    \caption{State evolution diagram of (a) Bi-VAMP where only an assignment prior on $\bm{U}^-$ is enforced, and (b) UCS where a row-wise prior on $\bm{U}^-$ and a column-wise prior on $\bm{U}^+$ are enforced. The color of each MSE function in (b) matches the color of the corresponding module in the block diagram in Fig.~\ref{fig:block-diagram} in which it is involved.}
    \label{fig:SE-bloc-diagrams}
    \vspace{-0.2cm}
\end{figure*}

\noindent Because UCS extends the Bi-VAMP algorithm \cite{akrout2020bilinear} to account for the known sensing matrix $\bm{A}$ and the two-stage denoising step of $\bm{U}$, its SE equations should be derived anew. To accommodate those key algorithmic changes between UCS and Bi-VAMP, the following modifications in the derivation of the LLR and MSE functions should be implemented as shown in Fig.~\ref{fig:UCS-SE-bloc-diagram}:
\begin{itemize}
    \item[1)] $\mathcal{E}_{x^-}(\cdot)$ and $\mathcal{E}_{x^+}(\cdot)$ should incorporate the known sensing matrix $\bm{A}$ and the fact that $\bm{U}$ is a permutation matrix in the estimation steps of $\bm{X}$.
    \item[2)] $\mathcal{E}_{u^-}(\cdot)$ pertaining to the LMMSE estimator of $\bm{U}$ has to account for the row-wise assignment prior on $\bm{U}^-$,
    \item[3)] {$\mathcal{L}_{u^+}(\cdot)$ and $\mathcal{L}_{u^-}(\cdot)$ account for the exchanged LLR values between the column-wise MMSE estimator of $\bm{U}^+$ and the row-wise MMSE estimator of $\bm{U}^-$.}
\end{itemize}

\noindent Next, we state the assumptions of the SE analysis when $\bm{U}$ is a permutation matrix and $\bm{X}$ has an entry-wise i.i.d. prior.

\subsection{State evolution assumptions}\label{sec:SE-assumptions}
The asymptotic regime analysis of (\ref{eq:bilinear-recovery}) refers to the case where $ R, N, M  \rightarrow +\infty$ with $\frac{R}{N} \rightarrow \beta_{u} = \mathcal{O}(1)$ and  $\frac{R}{M} \rightarrow \beta_{x} = \mathcal{O}(1)$ for some fixed ratios $\beta_{u}\leq 1$ and $\beta_{x}\leq 1$. More specifically, when it comes to the observation model of UCS in (\ref{eq:unlabeled-sensing}) with $N=R$, the asymptotic regime analysis corresponds to the special case $\beta_{u} = 1$ and  $\frac{N}{M} \rightarrow \beta_{x} \leq 1$.
We first start by considering the following approximations:
\begin{equation}\label{eq:var_y}
    \langle\boldsymbol{Y} \odot\boldsymbol{Y}\rangle~\approx~\bar{\sigma}_u^2\,\bar{\sigma}_A^2\,\bar{\sigma}_x^2\,N\,R\,+\,\sqrt{MN}\,\gamma_w^{-1},\vspace{-0.1cm}
\end{equation}
where
\begin{subequations}
    \begin{align}
    \sigma_u^2\triangleq\langle \bm{U}\odot\bm{U} \rangle\,&\approx\,\bar{\sigma}_u^2\,\triangleq\,\mathbb{E}[\mathsf{u}^2_{i,\ell}|p_{\mathsf{u}}(u)]~\forall i,\ell,\\
    \sigma_A^2\triangleq\langle  \bm{A}\odot\bm{A}\rangle\,&\approx\,\bar{\sigma}_A^2\,\triangleq\,\mathbb{E}[\mathsf{a}^2_{k\ell}|p_{\mathsf{a}}(a)] ~\forall k,\ell,\\
    \sigma_x^2\triangleq\langle \bm{X}\odot\bm{X} \rangle\,&\approx\,\bar{\sigma}_x^2\,\triangleq\,\mathbb{E}[\mathsf{x}^2_{j,\ell}|p_{\mathsf{x}}(x)]~\forall j,\ell.\vspace{-0.2cm}
    \end{align}
\end{subequations}
\noindent In approximate message passing practices, a state evolution ansatz is based on the following concentration of measure for the precision and LLR variables in the asymptotic regime:
\begin{subequations}\label{eq:asymptotic-convergent-precisions}
\begin{align}
\lim _{M \rightarrow \infty}\left(\gamma_{\boldsymbol{X}_{ \mathsf{p}}^{+}}, \gamma_{\boldsymbol{X}_{ \mathsf{e}}^{+}}\right)&=\left(\bar\gamma_{\boldsymbol{X}_{ \mathsf{p}}^{+}}, \bar\gamma_{\boldsymbol{X}_{ \mathsf{e}}^{+}}\right), \\
\lim _{M  \rightarrow \infty}\left(\gamma_{\boldsymbol{X}_{ \mathsf{p}}^{-}}, \gamma_{\boldsymbol{X}_{ \mathsf{e}}^{-}}\right)&=\left(\bar\gamma_{\boldsymbol{X}_{ \mathsf{p}}^{-}}, \bar\gamma_{\boldsymbol{X}_{ \mathsf{e}}^{-}}\right), \\
\lim _{N \rightarrow \infty}\left(\gamma_{\boldsymbol{U}_{ \mathsf{p}}^{-}}, \gamma_{\boldsymbol{U}_{ \mathsf{e}}^{-}}\right)&=\left(\bar\gamma_{\boldsymbol{U}_{ \mathsf{p}}^{-}}, \bar\gamma_{\boldsymbol{U}_{ \mathsf{e}}^{-}}\right), \\
\scriptsize{\forall\,k,i,}\, \lim _{N \rightarrow \infty}\left(\textrm{LLR}_{u^{+}_{ik},\textsf{e}},\textrm{LLR}_{u^{-}_{ik},\textsf{e}}\right) &=  \Big(\overbar{\textrm{LLR}}^+_{ik},\overbar{\textrm{LLR}}^-_{ik}\Big).\label{eq:convergent-LLR-definition}
\end{align}
\end{subequations}

\noindent Note here that at convergence, one must have equality between the extrinsic means $\widehat{\bm{\mathsf{u}}}_{i,\mathsf{e}}^+ = \widehat{\bm{\mathsf{u}}}_{i,\mathsf{e}}^- \triangleq\widehat{\bm{\mathsf{u}}}_{i,\mathsf{e}}$. Moreover, we assume that the extrinsic mean $\widehat{\bm{u}}_{i,\mathsf{e}}$ is an AWGN-corrupted observation of the ``true'' random vector $\bm{\mathsf{u}}_i^*$, i.e., $\widehat{\bm{u}}_{i,\mathsf{e}}$ is a realization of the random vector\footnote{This is a common assumption in approximate message passing practices, e.g., see AMP \cite{donoho2011design}, VAMP \cite{rangan2019vector}, and BiG-AMP \cite{parker2014bilinear} algorithms.}:
\begin{equation}\label{eq:approx-b}
\begin{aligned}[b]
    \widehat{\bm{\mathsf{u}}}_{i,\mathsf{e}} &= \bm{\mathsf{u}}_i^* + \bm{\mathsf{w}}_{\bm{\mathsf{u}}_i},\hspace{1.92cm} \bm{\mathsf{w}}_{\mathsf{u}_i}\sim\mathcal{N}\left(\bm{w};\bm{0},\bm{\Lambda}_{\bm{U}}^{-1}\right),\\
    & = \bm{\Lambda}_{\bm{U}}^{-1}\,\bm{\mathsf{b}}^*_{\bm{{u}}_i} + \bm{\Lambda}_{\bm{{U}}}^{-\frac{1}{2}} \,\bm{\mathsf{n}},\hspace{1cm} \bm{\mathsf{n}}\sim\mathcal{N}\left(\bm{n};\bm{0},\mathbf{I}_N\right),
    \end{aligned}
\end{equation}

\noindent where the last equality follows from the mean of the message {\scriptsize \protect\tikz[inner sep=0.4ex,baseline=.4ex] \protect\node[circle,draw,yshift=0.15cm] {{$1$}};} depicted in Fig. \ref{fig:factor-graph-bivamp}. Note that (\ref{eq:approx-b}) can be rewritten as a function of the random vector $\bm{\mathsf{b}}_{\bm{{u}}_i}$ as follows:
\begin{equation}\label{eq:approx-b2}
\bm{\mathsf{b}}_{\bm{{u}}_i} = \bm{\Lambda}_{\bm{U}}\,\widehat{\bm{\mathsf{u}}}_{i,\mathsf{e}}
=\bm{\mathsf{b}}^*_{\bm{{u}}_i} + \bm{\Lambda}_{\bm{U}}^{\frac{1}{2}}\,\bm{\mathsf{n}}.
\end{equation}

Without loss of generality, we substitute $\bm{\mathsf{u}}^*$ with $\boldsymbol{e}_1$ in (\ref{eq:approx-b})  and (\ref{eq:approx-b2}) because the SE analysis does not depend on any specific canonical basis vector for a proper characterization of the empirical MSE. By doing so, we get:
\begin{subequations}\label{eq:ui-extrinsic}
    \begin{align}
    \widehat{\bm{\mathsf{u}}}_{i,\mathsf{e}} &=  \boldsymbol{e}_1 + \bm{\Lambda}_{\bm{U}}^{-\frac{1}{2}}\,\bm{\mathsf{n}},\\
    \bm{\mathsf{b}}_{\bm{{u}}_i} &=  \bm{\Lambda}_{\bm{U}}\,\boldsymbol{e}_1 + \bm{\Lambda}_{\bm{U}}^{\frac{1}{2}}\,\bm{\mathsf{n}}.
    \end{align}
\end{subequations}
\noindent As $M$ and $N$ grow large, both the diagonal and off-diagonal elements of the matrix $\bm{\Lambda}_{\bm{U}}$ in (\ref{eq:ui-extrinsic}) converge to finite values determined by the law of large numbers. Using the expression of $\boldsymbol{\Lambda}_{\bm{U}}$ from line \ref{eq:algo-Au-non-low-rank} of Algorithm~\ref{algo:big-vamp}, we find the concentrated matrix $\bm{\bar\Lambda}_{\bm{U}}\,=\,\mathbb{E}\left[\boldsymbol{\Lambda}_{\bm{U}}|\,p_{\mathsf{x}_{ij}}(x),\,p_{\mathsf{a}_{ij}}(a_{ij})\right]$ as follows:
\begin{subequations}
    \begin{align}    \hspace{-0.2cm}\bm{\bar\Lambda}_{\bm{U}}&\stackrel{\textrm{(a)}}{\approx} \frac{\gamma_{w}}{\sqrt{MN}}~\mathbb{E}\left[\bm{A}\,\widehat{\boldsymbol{X}}_{ \textsf{p}}^{-} \,\widehat{\boldsymbol{X}}_{ \textsf{p}}^{-\top}\bm{A}^{\top}\bigg|\,p_{\mathsf{x}_{\textsf{p}, ij}}(x),\,p_{\mathsf{a}_{ij}}(a_{ij})\right]\label{eq:SE-expectation-lambda-u-with-posterior}\\
    &= \frac{\gamma_{w}\,\bar\sigma_x^2\,\bar\sigma_A^2}{\sqrt{MN}}\,\mathbf{I}_N \nonumber\\
    &\triangleq \widetilde\gamma_{\boldsymbol{U}_{ \mathsf{e}}}\,\mathbf{I}_N.\label{eq:SE-expectation-lambda-u}
    \end{align}
\end{subequations}

\begin{figure*}[!b]
\rule{\textwidth}{0.4 pt}
\vspace{-0.3cm}
\begin{subequations}
    \begin{align}
        \mathcal{E}_{x^-}(\bar\gamma_{\boldsymbol{X}_{ \textsf{e}}^{-}})&=\lim _{R \rightarrow \infty} \frac{1}{R} \operatorname{Tr}\left(\left[\bar\gamma_{\boldsymbol{X}^{-}_{ \mathsf{e}}} \, \mathbf{I}_R + \frac{\gamma_{w}}{\sqrt{MN}}\,\boldsymbol{A}^\top\left({\widehat{\boldsymbol{U}}_{ \mathsf{p}}^{-\top}}\widehat{\boldsymbol{U}}_{ \mathsf{p}}^{-}\,+\, N\,\boldsymbol{R}_{\boldsymbol{U}_{ \mathsf{p}}^{-}} -\, \frac{N\gamma_{w}}{\sqrt{MN}} \,\boldsymbol{R}_{\boldsymbol{U}_{ \mathsf{p}}^{-}} \,\langle\boldsymbol{Y} \odot\boldsymbol{Y}\rangle\right)\boldsymbol{A}\,\right]^{-1}\right).\label{eq:v2-final-error-function}\\
        \mathcal{E}_{x^-}(\bar\gamma_{\boldsymbol{X}_{ \textsf{e}}^{-}})&= \lim _{R \rightarrow \infty} \frac{1}{R}\operatorname{Tr}\left(\left[\bar\gamma_{\boldsymbol{X}_{ \textsf{e}}^{-}} \, \mathbf{I}_R+\, \gamma_{w}\,\sqrt{\frac{\beta_x}{\beta_u}}\,\left(\frac{1}{N}-\bar\gamma_{\boldsymbol{U}_{ \mathsf{p}}^{-}}^{-1}\,\widetilde\gamma_{\boldsymbol{X}_{ \textsf{e}}^{-}}\,\left(1+\frac{1}{N-1}\right)\right)\boldsymbol{A}^\top\,\boldsymbol{A} \right.\right. \nonumber\\
        &\left.\left.\hspace{8.1cm}+\,\,\gamma_w\,\sqrt{\frac{\beta_x}{\beta_u}}\left(\frac{\bar\gamma_{\boldsymbol{U}_{ \mathsf{p}}^{-}}^{-1}\,\widetilde\gamma_{\boldsymbol{X}_{ \textsf{e}}^{-}}}{N-1}\right)\,\bm{A}^\top\,\bm{J}_N\,\bm{A}\right]^{-1}\right).\label{eq:v2-final-error-function-simplification}
    \end{align}
\end{subequations}
\end{figure*}

\noindent where (a) follows from $i$) neglecting the small Onsager correction terms of order $O(M^{-1/2})$  in line \ref{eq:algo-Au-non-low-rank} of Algorithm~\ref{algo:big-vamp} and $ii)$ the Bayes optimality assumption \cite{lesieur2015phase}. The latter stipulates that under the expectation in (\ref{eq:SE-expectation-lambda-u-with-posterior}), there is no statistical difference between the ground-truth signal $\bm{X}$ and the random sample $\widehat{\boldsymbol{X}}_{ \textsf{p}}$ from the posterior distribution.

\noindent By injecting (\ref{eq:SE-expectation-lambda-u}) back into (\ref{eq:ui-extrinsic}), we obtain:
\begin{subequations}\label{eq:ui-bui-pdf}
\begin{align}
    \widehat{\bm{\mathsf{u}}}_{i,\mathsf{e}} &= \bm{e}_1 + \frac{1}{\sqrt{\widetilde\gamma_{\boldsymbol{U}_{ \mathsf{e}}}}}\, \bm{\mathsf{n}},\label{eq:ui-pdf}\\
    \bm{\mathsf{b}}_{\bm{u}_i} &= \widetilde\gamma_{\boldsymbol{U}_{ \mathsf{e}}}\,\bm{e}_1 +\sqrt{\widetilde\gamma_{\boldsymbol{U}_{\mathsf{e}}}}\, \bm{\mathsf{n}}.\label{eq:bui-pdf}
    \end{align}
\end{subequations}
From (\ref{eq:ui-bui-pdf}), it is seen that the distribution of the $k$th element of $\widehat{\bm{\mathsf{u}}}_{i,\mathsf{e}}$ \big(resp. $\bm{\mathsf{b}}_{\bm{u}_i}$\big) depends on whether $\widehat{\mathsf{u}}_{ik,\textsf{e}}$ \big(resp. $[\bm{\mathsf{b}}_{\bm{u}_i}]_k$\big) is a diagonal or an off-diagonal element.  For this reason, the assumed LLR convergence in (\ref{eq:convergent-LLR-definition}) can be rewritten for diagonal and off-diagonal LLRs as:
\begin{subequations}
\label{eq:convergent-LLR-diag-offdiag}
\begin{align}
    \lim _{N \rightarrow \infty}\left(\textrm{LLR}_{u^{+}_{ii},\textsf{e}},\textrm{LLR}_{u^{-}_{ii},\textsf{e}}\right) &=  \Big(\overbar{\textrm{LLR}}^+_{\textrm{diag}},\overbar{\textrm{LLR}}_{\textrm{diag}}^-\Big),\,\forall i,\label{eq:convergent-LLR-diag}\\
    \hspace{-0.2cm}\lim _{N \rightarrow \infty}\left(\textrm{LLR}_{u^{+}_{ik},\textsf{e}},\textrm{LLR}_{u^{-}_{ik},\textsf{e}}\right) &=\Big(\overbar{\textrm{LLR}}^+_{{\textrm{off-diag}}},\overbar{\textrm{LLR}}_{{\textrm{off-diag}}}^-\Big), \footnotesize{i\neq k}.\label{eq:convergent-LLR-offdiag}
\end{align}
\end{subequations}
\noindent This fact will be employed later in the derivation by distinguishing the component-wise MSE functions of diagonal and off-diagonal elements.

For large enough $M$ and $N$, we also assume that the element-wise errors in the matrix updates, $\widehat{\bm{U}}_{ \mathsf{p}}^-$ and  $\widehat{\bm{X}}_{ \mathsf{p}}^-$, are with zero mean thereby leading to:
\begin{subequations}
    \begin{align}
        \widehat{\bm{U}}_{ \mathsf{p}}^- &= \bm{U}^* + \bm{W}_{\bm{U}},\label{eq:SE-decomposition-posterior-U}\\
        \widehat{\bm{X}}_{ \mathsf{p}}^- &= \bm{X}^* + \bm{W}_{\bm{X}},\label{eq:SE-decomposition-posterior-X}
    \end{align}
\end{subequations}
and covariance matrices given by: \vspace{-0.2cm}
\begin{subequations}\label{eq:limit 11-13-19-1-new}
\begin{align}
 \boldsymbol{R}_{\boldsymbol{U}_{ \mathsf{p}}^{-}}=\mathbb{E}\Big[\bm{W}_{\bm{U}}^{\top}\,\bm{W}_{\bm{U}}\Big]&\approx\hspace{-0.3cm}\underbrace{\bar\gamma_{\boldsymbol{U}_{ \mathsf{p}}^{-}}^{-1}\mathbf{I}_{N}}_{\substack{\textrm{error correlation}\\
 \textrm{on diagonal elements}}}\hspace{-0.3cm}-\underbrace{\, \frac{\bar\gamma_{\boldsymbol{U}_{ \mathsf{p}}^{-}}^{-1}}{N-1} \,(\boldsymbol{J}_N - \mathbf{I}_N)}_{{\substack{\textrm{error correlation}\\
 \textrm{on off-diagonal elements}}}},\label{eq:SE-covariance-U}\\ \boldsymbol{R}_{\boldsymbol{X}_{ \mathsf{p}}^{-}} =\mathbb{E}\Big[\bm{W}_{\bm{X}}^{\top}\,\bm{W}_{\bm{X}}\Big]&\approx~\bar\gamma_{\boldsymbol{X}_{ \mathsf{p}}^{-}}^{-1}\,\mathbf{I}_{R}.\label{eq:SE-covariance-X}
 \end{align}
\end{subequations}
In (\ref{eq:SE-covariance-U}), similarly to \cite{lesieur2016phase}, we have separated the correlation error $\boldsymbol{R}_{\boldsymbol{U}_{ \mathsf{p}}^{-}}$ pertaining to $\widehat{\bm{U}}_{ \mathsf{p}}^-$ into a diagonal and off-diagonal elements' contributions for the sake of analytical tractability. In addition, we use (\ref{eq:SE-covariance-U}) to obtain the correlation of the posterior estimate $\widehat{\bm{U}}_{ \mathsf{p}}^-$ given in (\ref{eq:SE-decomposition-posterior-U}) as follows:

\begin{equation}
\begin{aligned}[b]
    &\mathbb{E}\left[\left(\widehat{\bm{U}}_{ \mathsf{p}}^-\right)^\top\,\widehat{\bm{U}}_{ \mathsf{p}}^-\right]\\
    &\hspace{0.1cm}= \mathbf{I}_N - N\,\boldsymbol{R}_{\boldsymbol{U}_{ \mathsf{p}}^{-}}\\
    &\hspace{0.1cm}=\underbrace{\left(1-\bar\gamma_{\boldsymbol{U}_{ \mathsf{p}}^{-}}^{-1}\left(N+\frac{N}{N-1}\right)\right)\mathbf{I}_N}_{_{{\substack{\textrm{posterior correlation}\\
 \textrm{on diagonal elements}}}}} + \hspace{-0.2cm}\underbrace{\,\frac{N}{N-1}\,\bar\gamma_{\boldsymbol{U}_{ \mathsf{p}}^{-}}^{-1}\,\bm{J}_N}_{_{{\substack{\textrm{posterior correlation}\\
 \textrm{on off-diagonal elements}}}}}.
\end{aligned}
\end{equation}

\vspace{-0.4cm}
\subsection{State evolution equations}\label{sec:SE-equations}
Now, we incorporate the key changes for the MSE functions described in Section \ref{subsec:diff-SE-UCS-BiVAMP}. We do so by in separate sections for each of those functions.

\subsubsection{Derivation of the MSE function $\mathcal{E}_{x^-}(\cdot)$}

\noindent From the update equation in line \ref{eq:posterior-covariance-v-gaussian-approx} of Algorithm \ref{algo:big-vamp}, the component-wise MSE of the bi-LMMSE denoiser of $\boldsymbol{X}$ in the large system limit is given by:
\begin{equation}\label{eq:definition-LMMSE-X}
    \mathcal{E}_{x^-}= \lim _{R \rightarrow \infty} \frac{1}{R} \operatorname{Tr}\left(\boldsymbol{R}_{\boldsymbol{X}^{-}_{ \mathsf{p}}}\right).
\end{equation}

\noindent Using the update equation in line \ref{eq:algo-Av-non-low-rank} of Algorithm \ref{algo:big-vamp} into (\ref{eq:definition-LMMSE-X}) yields the component-wise MSE of the bi-LMMSE denoiser of $\boldsymbol{X}$ in (\ref{eq:v2-final-error-function}), displayed at the bottom of this page.

\noindent Using (\ref{eq:limit 11-13-19-1-new}) and (\ref{eq:var_y}),  (\ref{eq:v2-final-error-function}) can be rewritten as provided in (\ref{eq:v2-final-error-function-simplification}) where
\begin{equation}\label{eq:gamma-X-tilde}
    \widetilde\gamma_{\boldsymbol{X}_{ \mathsf{e}}^{-}}= \gamma_w\,\sqrt{\beta_x\,\beta_u}\,N\,\bar\sigma_u^2\,\bar\sigma_x^2\,\bar\sigma_A^2 + 1.
\end{equation}

\noindent Note that the last term in the matrix inverse involved in (\ref{eq:v2-final-error-function-simplification}) is a rank-one matrix due to the fact that $\boldsymbol{J}_N = \boldsymbol{1}_N\boldsymbol{1}_N^\top$. Using the Sherman–Morrison formula \cite{bartlett1951inverse}, it follows that the contribution of the rank-one term $\bm{A}^\top\,\bm{J}_N\,\bm{A}$ is also of rank one, thereby justifying neglecting its trace in (\ref{eq:v2-final-error-function-simplification}). That is to say:
\begin{equation}\label{eq:denoiser-X-after-one-rank}
    \mathcal{E}_{x^-}(\bar\gamma_{\boldsymbol{X}_{ \textsf{e}}^{-}})\approx \lim _{R \rightarrow \infty} \frac{\bar\gamma_{\boldsymbol{X}_{ \textsf{e}}^{-}}^{-1}}{R}\operatorname{Tr}\left(\left[ \mathbf{I}_R+\, \alpha_x\,\boldsymbol{A}^\top\,\boldsymbol{A} \right]^{-1}\right),
\end{equation}

where
\begin{equation*}
    \alpha_x = \frac{\gamma_{w}}{\bar\gamma_{\boldsymbol{X}_{ \textsf{e}}^{-}}}\,\sqrt{\frac{\beta_x}{\beta_u}}\,\left(\frac{1}{N}-\bar\gamma_{\boldsymbol{U}_{ \mathsf{p}}^{-}}^{-1}\,\widetilde\gamma_{\boldsymbol{X}_{ \textsf{e}}^{-}}\,\left(1+\frac{1}{N-1}\right)\right).
\end{equation*}
\begin{figure*}[!t]
\begin{subequations}\label{eq:LLR-plus-diag-offdiag}
    \begin{align}
        \textrm{LLR}_{u^{+}_{ii},\textsf{e}} &= \widetilde\gamma_{\boldsymbol{U}_{ \mathsf{e}}} + \sqrt{\widetilde\gamma_{\boldsymbol{U}_{ \mathsf{e}}}}\,n_1 -\frac{1}{2}\,[\mathbf{\Lambda}_{\bm{U}}]_{11}- \ln \sum\limits_{\substack{i'=1\\i' \neq i}}^{N} \exp\left(\widetilde\gamma_{\boldsymbol{U}_{ \mathsf{e}}} + \sqrt{\widetilde\gamma_{\boldsymbol{U}_{ \mathsf{e}}}}\,n_1-\frac{1}{2}\,[\mathbf{\Lambda}_{\bm{U}}]_{11}+\textrm{LLR}_{{u}^{-}_{i'k},\textsf{e}}\right).\label{eq:LLR-plus-diag}\\
        \textrm{LLR}_{u^{+}_{ik},\textsf{e}}&= \sqrt{\widetilde\gamma_{\boldsymbol{U}_{ \mathsf{e}}}}\,n_q -\frac{1}{2}\,[\mathbf{\Lambda}_{\bm{U}}]_{qq}- \ln \sum\limits_{\substack{i'=1\\i' \neq i}}^{N} \exp\left( \sqrt{\widetilde\gamma_{\boldsymbol{U}_{ \mathsf{e}}}}\,n_q -\frac{1}{2}\,[\mathbf{\Lambda}_{\bm{U}}]_{qq}+\textrm{LLR}_{{u}^{-}_{i'k},\textsf{e}}\right),~~\textrm{with} ~q\neq1~\textrm{and}~i\neq k.\label{eq:LLR-plus-offdiag}\\
        \overbar{\textrm{LLR}}_{{\textrm{off-diag}}}^{+}& =-\frac{1}{2}\,\widetilde\gamma_{\boldsymbol{U}_{ \mathsf{e}}}-\int_{-\infty}^{+\infty} \ln\left[(N-2)\,\exp\left( \sqrt{\widetilde\gamma_{\boldsymbol{U}_{ \mathsf{e}}}}\,n_q -\frac{1}{2}\,\widetilde\gamma_{\boldsymbol{U}_{ \mathsf{e}}}+\overbar{\textrm{LLR}}_{{\textrm{off-diag}}}^-\right) + \exp\left( \frac{1}{2}\,\widetilde\gamma_{\boldsymbol{U}_{ \mathsf{e}}}+\sqrt{\widetilde\gamma_{\boldsymbol{U}_{ \mathsf{e}}}}\,n_q +\overbar{\textrm{LLR}}_{{\textrm{diag}}}^-\right)\right]\textrm{d}n_q.\label{eq:LLR-plus-diag-convergent-integral}
    \end{align}
\end{subequations}
\vspace{-0.4cm}
\rule{\textwidth}{0.4 pt}
\vspace{-0.4cm}
\end{figure*}
\noindent Using a well-known result in random matrix theory (cf. eq. (1.16) in \cite{tulino2004random}), the limit in (\ref{eq:denoiser-X-after-one-rank}) is obtained as:
\begin{equation}\label{eq:bi-LMMSE-X}
\begin{aligned}[b]
    \mathcal{E}_{x^-}(\bar\gamma_{\boldsymbol{X}_{ \mathsf{e}}^{-}})=\bar{\gamma}_{\boldsymbol{X}^{-}_{ \mathsf{e}}}^{-1}\left(1-\frac{\mathcal{F}( \alpha_{x},\beta_{u})}{4 \,\beta_{u}\,  \alpha_{x}}\right),
\end{aligned}
\end{equation}
wherein
\begin{equation}
    \mathcal{F}(x, z)=\left(\sqrt{x(1+\sqrt{z})^{2}+1}-\sqrt{x(1-\sqrt{z})^{2}+1}\ \right)^{2}. \label{eq:tolino-function}
\end{equation}

\subsubsection{Derivation of the MSE function $\mathcal{E}_{x^+}(\cdot)$}

The variance of the MMSE denoiser of the $\bm{X}^+$ matrix entries is obtained from line \ref{eq:algo-diag-sigma-v-alpha} of Algorithm~\ref{algo:big-vamp} as follows:
\begin{equation}\label{eq:x+-sensitivity-function-sum}
    \mathcal{E}_{x^+}(\gamma_{\boldsymbol{X}_{ \mathsf{e}}^{+}})~\triangleq~\frac{1}{\gamma_{\boldsymbol{X}^{+}_{ \mathsf{p}}}} = \frac{1}{M\gamma_{\boldsymbol{X}^{+}_{ \mathsf{e}}}} \,\sum_{j=1}^{M} \langle \mathbf{g}^{\prime}_{\mathsf{x}}(\widehat{\boldsymbol{x}}^{+}_{j,\mathsf{e}}, \gamma^{-1}_{\boldsymbol{X}^{+}_{ \mathsf{e}}})\rangle,
\end{equation}

\noindent Using the concentrations of measures in (\ref{eq:asymptotic-convergent-precisions}) in the large system limit, the empirical averages involved in (\ref{eq:x+-sensitivity-function-sum}) can be approximated by the following statistical average:
\begin{equation}
    {\mathcal{E}_{x^+}(\bar\gamma_{\boldsymbol{X}_{ \mathsf{e}}^{+}})}{\,\triangleq\,\frac{1}{\bar\gamma_{\boldsymbol{X}_{ \mathsf{e}}^{+}}}\,\mathbb{E}\left[g^{\prime}_{\textrm{{\rv{x}}}}(\widehat{{x}}_{\mathsf{e}}^{\,+},\bar\gamma_{\boldsymbol{X}_{ \mathsf{e}}^{+}}^{-1})\,\middle\vert\,\mathit{p}_{{\mathsf{x}}}({x})\,p_{\widehat{{\mathsf{x}}}_{\mathsf{e}}^{\,+}|{\mathsf{x}}}(\widehat{{x}}_{\mathsf{e}}^{\,+}|{x})\right]}. \label{eq:x+-sensitivity-function}
\end{equation}
In (\ref{eq:x+-sensitivity-function}), $p_{\widehat{{\mathsf{x}}}_{\mathsf{e}}^{\,+}|{\mathsf{x}}}(\widehat{{x}}_{\mathsf{e}}^{\,+}|{x})$ corresponds to the scalar observation model $\widehat{\mathsf{x}}_{\mathsf{e}}^{+}\,=\,\mathsf{x}\,+\,\mathsf{w}_x$ where under the matched conditions, we have $\mathsf{w}_x\sim\mathcal{N}(w_x;0,\bar\gamma_{\boldsymbol{X}_{ \mathsf{e}}^{+}}^{-1})$. Moreover, $g^{\prime}_{\textrm{{\rv{x}}}}(\cdot)$ is the derivative of the column-wise denoiser $g_{\textrm{{\rv{x}}}}(\cdot)$ in (\ref{eq:turbo-mean}). The latter can be evaluated once the prior $p_{\bm{\mathsf{x}}}(\bm{x})$ is specified in the expression of the denoiser $g_{\textrm{{\rv{x}}}}(\cdot)$ in (\ref{eq:turbo-mean}).

\subsubsection{Derivation of the LLR function $\mathcal{L}_{u_{ik}^-}(\cdot)$}\label{sec:LLR-minus-derivation}
When it comes to the permutation prior, the column-wise denoiser of $\bm{U}^+$ exchanges the LLR message $\textrm{LLR}_{\bm{{U}}^{-},\textsf{e}}$ with the row-wise denoiser of $\bm{U}^-$ as depicted in Fig.~\ref{fig:block-diagram}. In this section, we derive the function $\mathcal{L}_{u_{ik}^-}(\cdot)$ which determines the concentrated extrinsic LLR value $\overbar{\textrm{LLR}}^-$ defined in (\ref{eq:convergent-LLR-definition}). Toward this goal, we first recall the expression of the exchanged $ik$th extrinsic LLR value established in (\ref{eq:LR-extrinsic-plus}):
\begin{equation}\label{eq:LR-values}
    \textrm{LLR}_{{u}^{-}_{ik},\textsf{e}} = -\ln\sum\limits_{\substack{k'=1\\k' \neq k}}^{N} \exp\left(\textrm{LLR}_{{u}^{+}_{ik'},\textsf{e}}\right).
\end{equation}
\noindent Note here that both LLR values $\textrm{LLR}_{u^{-}_{ik},\textsf{e}}$ and $\textrm{LLR}_{u^{+}_{ik},\textsf{e}}$ in (\ref{eq:LR-values}) involve the term $\widehat{u}^-_{ik, \mathsf{p}}$ given in (\ref{eq:posterior-mean-U-final}) which itself depends on $\widehat{\bm{{u}}}_{i,\mathsf{e}}$. Therefore, we resort to the pdf of $\widehat{\bm{\mathsf{u}}}_{i,\mathsf{e}}$ in (\ref{eq:ui-pdf}) to characterize the asymptotic LLR values in (\ref{eq:LR-values}) as $N \rightarrow \infty$. Indeed, the pdf of $\widehat{\bm{\mathsf{u}}}_{i,\mathsf{e}}$ in (\ref{eq:ui-pdf}) suggests distinguishing between the LLR value of a diagonal element, $\textrm{LLR}_{u^{+}_{ii},\textsf{e}}$ or $\textrm{LLR}_{u^{-}_{ii},\textsf{e}}$, and the LLR value of an off-diagonal element, $\textrm{LLR}_{u^{+}_{ik},\textsf{e}}$ or $\textrm{LLR}_{u^{-}_{ik},\textsf{e}}$ for $i\neq k$. Using the concentration of measure for the diagonal and off-diagonal elements in (\ref{eq:convergent-LLR-diag-offdiag}), the LLR values in (\ref{eq:LR-values}) converge to the following values:
\begin{subequations}\label{eq:LR-values-diag-off-diag}
\begin{align}
    \overbar{\textrm{LLR}}_{{\textrm{diag}}}^{-}&=  -\ln\left(N-1\right) - \overbar{\textrm{LLR}}_{{\textrm{off-diag}}}^{+}, \label{eq:LR-values-minus-diag}\\
    \overbar{\textrm{LLR}}_{{\textrm{off-diag}}}^{-}& =  -\ln \Big( (N-2)\,\mathlarger{e^{\overbar{\textrm{LLR}}_{{\textrm{off-diag}}}^{+}}} ~+~ \mathlarger{e^ {\overbar{\textrm{LLR}}_{\textrm{diag}}^{+}}} \Big). \label{eq:LR-values-minus-off-diag}
\end{align}
\end{subequations}
We then rewrite (\ref{eq:LR-values-diag-off-diag}) in a compact form as
 \begin{equation}
 \overbar{\textrm{LLR}}_{{\textrm{diag}}}^{-} = \mathcal{L}_{u_{ik}^-}\left(\overbar{\textrm{LLR}}_{\textrm{diag}}^+,\overbar{\textrm{LLR}}_{{\textrm{off-diag}}}^+\right),\label{eq:LR-values-minus-compact}
 \end{equation}
 where
\begin{equation}
\begin{aligned}[b]
    \mathcal{L}_{u_{ik}^-}\left(x,y\right) =
    \begin{cases}
      -\ln\left(N-1\right) -y,& \small{\text{if $k=1$}},\vspace{0.15cm}\\
      -\ln \Big( (N-2)\,\exp \left(y\right) + \exp \left(x\right)\Big), & \small{\text{if $k \neq 1$}}.
    \end{cases}
    \end{aligned}
\end{equation}

\subsubsection{Derivation of the LLR function $\mathcal{L}_{u_{ik}^+}(\cdot)$}
The row-wise denoiser of $\bm{U}^-$ exchanges with the column-wise denoiser the LLR message $\textrm{LLR}_{\bm{{U}}^{-},\textsf{e}}$ already obtained in (\ref{eq:LR-extrinsic-minus}) as:
\begin{equation}\label{eq:LR-extrinsic-minus-SE}
    \begin{aligned}[b]
    \textrm{LLR}_{{u}^{+}_{ik},\textsf{e}} &= [\bm{b}_{\bm{{u}}_i}]_k-\frac{1}{2}\,[\mathbf{\Lambda}_{\bm{U}}]_{kk}\\
    &~~~-\,\ln \sum\limits_{\substack{i'=1\\i' \neq i}}^{N} \exp\left([\bm{b}_{\bm{{u}}_i}]_k-\frac{1}{2}\,[\mathbf{\Lambda}_{\bm{U}}]_{kk}+\textrm{LLR}_{{u}^{-}_{i'k},\textsf{e}}\right).
    \end{aligned}
\end{equation}
In this section, we derive the LLR functions $\mathcal{L}_{u_{ik}^+}(\cdot)$ which determine the concentrated extrinsic LLR value $\overbar{\textrm{LLR}}^+$. Similarly to the derivation of $\overbar{\textrm{LLR}}^-$ in Section \ref{sec:LLR-minus-derivation}, we first distinguish here between diagonal and off-diagonal LLR values according the distribution of $\bm{b}_{\bm{{u}}_i}$ in (\ref{eq:bui-pdf}) to obtain the expressions of $\textrm{LLR}_{u^{+}_{ii},\textsf{e}}$ and $\textrm{LLR}_{u^{+}_{ik},\textsf{e}}$ (for $i\neq k$) given in (\ref{eq:LLR-plus-diag}) and (\ref{eq:LLR-plus-offdiag}), respectively, displayed at the top of the next page. Now, using the concentration of measure for the LLRs in (\ref{eq:convergent-LLR-diag}), the LLRs in (\ref{eq:LLR-plus-diag}) and (\ref{eq:LLR-plus-offdiag}) converge to
\begin{equation}\label{eq:LLR_conv_plus}
\begin{aligned}[b]
    \overbar{\textrm{LLR}}^{+} = \mathcal{L}_{u_{ik}^+}\left(\overbar{\textrm{LLR}}_{\textrm{diag}}^-,\overbar{\textrm{LLR}}_{{\textrm{off-diag}}}^-\right)\triangleq
    \begin{cases}
      \overbar{\textrm{LLR}}_{{\textrm{diag}}}^{+},& \hspace{-0.2cm}\small{\text{if $k=1$}},\vspace{0.15cm}\\
      \overbar{\textrm{LLR}}_{{\textrm{off-diag}}}^+, & \hspace{-0.2cm}\small{\text{if $k \neq 1$}},
    \end{cases}
    \end{aligned}
\end{equation}
while the concentrated diagonal LLR is given by:
\begin{equation}
    \begin{aligned}[b]        \overbar{\textrm{LLR}}_{{\textrm{diag}}}^{+}&= \mathbb{E}\Big[ \textrm{LLR}_{u^{+}_{ii},\textsf{e}}\Big\vert\,p_{{\mathsf{n}}_{1}}({{n}}_{1})\Big],\\
        &=\widetilde\gamma_{\boldsymbol{U}_{ \mathsf{e}}} -\ln(N-1)-\overbar{\textrm{LLR}}_{{\textrm{off-diag}}}^{-},\label{eq:LLR-plus-diag-convergent-integral-1}
    \end{aligned}
\end{equation}
and the off-diagonal LLR converges to
\begin{equation}
\overbar{\textrm{LLR}}_{{\textrm{off-diag}}}^{+}=  \mathbb{E}\Big[ \textrm{LLR}_{u^{+}_{ik},\textsf{e}}\Big\vert\,p_{{\mathsf{n}}_{q}}({{n}}_{q})\Big],~\textrm{with}~q\neq 1~\textrm{and}~i\neq k.\label{eq:LLR-plus-diag-convergent-integral-2}
\end{equation}
After some algebraic manipulations, we show that (\ref{eq:LLR-plus-diag-convergent-integral-2}) is explicitly expressed as in (\ref{eq:LLR-plus-diag-convergent-integral}).

\begin{figure*}[!b]
\vspace{-0.3cm}
\rule{\textwidth}{0.4 pt}
\vspace{-0.2cm}
    \begin{equation}        \hspace{-1.6cm}f_k\left(\bm{{b}}_{\bm{{u}}_i},\overbar{\textrm{LLR}}_{\textrm{diag}}^-,\overbar{\textrm{LLR}}_{{\textrm{off-diag}}}^-\right) = \frac{\delta(k-1)\cdot\exp \left([\boldsymbol{b}_{\bm{{u}}_i}]_1+\overbar{\textrm{LLR}}_{\textrm{diag}}^-\right) + (1-\delta(k-1)) \cdot \exp \left([\boldsymbol{b}_{\bm{{u}}_i}]_k+\overbar{\textrm{LLR}}_{{\textrm{off-diag}}}^-\right)}{\exp\left([\boldsymbol{b}_{\bm{{u}}_i}]_1+\overbar{\textrm{LLR}}_{\textrm{diag}}^-\right)+\exp\left(\overbar{\textrm{LLR}}_{{\textrm{off-diag}}}^-\right)\sum\limits_{q=2}^{N} \exp\left([\boldsymbol{b}_{\bm{{u}}_i}]_q\right)}.\label{eq:fk}
    \end{equation}
    \begin{equation}
g_k\left(\eta,\overbar{\textrm{LLR}}_{\textrm{diag}}^-,\overbar{\textrm{LLR}}_{{\textrm{off-diag}}}^-\right)=\frac{\delta(k-1)\cdot\exp \left(\eta+\sqrt{\eta}\,n_1+\overbar{\textrm{LLR}}_{\textrm{diag}}^-\right) + (1-\delta(k-1))\cdot\exp \left(\sqrt{\eta}\,n_1+\overbar{\textrm{LLR}}_{{\textrm{off-diag}}}^-\right)}{\exp \left(\eta+\sqrt{\eta}\,n_1 + \overbar{\textrm{LLR}}_{\textrm{diag}}^-\right)+(N-1)\,\exp\left(\eta/2+\overbar{\textrm{LLR}}_{{\textrm{off-diag}}}^-\right)}.\label{eq:gk}
    \end{equation}
\end{figure*}

\subsubsection{Derivation of the MSE function $\mathcal{E}_{u^-}(\cdot)$}
Similarly to (\ref{eq:definition-LMMSE-X}), the component-wise MSE of the bi-LMMSE denoiser of $\boldsymbol{U}^-$ in the large system limit is given by:
\begin{equation}\label{eq:definition-LMMSE-U}
\begin{aligned}[b]
    \mathcal{E}_{u^-}&= \lim _{N \rightarrow \infty} \frac{1}{N} \operatorname{Tr}\left(\boldsymbol{R}_{\boldsymbol{U}^{-}_{ \mathsf{p}}}\right)\\
    &\overset{(a)}{=}\lim _{N \rightarrow \infty} \frac{1}{N} \sum\limits_{i=1}^{N}\frac{1}{N}\operatorname{Tr}\left( \boldsymbol{R}_{\bm{u}_{i, \mathsf{p}}^{-}}\right)\\
    &\overset{(b)}{=}\lim _{N \rightarrow \infty} \frac{1}{N} \sum\limits_{i=1}^{N}\frac{1}{N}\sum\limits_{k=1}^{N} \sigma^{2}_{\mathsf{u}_{ik, \textsf{p}}^{-}}\\
    & \overset{(c)}{=}\lim _{N \rightarrow \infty} \frac{1}{N} \sum\limits_{i=1}^{N}\langle h_{\mathsf{u}_{ik}}^\prime\left(\boldsymbol{b}_{\bm{{u}}_i}, \mathbf{\Lambda}_{\bm{U}},\text{LLR}_{\bm{{U}}^{-}\hspace{-0.06cm},{\textsf{e}}}\right)\rangle,
    \end{aligned}
\end{equation}
where (a) and (b) follow from lines \ref{algo:sum-Ru} and \ref{algo:eq-varUp-minus-with-LRs} of Algorithm~\ref{algo:big-vamp}, respectively, while (c) results from the definition in (\ref{eq:posterior-variance-U-final}).

\noindent To ease the notation, we introduce the functions $f_k(\cdot,\cdot,\cdot)$ and $g_k(\cdot,\cdot,\cdot)$ given in (\ref{eq:fk}) and (\ref{eq:gk}), displayed at the bottom of the next page.

\noindent Using the concentrations of measures in (\ref{eq:SE-expectation-lambda-u}) and (\ref{eq:convergent-LLR-definition}) in the large system limit, the empirical averages involved in (\ref{eq:definition-LMMSE-U}) can be approximated by the following statistical average:
\begin{equation}\label{eq:u+-sensitivity-function-without-delta}
\mathcal{E}_{u^-}\big(\overbar{\text{LLR}}_{\bm{{U}}^{-}\hspace{-0.06cm},{\textsf{e}}}\big) = \mathbb{E}\Big[ h^{\prime}_{{\mathsf{u}}_{ik}}\left(\bm{{b}}_{\bm{{u}}_i},\bm{\bar{\Lambda}}_{\bm{U}},\overbar{\text{LLR}}_{\bm{{U}}^{-}\hspace{-0.06cm},{\textsf{e}}}\right)\Big\vert\,p_{\bm{\mathsf{b}}_{\bm{{u}}_i}}(\bm{{b}}_{\bm{{u}}_i})\Big],
\end{equation}
where the distribution $p_{\bm{\mathsf{b}}_{\bm{{u}}_i}}(\bm{{b}}_{\bm{{u}}_i})$ is given in (\ref{eq:bui-pdf}), and  $h^{\prime}_{{\mathsf{u}_{ik}}}(\cdot,\cdot,\cdot)$ is the component-wise derivative of the two-stage denoiser $h_{{\mathsf{u}_{ik}}}(\cdot,\cdot,\cdot)$ with respect to $[\boldsymbol{b}_{\bm{{u}}_i}]_k$ as defined in (\ref{eq:posterior-variance-U-final}).

\noindent Since the elements of the matrix $\overbar{\text{LLR}}_{\bm{{U}}^{-}\hspace{-0.06cm},{\textsf{e}}}$ converge to either $\overbar{\textrm{LLR}}_{\textrm{diag}}^-$ or $\overbar{\textrm{LLR}}_{{\textrm{off-diag}}}^-$, we reparametrize $h^{\prime}_{{\mathsf{u}_{ik}}}(\cdot,\cdot,\cdot)$ as follows:
\begin{equation}\label{eq:derivative-denoiser-uik-compact}
\begin{aligned}[b]    &h_{\mathsf{u}_{ik}}^\prime\left(\boldsymbol{b}_{\bm{{u}}_i}, \overbar{\textrm{LLR}}_{\textrm{diag}}^-,\overbar{\textrm{LLR}}_{{\textrm{off-diag}}}^-\right) \\
    &\hspace{0.5cm}= f_k\left(\bm{{b}}_{\bm{{u}}_i},\overbar{\textrm{LLR}}_{\textrm{diag}}^-,\overbar{\textrm{LLR}}_{{\textrm{off-diag}}}^-\right)\\
    &\hspace{1.5cm}\times \Big(1-f_k\left(\bm{{b}}_{\bm{{u}}_i},\overbar{\textrm{LLR}}_{\textrm{diag}}^-,\overbar{\textrm{LLR}}_{{\textrm{off-diag}}}^-\right)\Big).
    \end{aligned}
\end{equation}
\noindent In (\ref{eq:derivative-denoiser-uik-compact}), we use the delta function in the expression of $f_k(\cdot,\cdot,\cdot)$ given in (\ref{eq:fk}) to rewrite $h^{\prime}_{{\mathsf{u}_{ik}}}(\cdot,\cdot,\cdot)$ in a compact form for both diagonal and off-diagonal elements, and drop the parameter $\bm{\bar\Lambda}_{\bm{U}}$ as it gets simplified.

To evaluate (\ref{eq:u+-sensitivity-function-without-delta}) using (\ref{eq:derivative-denoiser-uik-compact}), we note that both the $k$th component of the denoiser derivative $\bm{h}^{\prime}_{{\mathsf{u}}}(\cdot, \cdot, \cdot)$ defined in (\ref{eq:derivative-denoiser-uik-compact}) and the distribution $p_{\bm{\mathsf{b}}_{\bm{{u}}_i}}(\bm{{b}}_{\bm{{u}}_i})$ given in (\ref{eq:bui-pdf}) depend on whether the $k$th element of $\bm{{b}}_{\bm{{u}}_i}$, $\left[\bm{{b}}_{\bm{{u}}_i}\right]_k$, corresponds to $k=i$ or $k \neq i$. For this reason, we add the subscript $k$ to the definition of the MSE function $\mathcal{E}_{u^-,k}(\cdot)$ to account for both diagonal and off-diagonal entries simultaneously.

\noindent Using the law of large numbers, we show in Appendix \ref{appendix:MSE-U-minus} that the MSE function $\mathcal{E}_{u^-,k}(\cdot)$ is given by:
\begin{equation}
\label{eq:final-Euk}
\begin{aligned}
&{\mathcal{E}_{u^-,k}\left(\widetilde\gamma_{\boldsymbol{U}_{ \mathsf{e}}},\overbar{\textrm{LLR}}_{\textrm{diag}}^-,\overbar{\textrm{LLR}}_{{\textrm{off-diag}}}^-\right)} \\
& =  \int\limits_{-\infty}^{+\infty} \hspace{-0.1cm}g_k\left(\widetilde\gamma_{\boldsymbol{U}_{ \mathsf{e}}},\overbar{\textrm{LLR}}_{\textrm{diag}}^-,\overbar{\textrm{LLR}}_{{\textrm{off-diag}}}^-\right) \\
&\hspace{0.4cm}\times\left(1-g_k\left(\widetilde\gamma_{\boldsymbol{U}_{ \mathsf{e}}},\overbar{\textrm{LLR}}_{\textrm{diag}}^-,\overbar{\textrm{LLR}}_{{\textrm{off-diag}}}^-\right)\right)\,\frac{\exp\left(-n_{1}^2/2\right)}{\sqrt{2\pi}}\,\text{d}n_{1}.
\end{aligned}
\end{equation}

Now that the MSE function of the $k$th entry of $\bm{u}^-_i$, $\mathcal{E}_{u^-,k}(\cdot,\cdot,\cdot)$, is established in (\ref{eq:final-Euk}), the overall MSE function pertaining to the vector $\bm{u}^-_i$ follows by simply averaging over its element-wise MSEs, i.e.:
\begin{equation}\label{eq:SE-Eu--final}
\begin{aligned}[b]
    &\mathcal{E}_{u^-}\left(\widetilde\gamma_{\boldsymbol{U}_{ \mathsf{e}}},\overbar{\textrm{LLR}}_{\textrm{diag}}^-,\overbar{\textrm{LLR}}_{{\textrm{off-diag}}}^-\right) \\
    &= \frac{1}{N}\,\sum\limits_{k=1}^{N} \mathcal{E}_{u^-,k}\left(\widetilde\gamma_{\boldsymbol{U}_{ \mathsf{e}}},\overbar{\textrm{LLR}}_{\textrm{diag}}^-,\overbar{\textrm{LLR}}_{{\textrm{off-diag}}}^-\right).
    \end{aligned}
\end{equation}
\subsection{State evolution pseudo-code}
We can now summarize in Algorithm~\ref{algo:u-bi-vamp-SE} our main result, which is the SE equations for UCS.
\subfile{SE-algorithm-LLR}
\section{Simulation Results}\label{sec:simulation-results}

In this section, we assess the performance of the proposed UCS algorithm for solving the unlabeled sensing and unlabeled compressed sensing problems. In all simulations, we set the precision tolerance to $\xi=10^{-6}$ without specifying any value for $T_{\textrm{max}}$. The code of the UCS algorithm is available at {\scriptsize{\href{https://github.com/makrout/Unlabeled-Compressed-Sensing}{\textcolor{github-link}{\texttt{https://github.com/makrout/Unlabeled-Compressed-Sensing
}}}}}.

\subsection{Evaluation metrics}
To measure the performance of UCS for bilinear recovery, we use the following metrics:
 \begin{itemize}[leftmargin=*]
     \item the normalized root MSE (NRMSE) between the ground-truth and estimated matrices $\bm{X}$ and $\widehat{\bm{X}}$. The NRMSE metric is defined for any matrix $\bm{P}$ and its estimate $\widehat{\bm{P}}$ as:
     \begin{equation*}
    \textrm{NRMSE}\Big(\bm{P}, \widehat{\bm{P}}\Big) ~=~  \frac{\Big\|\bm{P}-\widehat{\bm{P}}\Big\|_{\textrm{F}}}{\|\bm{P}\|_{\textrm{F}}}.
\end{equation*}
\item the Hamming distortion (HD) between the ground-truth and estimated permutation matrix $\bm{U}$ and $\widehat{\bm{U}}$, which counts the number of their element-wise mismatches. HD is defined as:
\begin{equation*}
    \textrm{HD}\Big(\bm{U}, \widehat{\bm{U}}\Big) ~=~ \frac{1}{2\,N}\,\sum_{i=1}^{N}\sum_{j=1}^{N}\delta(u_{ij} \neq \widehat{u}_{ij} ).
\end{equation*}
 \end{itemize}

\subsection{Baselines}
\noindent All the existing algorithms described in Table~\ref{tab:algorithms-review} were not designed to recover a large-size (i.e., $N > 10$) fully shuffled permutation matrix $\bm{U}$. For this reason, we specify an additional constraint to restrict the permutation matrix to be $p$-local (i.e., $p \times p$ block diagonal) to ensure a fair comparison of UCS against the following benchmarks:
\begin{itemize}[leftmargin=*]
    \item \texttt{Biconvex}: an ADMM algorithm devised in \cite{zhang2019permutation} to minimize the biconvex relaxation of the $L_2$-norm objective function $\|\bm{Y} - \bm{U}\,\bm{A}\,\bm{X}\|_{\textrm{F}}^2$. Its penalty parameter $\rho$ was tuned between $10^{−3}$ and $10^{−5}$ and the best parameter was retained,
    \item \texttt{Levsort}: the Levsort algorithm \cite{pananjady2017denoising} involving spectral computations on the matrices $\bm{U}$ and $\bm{A}$ based on their singular value decomposition and a matching step using the pseudo-inverse of the matrix $\widehat{\bm{U}}\bm{A}$,
    \item \texttt{One-step}: the alternating minimization method proposed in \cite{zhang2020optimal} which exploits the fact that $\bm{A}$ is a Gaussian matrix to approximate the same minimization of \texttt{Biconvex},
    \item \texttt{RLUS}: an algorithm that finds the permutation matrix $\bm{U}$ for the local unlabeled sensing problem proposed in \cite{abbasi2021r}.
    \item \texttt{Prox-alt-min}: an improved version of \texttt{RLUS} using a proximal operator proposed in \cite{abbasi2021r2} that converges to a first-order stationary point of the $L_2$-norm objective function $\|\bm{Y} - \bm{U}\,\bm{A}\,\bm{X}\|_{\textrm{F}}^2$.
\end{itemize}
For non-local unlabeled sensing problems, we only compare UCS to its state evolution because we are not aware of any existing algorithm designed to recover arbitrary permutation matrices for $N\geq 10$. As a matter of fact, we tried to run the branch-and-bound algorithm \cite{emiya2014compressed} on the unlabeled sensing problem with $N=10$, which did not provide any meaningful results after a few days of running time, even though the number of explored subproblems by the branch-and-bound algorithm gets dramatically low as $N\rightarrow\infty$.
Similarly, for the non-local unlabeled compressed sensing problem, we only report the reconstruction performance of UCS since it is the first algorithm designed to solve it.

\subsection{Experiments}
\subsubsection{$p$-local unlabeled sensing}
\begin{figure}[h!]
\centering
\includegraphics[scale=0.35]{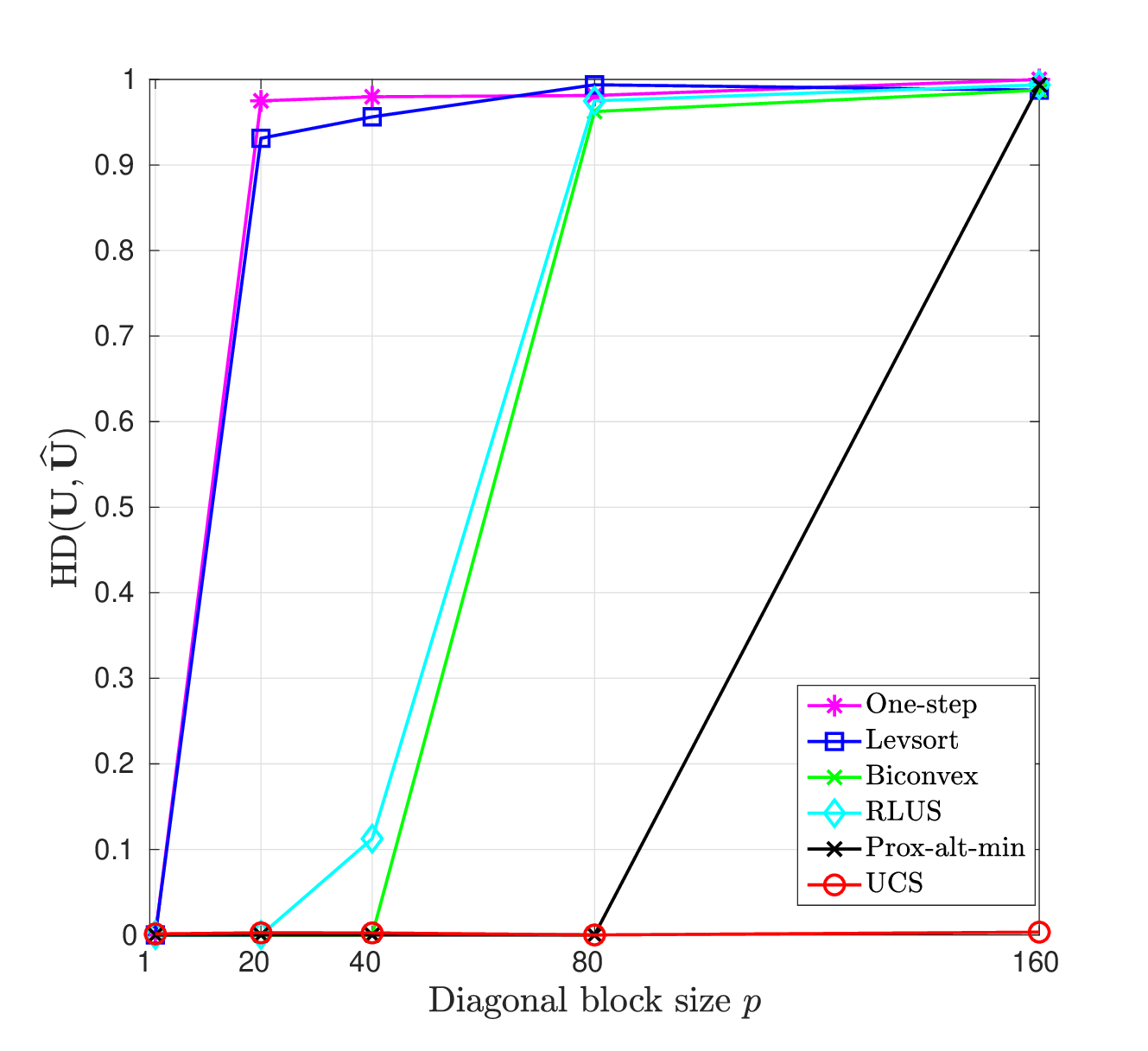}
\caption{HD of UCS vs. the size $p$ of the diagonal blocks of $\bm{U}$ for the $p$-local unlabeled sensing problem at SNR = 20 dB: Gaussian prior on $\bm{X}$ with $N=160$, $M=100$, and $R=10$.}
\label{fig:p-local-experiment}
\end{figure}

In Fig. \ref{fig:p-local-experiment}, we compare UCS against the aforementioned baselines on the $p$-local unlabeled sensing problem.

\noindent We observe that most of the existing benchmarks fail to recover $\bm{U}$ even for small values of $p$ with \texttt{Prox-alt-min} being the best benchmark that recovers $\boldsymbol{U}$ when $p\leq N/2$. Because UCS is designed to recover arbitrary permutation matrices (i.e., $p=N$), it outperforms \texttt{Prox-alt-min} for high $p$ values even without knowing the value of $p$ as opposed to \texttt{Prox-alt-min} which is designed specifically for $p$-local unlabeled sensing recovery.

\subsubsection{Unlabeled sensing}
\begin{figure}[htb]
\begin{minipage}[b]{.48\linewidth}
  \centering
  {\includegraphics[scale=0.28]{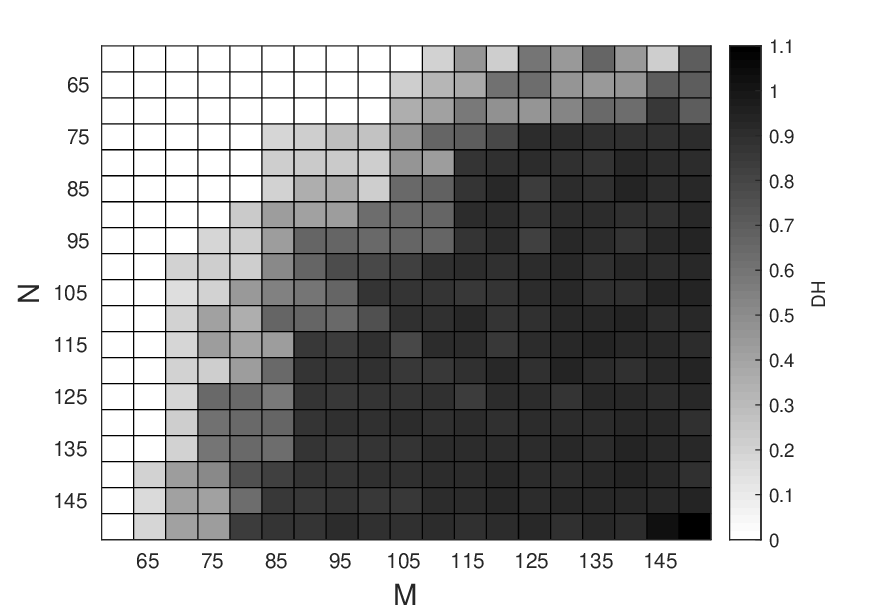}\label{fig:US-phase-transition-DH-U}}
  \centerline{\small{(a) HD($\bm{U}$, $\widehat{\bm{U}}$)}}\medskip
\end{minipage}
\hfill
\begin{minipage}[b]{0.48\linewidth}
  \centering
  \centerline{\includegraphics[scale=0.28]{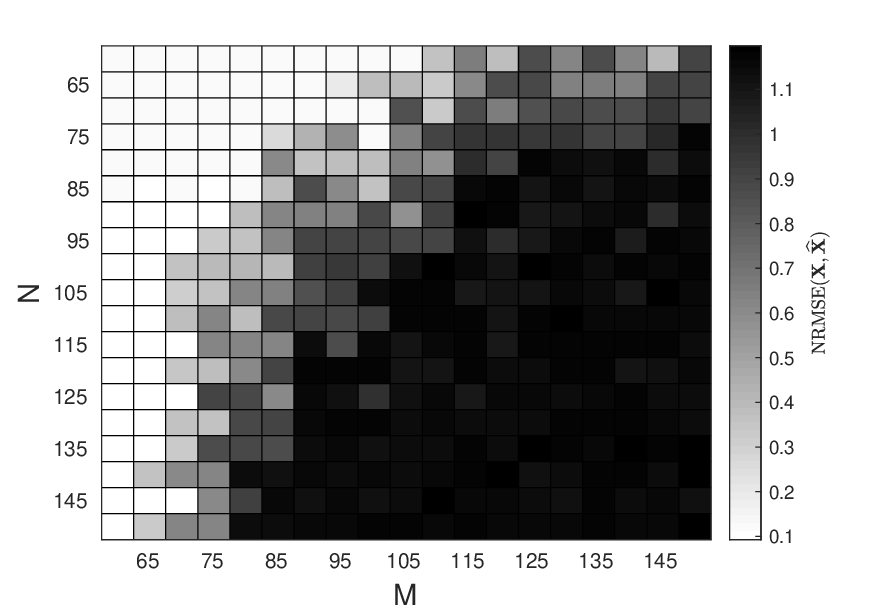}\label{fig:US-phase-transition-NRMSE-X}}
  \centerline{\small{(b) NRMSE($\bm{X}$, $\widehat{\bm{X}}$)}}\medskip
\end{minipage}
\vspace{-0.3cm}
\caption{UCS reconstruction performance of the matrices $\bm{U}$ and $\bm{X}$ for the unlabeled sensing problem over a grid of $N$ and $M$ values between 60 and 150 at SNR = 30 dB: Gaussian prior on $\bm{X}$ and $R=10$.}
\label{fig:phase-transition-U-X}
\end{figure}

The unlabeled sensing problem in (\ref{eq:unlabeled-sensing}) considers the entries $x_{ij}$ of the signal matrix $\bm{X}$ to be i.i.d. zero-mean Gaussian random variables with unit variance, i.e., $x_{ij} \sim$ $\mathcal{N}({x}_{ij};0, 1)$. Fig.~\ref{fig:phase-transition-U-X} depicts the UCS reconstruction performance of $\bm{U}$ and $\bm{X}$. Such a reconstruction undergoes a sharp phase transition as a function of the permutation size $N$. At a fixed SNR, this result is consistent with the analysis previously reported in \cite{pananjady2016linear}.

\begin{figure}[h!]
\centering
\includegraphics[scale=0.5]{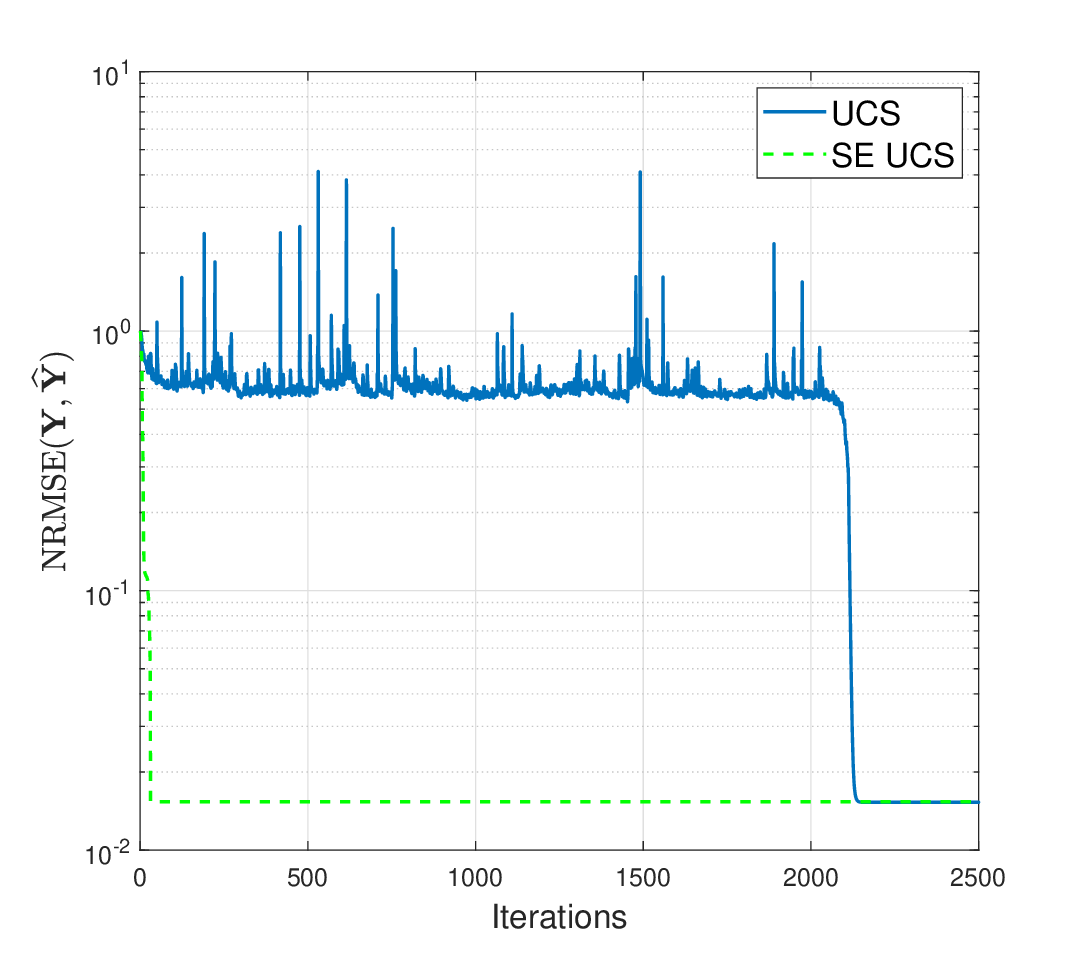}
\caption{NRMSE of UCS and its SE vs. iterations for the unlabeled sensing problem at SNR = 30 dB: Gaussian prior on $\bm{X}$ with $N=50$, $M=100$, and $R=10$.}
\label{fig:ubivamp-SE-Gauss}
\end{figure}

\noindent In Fig.~\ref{fig:ubivamp-SE-Gauss}, it is seen that the empirical NRMSE of UCS is accurately predicted by the analytical (yet non-rigorous) state evolution recursion thereby corroborating the theoretical analysis we conducted in Section \ref{sec:SE}. Moreover, the time evolution of the NRMSE brings to mind the search trajectories of optimization algorithms whose transiently chaotic behavior is known to be connected to the optimization hardness of the problem instance, and more generally to dynamical system and computational complexity theories \cite{ercsey2011optimization, sahai2020dynamical}.

\begin{figure}[h!]
\centering
\includegraphics[scale=0.49]{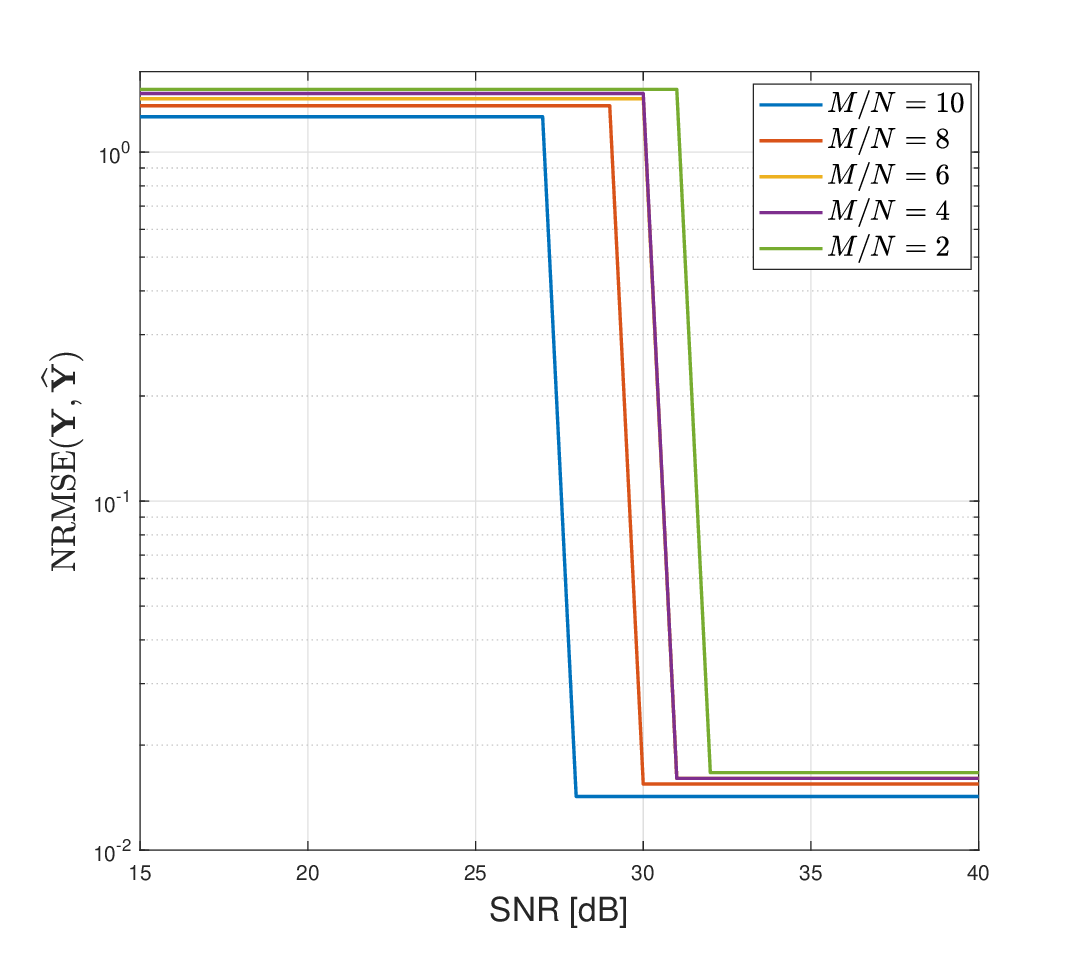}
\caption{NRMSE of SE vs. SNR for the unlabeled sensing problem for different values of the sampling rate $\frac{M}{N} \in \{2,4,6,8,10\}$: Gaussian prior on $\bm{X}$ with $N=50$ and $R=10$.}
\label{fig:UCS-SE-ratio-MN}
\end{figure}

To gain more insights into the recovery limits of UCS, we run the SE algorithm presented in Algorithm \ref{algo:u-bi-vamp-SE} over the SNR range [$15$ dB, $40$ dB] for multiple sampling rates $M/N \in \{10,8,6,4,2\}$ as shown in Fig.~\ref{fig:UCS-SE-ratio-MN}. It is observed that the sharp phase transition of the NRMSE occurs at a lower SNR threshold as the sampling rate increases. This highlights the fact that increasing the number of measurements enables better UCS recovery at lower SNR levels. We also note that all phase transitions happen at the high-SNR regime, thereby confirming the theoretical difficulty of solving the unlabeled sensing problem at medium-to-low SNRs \cite{pananjady2016linear}.

\subsubsection{Unlabeled compressed sensing} Unlike the proposed UCS technique, none of the existing algorithms for the unlabeled sensing problem is able to accommodate a sparse unknown signal matrix $\bm{X}$. In this context, we assume here that the entries $x_{ij}$ of the signal matrix $\bm{X}$ follow an i.i.d. Bernoulli-Gaussian prior, i.e., $x_{ij} \sim \rho\,\delta(x_{ij}) + (1-\rho)\,\mathcal{N}(x_{ij};0, 1)$ where $\rho\in ]0,1[$ is the sparsity ratio (i.e., the percentage of the zero components). 
\begin{figure}[h!]
\vspace{-0.3cm}
\centering
\includegraphics[scale=0.5]{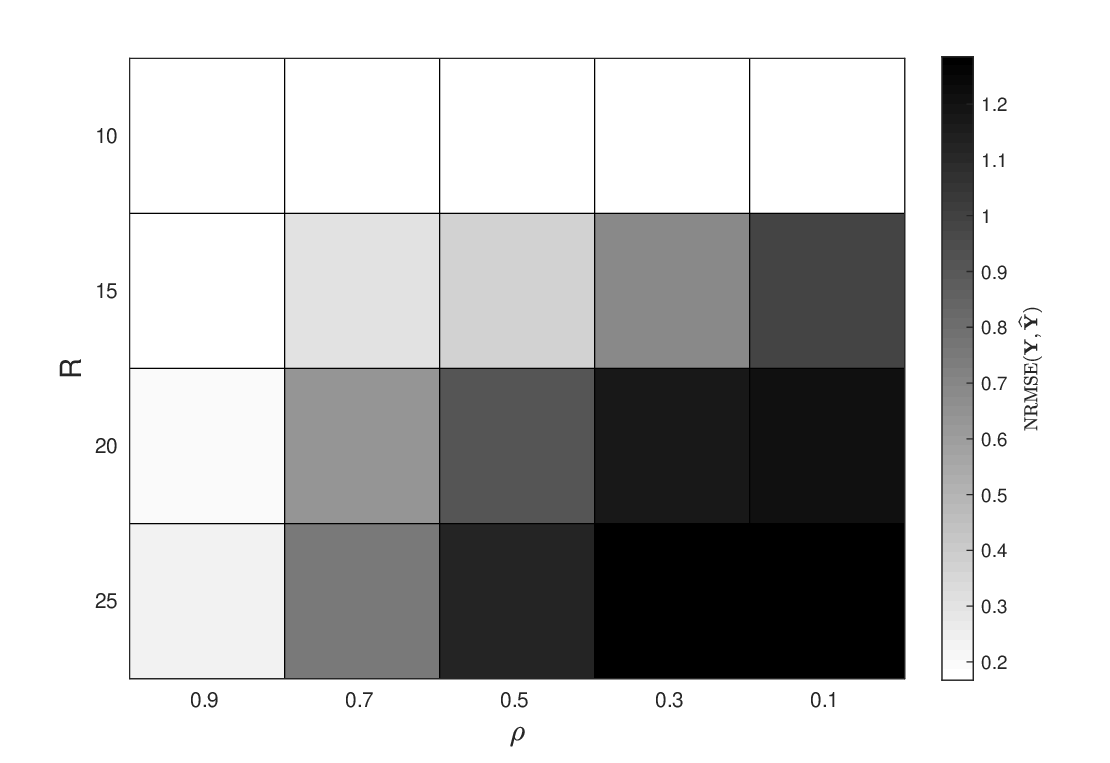}
\caption{NRMSE of UCS over a grid of ranks $R$ and sparsity ratios $\rho$ for the unlabeled sensing problem at SNR = 30 dB: Bernoulli-Gaussian prior on $\bm{X}$ with $N=50$, $M=100$.}
\label{fig:ucs-bg-phase-transition}
\end{figure}
\begin{figure*}[!b]
\vspace{-0.2cm}
\rule{\textwidth}{0.4 pt}
\begin{equation}\label{eq:tk}
t_k\left(\eta,\overbar{\textrm{LLR}}_{\textrm{diag}}^-,\overbar{\textrm{LLR}}_{{\textrm{off-diag}}}^-\right)=\frac{\delta(k-1)\cdot\exp \left(\eta+\sqrt{\eta}\,n_1+\overbar{\textrm{LLR}}_{\textrm{diag}}^-\right) + (1-\delta(k-1))\cdot\exp \left(\sqrt{\eta}\,n_1+\overbar{\textrm{LLR}}_{{\textrm{off-diag}}}^-\right)}{\exp \left(\eta+\sqrt{\eta}\,n_1 + \overbar{\textrm{LLR}}_{\textrm{diag}}^-\right)+\exp\left(\overbar{\textrm{LLR}}_{{\textrm{off-diag}}}^-\right)\sum\limits_{q=2}^{N} \exp \left(\sqrt{\eta}\,n_q\right)}
\end{equation}
\end{figure*}

\noindent Fig.~\ref{fig:ucs-bg-phase-transition} depicts the NRMSE of recovering the matrix $\bm{X}$ over a two-dimensional grid of sparsity ratios and rank values. There, it is seen that the higher the sparsity of the matrix (i.e., the fewer variables to recover), the lower the NRMSE becomes. This is in agreement with the compressed sensing theory which relies on the sparsity of the signal $\boldsymbol{X}$ to recover it from much fewer samples than what is required by the Nyquist–Shannon sampling theorem.

\section{Conclusion}\label{sec:conclusion}
In this work, we devised a new algorithm, coined unlabeled compressed sensing (UCS), to solve the unlabeled compressed sensing problem using the AMP framework. We described how UCS handles the permutation prior by putting forward a belief propagation algorithm to jointly denoise the columns and rows of the permutation matrix. Additionally, we derived the state evolution of UCS to precisely track its asymptotic behavior for large permutation and i.i.d. signal matrices. While our work extends the unlabeled sensing problem to the compressed sensing settings, one limitation of the proposed approach is that it assumes that the dimensions of the estimated matrices ($N$ and $M$ ) are large enough to have negligible error correction terms in the expectation propagation approximation. Closing this gap is a worthwhile endeavor for future research.

\begin{appendices}
\renewcommand{\thesectiondis}[2]{\Alph{section}:}
\section{Derivation of the MSE function $\mathcal{E}_{u^-}(\cdot)$}\label{appendix:MSE-U-minus}

The goal of this appendix is to show how to find the expression of $\mathcal{E}_{u^-}(\cdot)$ in (\ref{eq:final-Euk}) starting from the following expression given in (\ref{eq:u+-sensitivity-function-without-delta})--(\ref{eq:derivative-denoiser-uik-compact}):
\begin{equation}
\label{appendix:u+-sensitivity-function-without-delta}
\begin{aligned}
&\mathcal{E}_{u^-}\left(\overbar{\textrm{LLR}}_{\textrm{diag}}^-,\overbar{\textrm{LLR}}_{{\textrm{off-diag}}}^-\right) \\
&= \mathbb{E}\Big[ h^{\prime}_{{\mathsf{u}}_{ik}}\left(\bm{{b}}_{\bm{{u}}_i},\overbar{\textrm{LLR}}_{\textrm{diag}}^-,\overbar{\textrm{LLR}}_{{\textrm{off-diag}}}^-\right)\Big\vert\,p_{\bm{\mathsf{b}}_{\bm{{u}}_i}}(\bm{{b}}_{\bm{{u}}_i})\Big],
\end{aligned}
\end{equation}
with
\begin{equation}\label{appendix:derivative-denoiser-uik-compact}
\begin{aligned}[b]
&h_{\mathsf{u}_{ik}}^\prime\left(\boldsymbol{b}_{\bm{{u}}_i}, \overbar{\textrm{LLR}}_{\textrm{diag}}^-,\overbar{\textrm{LLR}}_{{\textrm{off-diag}}}^-\right) \\
    &\hspace{0.5cm}= f_k\left(\bm{{b}}_{\bm{{u}}_i},\overbar{\textrm{LLR}}_{\textrm{diag}}^-,\overbar{\textrm{LLR}}_{{\textrm{off-diag}}}^-\right)\\
    &\hspace{1.5cm}\times \Big(1-f_k\left(\bm{{b}}_{\bm{{u}}_i},\overbar{\textrm{LLR}}_{\textrm{diag}}^-,\overbar{\textrm{LLR}}_{{\textrm{off-diag}}}^-\right)\Big).
    \end{aligned}
\end{equation}

\noindent It is noteworthy that the expectation in (\ref{appendix:u+-sensitivity-function-without-delta}) involves an $N$-dimensional integral owing to the inseparable expression of $h_{\mathsf{u}_{ik}}^\prime(\cdot,\cdot,\cdot)$. To transform the $N$-dimensional integral to a $1$-dimensional integral, we first recall that the distribution $p_{\bm{\mathsf{b}}_{\bm{{u}}_i}}(\bm{{b}}_{\bm{{u}}_i})$ given in (\ref{eq:bui-pdf}) depends on whether the $k$th element of $\bm{{b}}_{\bm{{u}}_i}$, $\left[\bm{{b}}_{\bm{{u}}_i}\right]_k$, corresponds to $k=i$ or $k \neq i$. As result, we re-express/reparametrize the function $f_k\left(\cdot, \cdot,\cdot\right)$ obtained in (\ref{eq:fk}) as follows:
\begin{equation}    f_k\left(\bm{{b}}_{\bm{{u}}_i},\overbar{\textrm{LLR}}_{\textrm{diag}}^-,\overbar{\textrm{LLR}}_{{\textrm{off-diag}}}^-\right) = t_k\left(\widetilde\gamma_{\boldsymbol{U}_{ \mathsf{e}}},\overbar{\textrm{LLR}}_{\textrm{diag}}^-,\overbar{\textrm{LLR}}_{{\textrm{off-diag}}}^-\right),
\end{equation}
where $t_k(\cdot,\cdot,\cdot)$ is given in (\ref{eq:tk}). We then invoke the law of large numbers to find the limiting value of the summation $\sum\limits_{q=2}^{N} \exp \left(\sqrt{\eta}\,n_q\right)$ involved in the denominator of $t_k\left(\cdot,\cdot,\cdot\right)$ in (\ref{eq:tk}). We do so by recalling the fact that, as $N$ grows large, we have:
\begin{equation}\label{appendix:law-large-number-sum-M-function}
\begin{aligned}
    \frac{1}{N-1}\sum\limits_{q=2}^{N} \exp \left(\sqrt{\eta}\,n_q\right) &\rightarrow \hspace{-0.1cm}\int\limits_{-\infty}^{+\infty} \hspace{-0.1cm}\exp \left(\sqrt{\eta}\,n_q\right)\frac{\exp\left(-n_{q}^2/2\right)}{\sqrt{2\pi}}\,\,\text{d}n_{q}\\
    &= \hspace{0.1cm}\exp \left(\eta/2\right),
\end{aligned}
\end{equation}

\noindent where the last equality follows from the following identity:
\begin{equation*}
    \int_{-\infty}^{+\infty}\exp\Big(-(a\,x^2 + b\,x)\Big)\,\textrm{d}x = \sqrt{\frac{\pi}{a}}\,\exp\Bigg(\frac{b^2}{4a}\Bigg), ~\textrm{for}~ a>0.
\end{equation*}
\noindent Injecting (\ref{appendix:law-large-number-sum-M-function}) back into $t_k(\cdot,\cdot,\cdot)$ in (\ref{eq:tk}) yields the expression of  $g_k\left(\cdot,\cdot,\cdot\right)$ in (\ref{eq:gk}). Finally, we obtain the expression of the MSE function $\mathcal{E}_{u^-,k}(\cdot,\cdot,\cdot)$ in (\ref{eq:final-Euk}) by plugging the expression of   $g_k\left(\cdot,\cdot,\cdot\right)$ in (\ref{appendix:u+-sensitivity-function-without-delta}).
\end{appendices}
\balance
\bibliographystyle{unsrt}
\bibliography{IEEEabrv,references.bib}
 
\end{document}

%% file: CLT_approx_factor_graph.tex
\begin{tikzpicture}[thick,scale=0.75, every node/.style={transform shape}]
  \node[latent, xshift=-0.5cm, yshift=1cm]  (vj)   {$\textrm{\textbf{{\rv{v}}}}_j$};
  \node[latent, xshift=-3cm, yshift=-1.5cm]  (ui)   {$\textrm{\textbf{{\rv{u}}}}_i$};
  \node[latent, right=of ui, xshift=2cm, yshift=0cm, fill=light-gray]  (yij)   {$\bm{\mathsf{Y}}$};
  \node (v1) at ($(vj) + (-0.2,-1.5)$) {};
  \node (vM) at ($(vj) + (0.2,-1.5)$) {};
  \node (u1) at ($(ui) + (1.5,-0.2)$) {};
  \node (uN) at ($(ui) + (1.5,0.2)$) {};
  %
  \factor[right=of u1, xshift=0.2cm,yshift=0.2cm, xscale=2.5, yscale=2.5] {fij}  {below:
  } {ui, vj, yij, v1, vM, u1, uN}{};
  \node at ($(fij) + (-0.05,-0.8)$) {{$p_{\bm{\mathsf{Y}}|\bm{\mathsf{U}}, \bm{\mathsf{V}}}(\bm{Y}|\bm{U},\bm{V})$}};
  \node at ($(fij) + (-0.001,0.92)$) {\yourtightDots};
  \node at ($(fij) + (-0.94,-0.01)$) [rotate=90]{\yourtightDots};
  %
  \node (u_f_right) at ($(ui)!0.5!(fij) + (0,0.25)$) {};
  \node (u_f_left) at ($(ui)!0.5!(fij) + (-0.8,0.25)$) {};
  \draw [-latex,very thick] (u_f_right.center) -- (u_f_left.center);
  \draw[] (ui) to node[pos=0.3,above=0.85em,align=center]{
{\scriptsize \protect\tikz[inner sep=0.4ex,baseline=.4ex] \protect\node[circle,draw] {\large{$1$}};}
  } (fij);
  %
  \node (v_f_right) at ($(vj)!0.5!(fij) + (-0.25,0)$) {}; 
  \node (v_f_left) at ($(vj)!0.5!(fij) + (-0.25,0.8)$) {};
  \draw [-latex,very thick] (v_f_right.center) -- (v_f_left.center);
  \draw[] (vj) to node[pos=0.3,right=0.1em,align=center,xshift=-1cm, yshift=0cm]{
{\scriptsize \protect\tikz[inner sep=.4ex,baseline=-.75ex] \protect\node[circle,draw] {\large{$2$}};}
  } (fij);
  \factor[left=of ui, xscale=2.5, yscale=2.5] {pui} {left:\large{$p_{\textrm{\textbf{{\rv{u}}}}_{i}}(\bm{u}_i)$}} {ui} {};
  \factor[above=of vj, xscale=2.5, yscale=2.5] {pvj} {right:\large{$p_{\textrm{\textbf{{\rv{v}}}}_{j}}(\bm{v}_j)$}} {vj} {};
\end{tikzpicture}

%% file: algorithm2-new.tex
\begin{algorithm*}[h!]
\footnotesize
\caption{UCS Recovery  Algorithm}\label{algo:big-vamp}
\begin{multicols}{2}
\begin{algorithmic}[1]
\Statex $\mathbf{Require:}$ Measurement matrix $\boldsymbol{Y}$ $\in$ $\mathbb{R}^{N \times M}$; sensing matrix $\boldsymbol{A}$ $\in$ $\mathbb{R}^{N \times R}$; temperature parameter $\beta$ set to 1; precision tolerance ($\xi=10^{-6}$);  maximum number of iterations ($T_\textrm{max}$);  one denoiser $\mathbf{g}_{\mathsf{x}}(\cdot)$; noise precision $\gamma_w$.\vspace{0.1cm}  
\State $\mathbf{Initialize}$
\State $t\gets 1$
\Statex\LeftCommentNoIntent{posterior means, covariances and precisions} 
\Statex $\boldsymbol{\widehat{U}}^{-}_{ \mathsf{p}, 0}$, $\boldsymbol{\widehat{X}}^{-}_{ \mathsf{p}, 0}$,
$\boldsymbol{\widehat{U}}_{ \mathsf{p}, 1}^{-}$, $\boldsymbol{\widehat{V}}_{  \mathsf{p}, 1}^{-}$,
$\boldsymbol{\widehat{X}}^{-}_{ \mathsf{p}, 1}$, $\boldsymbol{R}_{\boldsymbol{U}_{ \mathsf{p}}^{-}, 1}$, $\boldsymbol{R}_{\boldsymbol{V}^{-}_{ \mathsf{p}}, 1}$, $\boldsymbol{R}_{\boldsymbol{X}^{-}_{ \mathsf{p}}, 1}$
\Statex $\boldsymbol{\widehat{U}}_{ \mathsf{p}, 1}^{+}$, $\boldsymbol{\widehat{X}}_{ \mathsf{p}, 1}^{+}$, $\gamma_{\boldsymbol{U}^{-}_{ \mathsf{p}}, 1}$, $\gamma_{\boldsymbol{X}^{-}_{ \mathsf{p}}, 1}$  
\Statex\LeftCommentNoIntent{extrinsic means and precisions}
\Statex $\boldsymbol{\widehat{U}}^{-}_{ \mathsf{e}, 1}$, $\boldsymbol{\widehat{X}}^{-}_{ \mathsf{e}, 1}$, $\gamma_{\boldsymbol{U}^{-}_{ \mathsf{e}}, 1}$, $\gamma_{\boldsymbol{X}^{-}_{ \mathsf{e}}, 1}$
\Statex  $\boldsymbol{\widehat{X}}^{+}_{ \mathsf{e}, 1}$, $\boldsymbol{\widehat{X}}^{+}_{ \mathsf{e}, 1}$, $\gamma_{\boldsymbol{U}^{+}_{ \mathsf{e}}, 1}$, $\gamma_{\boldsymbol{X}^{+}_{ \mathsf{e}}, 1}$ \vspace{0.1cm}
\Statex\LeftCommentNoIntent{means and precisions of the Bi-VAMP module}
\Statex $\boldsymbol{B}_{\boldsymbol{U}, 1}$, $\boldsymbol{\Lambda}_{\boldsymbol{U}, 1}$, $\boldsymbol{B}_{\boldsymbol{V}, 1}$, $\boldsymbol{\Lambda}_{\boldsymbol{V}, 1}$, $\boldsymbol{B}_{\boldsymbol{X}, 1}$, $\boldsymbol{\Lambda}_{\boldsymbol{X}, 1}$,\vspace{0.1cm}
\Statex\LeftCommentNoIntent{extrinsic LLR of $\widehat{\boldsymbol{u}}^{-}_{ik,\mathsf{e}}$, $\textrm{LLR}_{{u}_{ik}^-, \textsf{e},1}$, $\,\forall i\,\,\forall k$}
\Repeat
\Statex\LeftComment{1}{Compute the change of variable in (\ref{eq:change-variable-V-AX}) (from $\boldsymbol{X}$ to $\boldsymbol{V}$)}
\State $ {{}\widehat{\boldsymbol{V}}^{-}_{ \mathsf{p}, t}} = \left(\boldsymbol{A}\,\boldsymbol{\widehat{X}}^{-}_{ \mathsf{p}, t}\right)^{\top} ,  ~~\boldsymbol{R}_{\boldsymbol{V}_{ \mathsf{p}}^{-}, t} = \boldsymbol{A}\,\boldsymbol{R}_{\boldsymbol{X}_{ \mathsf{p}}^{-}, t}\, \boldsymbol{A}^{\top} $\vspace{0.15cm}
\Statex\LeftCommentNoTriangle{1}{\textcolor{blue-violet}{\underline{I. Bi-VAMP module to denoise of $\boldsymbol{X}^-$ and $\boldsymbol{U}^-$}}}
\Statex\LeftComment{1}{Compute an approximation of all incoming messages to all $\textrm{\textbf{{\rv{u}}}}_i$}
\StateBlue $\boldsymbol{B}_{\boldsymbol{U}, t} = \gamma_{w} \Big(
{\boldsymbol{Y}} \,{{}\widehat{\boldsymbol{V}}^{-}_{ \mathsf{p}, t}}\, - M\,\gamma_{w}\,{}\widehat{\boldsymbol{U}}^{-}_{ \mathsf{p}, t-1} \,\boldsymbol{R}_{\boldsymbol{V}_{ \mathsf{p}}^{-}, t}  \,\langle\boldsymbol{Y} \odot\boldsymbol{Y}\rangle\Big)$ \label{eq:algo-Bu-non-low-rank}
\State $\boldsymbol{\Lambda}_{\boldsymbol{U}, t} = \gamma_{w}\Big(\boldsymbol{\widehat{V}}^{- \top}_{ \mathsf{p}, t} \boldsymbol{\widehat{V}}^{-}_{ \mathsf{p}, t}  + \frac{M}{\beta}\,\boldsymbol{R}_{\boldsymbol{V}_{ \mathsf{p}}^{-}, t} -\, M\,\gamma_{w} \,\boldsymbol{R}_{\boldsymbol{V}_{ \mathsf{p}}^{-}, t} \,\langle\boldsymbol{Y} \odot\boldsymbol{Y}\rangle\Big)$ \label{eq:algo-Au-non-low-rank}\vspace{0.1cm}
\Statex\LeftComment{1}{Compute an approximation to the incoming messages to all $\textrm{\textbf{{\rv{v}}}}_j$}
\State $\boldsymbol{B}_{\boldsymbol{V}, t} = \gamma_{w} \Big(
{\boldsymbol{Y}}^{\top} \,{{}\widehat{\boldsymbol{U}}^{-}_{ \mathsf{p}, t}}\, - N\,\gamma_{w}\,{}\widehat{\boldsymbol{V}}^{-}_{ \mathsf{p}, t-1} \, \boldsymbol{R}_{\boldsymbol{U}_{ \mathsf{p}}^{-}, t}\,\langle\boldsymbol{Y} \odot\boldsymbol{Y}\rangle\Big)$\label{eq:algo-Bv-non-low-rank}\vspace{0.05cm}
\State $\boldsymbol{\Lambda}_{\boldsymbol{V}, t} = \gamma_{w}\Big(\boldsymbol{\widehat{U}}^{-\top}_{ \mathsf{p}, t} \, \boldsymbol{\widehat{U}}^{-}_{ \mathsf{p}, t}  + \frac{N}{\beta}\,\boldsymbol{R}_{\boldsymbol{U}_{ \mathsf{p}}^{-}, t} -\, N\,\gamma_{w} \,\boldsymbol{R}_{\boldsymbol{U}_{ \mathsf{p}}^{-}, t}\,\langle\boldsymbol{Y} \odot\boldsymbol{Y}\rangle\Big)$
\label{eq:algo-Av-non-low-rank}\vspace{0.1cm}
\Statex\LeftComment{1}{Compute the change of variable in (\ref{eq:change-variable-X-to-V}) (from $\boldsymbol{V}$ to $\boldsymbol{X}$)}
\State $ \boldsymbol{B}_{\boldsymbol{X}, t} = \,\boldsymbol{A}^\top\,\boldsymbol{B}_{\boldsymbol{V}, t} ~~,  ~~\boldsymbol{\Lambda}_{\boldsymbol{X}, t} = \boldsymbol{A}^\top\,\boldsymbol{\Lambda}_{\boldsymbol{V}, t}\, \boldsymbol{A} $\vspace{0.1cm} \label{eq:algo-change-variable-V-to-X}

\Statex\LeftComment{1}{Update the posterior statistics $\boldsymbol{\widehat{X}}_{ \mathsf{p},t+1}^{-}$, $\boldsymbol{R}_{\boldsymbol{X}_{ \mathsf{p}}^{-}, t+1}$}
\State $\boldsymbol{R}_{\boldsymbol{X}_{ \mathsf{p}}^{-}, t+1} = (\gamma_{\boldsymbol{X}^{-}_{ \mathsf{e}}, t} \, \mathbf{I}_R + \boldsymbol{\Lambda}_{\boldsymbol{X},t})^{-1}$
\label{eq:posterior-covariance-v-gaussian-approx}
\State $\boldsymbol{\widehat{X}}_{ \mathsf{p},t+1}^{-}\;\;=(\boldsymbol{B}_{\boldsymbol{X},t}+\gamma_{\boldsymbol{X}^{-}_{ \mathsf{e}},t}\boldsymbol{\widehat{X}}_{ \mathsf{e},t}^{-}) \,\boldsymbol{R}_{\boldsymbol{X}_{ \mathsf{p}}^{-}, t+1}$\label{eq:posterior-mean-v-gaussian-approx}
\Statex\LeftComment{1}{Update the posterior statistics $\boldsymbol{\widehat{U}}_{ \mathsf{p}, t+1}^{-}$ and $\boldsymbol{R}_{\boldsymbol{U}_{ \mathsf{p}}^{-}, t+1}$ using (\ref{eq:posterior-mean-variance-U-final})}
\StateBlue $\forall i \,\,\forall k\,, \widehat{u}^{-}_{ik, \mathsf{p}, t+1}=\frac{\exp \left(\left[\boldsymbol{b}_{\bm{u}_i,t}\right]_k-\frac{1}{2}\,[\mathbf{\Lambda}_{\bm{U},t}]_{kk} +\, \textrm{LLR}_{{u}_{ik}^-, \textsf{e},t}\right)}{\sum\limits_{l=1}^{N}\,\exp \left(\left[\boldsymbol{b}_{\bm{u}_i,t}\right]_l-\frac{1}{2}\,[\mathbf{\Lambda}_{\bm{U},t}]_{ll} +\, \textrm{LLR}_{{u}_{il}^-, \textsf{e},t}\right)}$\vspace{0.1cm}\label{algo:eq-Up-minus-with-LRs}
\StateBlue $\forall i \,\,\forall k,\, \textrm{LLR}_{{u}_{ik}^-, \textsf{p},t+1} = \ln\left(\widehat{u}^{-}_{ik, \mathsf{p},t+1}\Big/\big(1-\widehat{u}^{-}_{ik, \mathsf{p},t+1}\big)\right)$\vspace{0.1cm}
\StateBlue $\forall i \,\,\forall k\,,\left[\boldsymbol{R}_{\boldsymbol{u}_{\mathsf{p}}^{-},t+1}\right]_{ik} = \widehat{u}^{-}_{ik,\mathsf{p},t+1} - \big(\widehat{u}^{-}_{ik,\mathsf{p},t+1}\big)^2$\label{algo:eq-varUp-minus-with-LRs}\vspace{0.05cm}
\StateBlue $\boldsymbol{R}_{\boldsymbol{U}_{ \mathsf{p}}^{-}, t+1} = (1/N)\,\cdot\,\sum\limits_{i=1}^{N} \boldsymbol{R}_{\bm{u}_{\mathsf{p}}^{-},t+1}$\label{algo:sum-Ru}
\vspace{0.05cm}
\Statex\LeftComment{1}{compute the extrinsic LLR of $\boldsymbol{{U}}_{t+1}^{+}$ in (\ref{eq:extrinsic-U-final})}\vspace{0.1cm}
\StateBlue $\forall i \,\,\forall k,\, \textrm{LLR}_{{u}_{ik}^+, \textsf{e},t+1} = \textrm{LLR}_{{u}_{ik}^-, \textsf{p},t+1} - \textrm{LLR}_{{u}_{ik}^-, \textsf{e},t}$\label{algo:eq-LRe-from-LRp-minus}\vspace{0.1cm}
\Statex\LeftComment{1}{Update the extrinsic statistics $\boldsymbol{\widehat{X}}_{ \mathsf{e},t+1}^{+}$ and $\gamma_{\boldsymbol{X}^{+}_{ \mathsf{e}}, t+1}$}
\StateBlack $\gamma_{\boldsymbol{X}_{ \mathsf{p}}^{-},t+1}\,=\,\big(\frac{1}{R}\text{Tr}(\boldsymbol{R}_{\boldsymbol{X}_{ \mathsf{p}}^{-}, t+1})\big)^{-1}$ \label{eq:algo-diag-sigma-v}
\State $\gamma_{\boldsymbol{X}^{+}_{ \mathsf{e}},t+1} = \gamma_{\boldsymbol{X}_{ \mathsf{p}}^{-},t+1} -\, \gamma_{\boldsymbol{X}^{-}_{ \mathsf{e}}, t}$ \label{eq:algo-diag-sigma-v-e}
\State $\boldsymbol{\widehat{X}}_{ \mathsf{e},t+1}^{+} \;\,= \gamma^{-1}_{\boldsymbol{X}^{+}_{ \mathsf{e}},t+1}\,\big(\gamma_{\boldsymbol{X}_{ \mathsf{p}}^{-},t+1}\boldsymbol{\widehat{X}}_{ \mathsf{p},t+1}^{-} \, - \,\gamma_{\boldsymbol{X}^{-}_{ \mathsf{e}},t}\boldsymbol{\widehat{X}}_{ \mathsf{e},t}^{-} \big)$\label{eq:algo-diag-mean-v}\vspace{0.1cm}
\Statex\LeftCommentNoTriangle{1}{\textcolor{carnelian}{\underline{II. denoising module of $\boldsymbol{X}^+$}}}\vspace{0.05cm}
\Statex\LeftCommentNoIntent{$\forall j:$ update the $j$th column of $\boldsymbol{\widehat{X}}_{ \mathsf{p},t+1}^{+}$}
\StateRed $\widehat{\boldsymbol{x}}^{+}_{j,\mathsf{p},t+1}\,=\, \mathbf{g}_{\mathsf{x}}(\widehat{\boldsymbol{x}}^{+}_{j,\mathsf{e},t+1}, \gamma^{-1}_{\boldsymbol{X}^{+}_{ \mathsf{e}},t+1}),$ \label{denoiser_v}
\StateRed $\gamma_{\boldsymbol{X}^{+}_{ \mathsf{p}},t+1} = \gamma_{\boldsymbol{X}^{+}_{ \mathsf{e}},t+1} \,\left(\frac{1}{M}\sum_{j=1}^M\big\langle \mathbf{g}^{\prime}_{\mathsf{x}}(\widehat{\boldsymbol{x}}^{+}_{j,\mathsf{e},t+1}, \gamma^{-1}_{\boldsymbol{X}^{+}_{ \mathsf{e}},t+1})\big\rangle\right)^{-1} $\label{eq:algo-diag-sigma-v-alpha}\vspace{0.1cm}
\Statex\LeftComment{1}{update the extrinsic statistics $\boldsymbol{\widehat{X}}_{ \mathsf{e},t+1}^{-}$, $\gamma_{\boldsymbol{X}^{-}_{ \mathsf{e}},t+1}$} 
\StateBlack $\gamma_{\mathbf{X}^{-}_{ \mathsf{e}},t+1} = \gamma_{\boldsymbol{X}^{+}_{ \mathsf{p}},t+1}- \gamma_{\boldsymbol{X}^{+}_{ \mathsf{e}},t+1}$\label{algo:eq-extrinsic-precision}\vspace{0.1cm}
\State $\boldsymbol{\widehat{X}}_{ \mathsf{e},t+1}^{-} \;\,= \gamma^{-1}_{\boldsymbol{X}^{-}_{ \mathsf{e}},t+1}\,\big(\gamma_{\boldsymbol{X}^{+}_{ \mathsf{p}},t+1}\,\boldsymbol{\widehat{X}}_{ \mathsf{p},t+1}^{+} - \gamma_{\boldsymbol{X}^{+}_{ \mathsf{e}},t+1}\boldsymbol{\widehat{X}}_{ \mathsf{e},t+1}^{+} \big)$ \label{eq:algo-mean-Ve+}\vspace{0.15cm}
\Statex\LeftCommentNoTriangle{1}{\textcolor{darkpastelgreen}{\underline{IV. denoising module of $\boldsymbol{U}^+$}}}\vspace{0.005cm}
\Statex\LeftComment{1}{Update the posterior mean $\boldsymbol{\widehat{U}}_{ \mathsf{p},t+1}^{+}$ using (\ref{eq:posterior-U-transpose-final})}
 \StateGreen $\forall i \,\,\forall k\,,\widehat{u}^{+}_{ik,\mathsf{p},t+1} = \small{\left(1+\sum\limits_{\substack{k'=1\\k' \neq k}}^{N}\exp\Big(\textrm{LLR}_{{u}^{+}_{ik'},\textsf{e},t+1}-\,\textrm{LLR}_{{u}^{+}_{ik},\textsf{e},t+1}\Big)\right)^{-1}}$\label{algo:eq-Up-plus-LRR}
 \StateGreen $\forall i \,\,\forall k\,, \textrm{LLR}_{{u}_{ik}^+, \textsf{p},t+1} = \ln\left(\widehat{u}^{+}_{ik,\mathsf{p}, t+1}\Big/\big(1-\widehat{u}^{+}_{ik,\mathsf{p}, t+1}\big)\right)$\vspace{0.1cm}
 \Statex\LeftComment{1}{compute the extrinsic LLR of $\boldsymbol{{U}}_{t+1}^{-}$ in (\ref{eq:extrinsic-U-final})}\vspace{0.1cm}
\StateGreen $\forall i \,\,\forall k\,, \textrm{LLR}_{{u}_{ik}^-, \textsf{e},t+1} = \textrm{LLR}_{{u}_{ik}^+, \textsf{p},t+1} - \textrm{LLR}_{{u}_{ik}^+, \textsf{e},t+1}$\label{algo:eq-LRe-from-LRp-plus}\vspace{0.1cm}
 \StateBlack $t \gets t + 1$\vspace{0.1cm}
 \Until{$\Big(\big|\!\big|\boldsymbol{\widehat{U}}_{ \mathsf{p},t+1}^{+}-\boldsymbol{\widehat{U}}_{ \mathsf{p},t}^{+}\big|\!\big|^2_{\textrm{F}} + \big|\!\big|\boldsymbol{\widehat{X}}_{ \mathsf{p},t+1}^{+}-\boldsymbol{\widehat{X}}_{ \mathsf{p},t}^{+}\big|\!\big|^2_{\textrm{F}}\Big)$}
 \Statex \hspace{2cm}$\leq\xi\Big( \big|\!\big|\boldsymbol{\widehat{U}}_{ \mathsf{p},t}^{+}\big|\!\big|^2_{\textrm{F}} + \big|\!\big|\boldsymbol{\widehat{X}}_{ \mathsf{p},t}^{+}\big|\!\big|^2_{\textrm{F}}$\Big)~~\textsf{or}~~ \Big($t>T_\textrm{max}$\Big) \vspace{0.1cm}
\State \textbf{return} $\boldsymbol{\widehat{U}}_{ \mathsf{p}, T_{\textrm{max}}+1}^{-}$, $\boldsymbol{\widehat{X}}_{ \mathsf{p}, T_{\textrm{max}}+1}^{-}$
\end{algorithmic}
\end{multicols}
\end{algorithm*}

%% file: block-diagram-with-transpose.tex
\begin{figure*}[h!]
\centering
\begin{tikzpicture}[thick,scale=0.9, every node/.style={transform shape}]
  \node[block, fill=green!15, text width=2.5cm] (p_u) {\large Column-wise\\Denoiser \\\vspace {0.2 cm} $p_{\textrm{\textbf{{\rv{U}}}}^+}(\bm{U}^+)$};
  \node[block, fill=blue!15, right= 5cm of p_u,text width=4.5cm] (z_uv) {\large $\text{Bi-LMMSE}$\\\vspace {0.3 cm}
  $\bm{Y} = \bm{U}^-(\mathbf{A}\bm{X}^-)^{\top}+\bm{W}$\\\vspace {0.3cm}
  };
  \node[block, fill=carnelian!15, right= 5cm of z_uv,text width=2.5cm] (phi_z){\large Denoiser \\\vspace {0.2 cm} $p_{\textrm{\textbf{{\rv{X}}}}^+}(\bm{X}^+)$};
  \node[blockExt,right=of p_u, xshift=0.4cm, yshift=-1.5cm] (ext_pu_to_z_uv) {$\mathrm{\textbf{ext-BP}}$};
  \node[blockExt,left=of z_uv, xshift=-0.4cm, yshift=1.5cm] (ext_z_uv_to_p_u) {$\mathrm{\textbf{ext-BP}}$};
  \node[blockExt,right=of z_uv, xshift=0.4cm, yshift=-1.5cm] (ext_z_uv_to_y) {$\mathrm{\textbf{ext-EP}}$};
  \node[blockExt,left=of phi_z, xshift=-0.4cm, yshift=1.5cm] (ext_y_to_z_uv) {$\mathrm{\textbf{ext-EP}}$};
  \draw [-latex,very thick] ([yshift=-4.25em]p_u.east) -- 
  node [midway,below=0em,align=center ] { \footnotesize{$\text{LLR}_{\bm{{U}}^{+}\hspace{-0.06cm},{ \textsf{p}}}$}}
  (ext_pu_to_z_uv.west);
  \draw [-latex,very thick] (ext_pu_to_z_uv) --
  node [midway,below=0em,align=center ] { \footnotesize{$\text{LLR}_{\bm{{U}}^{-}\hspace{-0.06cm},{ \textsf{e}}}$}}
  ([yshift=-4.25em]z_uv.west)
  node [pos=0.125](ext_between_pu_z_uv){};
  \draw [-latex,very thick] ([yshift=-4.25em]z_uv.east) --
  node [midway,below=0em,align=center ] { $\boldsymbol{\widehat{X}}_{ \textsf{p}}^{-}$}
  node [midway,below=1.6em,align=center ] {$\gamma_{\boldsymbol{X}^{-}_{ \textsf{p}}}$}
  (ext_z_uv_to_y.west);
  \draw [-latex,very thick] (ext_z_uv_to_y) --
  node [midway,below=0em,align=center ] { $\boldsymbol{\widehat{X}}_{ \textsf{e}}^{+}$}
  node [midway,below=1.7em,align=center ] {$\gamma_{\boldsymbol{X}^{+}_{ \textsf{e}}}$}
  ([yshift=-4.25em]phi_z.west)
  node [pos=0.125](ext_between_z_uv_y){};
  \draw [-latex,very thick] ([yshift=4.25em]phi_z.west) --
  node [midway,above=0em,align=center ] { $\boldsymbol{\widehat{X}}_{ \textsf{p}}^{+}$}
  node [midway,above=1.6em,align=center ] {$\gamma_{\boldsymbol{X}^{+}_{ \textsf{p}}}$}
  (ext_y_to_z_uv.east);
  \draw [-latex,very thick] (ext_y_to_z_uv) --
  node [midway,above=0em,align=center ] { $\boldsymbol{\widehat{X}}_{ \textsf{e}}^{-}$}
  node [midway,above=1.7em,align=center ] {$\gamma_{\boldsymbol{X}^{-}_{ \textsf{e}}}$}
  ([yshift=4.25em]z_uv.east)
  node [pos=0.125](ext_between_y_z_uv){};
  \draw [-latex,very thick] ([yshift=4.25em]z_uv.west) --
  node [midway,above=0em,align=center ] { \footnotesize{$\text{LLR}_{\bm{{U}}^{-}\hspace{-0.06cm},{ \textsf{p}}}$}}
  (ext_z_uv_to_p_u.east);
  \draw [-latex,very thick] (ext_z_uv_to_p_u) --
  node [midway,above=0em,align=center ] { \footnotesize{$\text{LLR}_{\bm{{U}}^{+}\hspace{-0.06cm},{\textsf{e}}}$}}
  ([yshift=4.25em]p_u.east)
  node [pos=0.125](ext_between_z_uv_pu){};
  \draw [-latex,very thick] (ext_between_pu_z_uv.center) --
  (ext_z_uv_to_p_u.south);
  \draw [-latex,very thick] (ext_between_z_uv_y.center) --
  (ext_y_to_z_uv.south);
  \draw [-latex,very thick] (ext_between_y_z_uv.center) --
  (ext_z_uv_to_y.north);
  \draw [-latex,very thick] (ext_between_z_uv_pu.center) --
  (ext_pu_to_z_uv.north);
\end{tikzpicture}
\caption{Block diagram of the proposed UCS recovery algorithm with its three modules: the two prior modules $p_{\textrm{\textbf{{\rv{U}}}}^+}(.)$ and $p_{\textrm{\textbf{{\rv{X}}}}^+}(.)$, and the Bi-LMMSE module. The latter exchanges extrinsic information/messages with the prior modules $p_{\textrm{\textbf{{\rv{U}}}}^+}(.)$ and $p_{\textrm{\textbf{{\rv{X}}}}^+}(.)$ through the \protect\tikz[inner sep=.25ex,baseline=-.75ex] \protect\node[rectangle,draw,thick,minimum width=0.45cm,minimum height=0.45cm] {\footnotesize \textbf{ext-BP}}; and \protect\tikz[inner sep=.25ex,baseline=-.75ex] \protect\node[rectangle,draw,thick,minimum width=0.45cm,minimum height=0.45cm] {\footnotesize \textbf{ext-EP}}; blocks, respectively. The LLR messages (to the left of Bi-LMMSE) are calculated using belief propagation while denoising $\boldsymbol{U}^+$. The Gaussian messages (to the right of Bi-LMMSE) are calculated using the expectation propagation principle while denoising $\boldsymbol{X}^+$. The color of each module matches the color of the corresponding line numbers in Algorithm~\ref{algo:big-vamp}.
}
\label{fig:block-diagram}
\end{figure*}

%% file: block-diagram-SE-bivamp.tex
\begin{tikzpicture}[thick,scale=0.87, every node/.style={transform shape}]

\node[] at (3.0,-1.6) {};

\node[] at (3.0,-0.5) {\small{\textrm{bi-LMMSE module}}};
\node[] at (3,-0.9) {\small{\textrm{of $\bm{U}^-$ and $\bm{V}^-$}}};
\draw[dashdotted] (0.9,-1.2) -- (0.9,1.4);
\draw[dashdotted] (5,-1.2) -- (5,1.4);
\draw (1.4, 0.75) rectangle ++ (1.2, 0.55);
\node[] at (2,1) {{$\mathcal{E}_{u^-}(\cdot)$}};
\node[] at (2.0,0.4) {\scriptsize{\textrm{LMMSE of}}};
\node[] at (2.1,0.15) {\scriptsize{\textrm{i.i.d. $\bm{U}^-$}}};
\draw (3.45, 0.75) rectangle ++ (1.1, 0.55);
\node[] at (4,1) {{$\mathcal{E}_{v^-}(\cdot)$}};
\node[] at (4.0,0.4) {\scriptsize{\textrm{LMMSE of}}};
\node[] at (4.1,0.15) {\scriptsize{\textrm{i.i.d. $\bm{V}^-$}}};


\node[] at (-0.3,-0.5) {\small{\textrm{prior module}}};
\node[] at (-0.3,-0.9) {\small{\textrm{of $\bm{U}^+$}}};
\draw (-0.75, 0.75) rectangle ++ (1.1, 0.55);
\node[] at (-0.2,1) {{$\mathcal{E}_{u^+}(\cdot)$}};
\draw [-stealth] (1.4,0.85) -- (0.35,0.85);
\draw [-stealth] (0.35,1.15) -- (1.4,1.15);
\node[] at (-0.2,0.4) {\scriptsize{\textrm{MMSE of}}};
\node[] at (-0.1,0.15) {\scriptsize{\textrm{i.i.d. $\bm{U}^+$}}};

\node[] at (6.1,-0.5) {\small{\textrm{prior module}}};
\node[] at (6.1,-0.9) {\small{\textrm{of $\bm{V}^+$}}};
\draw (5.5, 0.75) rectangle ++ (1.1, 0.55);
\node[] at (6.05,1) {{$\mathcal{E}_{v^+}(\cdot)$}};

\draw [-stealth] (5.5,0.85) -- (4.55,0.85);

\draw [-stealth] (4.55,1.15) -- (5.5,1.15);
\node[] at (6.05,0.4) {\scriptsize{\textrm{MMSE of}}};
\node[] at (6.15,0.15) {\scriptsize{\textrm{i.i.d. $\bm{V}^+$}}};
\end{tikzpicture}

%% file: block-diagram-SE-permutation.tex
\begin{tikzpicture}[thick,scale=0.87, every node/.style={transform shape}]

\node[] at (3.0,-0.5) {\small{\textrm{bi-LMMSE module}}};
\node[] at (3,-0.9) {\small{\textrm{of $\bm{U}^-$ and $\bm{X}^-$ and}}};
\node[] at (3,-1.4) {\small{\textrm{row-wise denoising of $\bm{U}^-$}}};
\draw[dashdotted] (0.9,-1.6) -- (0.9,2.75);
\draw[dashdotted] (5,-1.6) -- (5,2.75);
\draw[fill=blue!15] (1.4, 0.75) rectangle ++ (1.2, 0.55);
\node[] at (2,1) {{$\mathcal{E}_{u^-}(\cdot)$}};
\node[] at (2.03,0.4) {\scriptsize{\textrm{LMMSE + row-wise}}};
\node[] at (2.1,0.15) {\scriptsize{\textrm{denoising of $\bm{U}^-$}}};
\draw[fill=blue!15] (3.45, 0.75) rectangle ++ (1.1, 0.55);
\node[] at (4,1) {{$\mathcal{E}_{x^-}(\cdot)$}};
\node[] at (4.0,0.4) {\scriptsize{\textrm{LMMSE of}}};
\node[] at (4.1,0.15) {\scriptsize{\textrm{i.i.d. $\bm{X}^-$}}};
%
\draw[fill=blue!15] (1.4, 2.05) rectangle ++ (1.2, 0.6);
\node[] at (2,2.35) {{$\mathcal{L}_{u^-_{ik}}(\cdot)$}};
\node[] at (2.03,1.85) {\scriptsize{~\textrm{LLR values of $\bm{U}^-$}}};
\draw[-stealth] (-0.5,2.5) --(1.4,2.5);
\draw (-0.5,2.515) --(-0.5,1.34);
\draw (-0.2,2.15) -- (1.4,2.15);
\draw [-stealth] (-0.2,2.165) -- (-0.2,1.34);

\node[] at (-0.4,-0.5) {\small{\textrm{column-wise prior }}};
\node[] at (-0.3,-0.9) {\small{\textrm{module of $\bm{U}^+$}}};
\draw[fill=green!15] (-0.84, 0.7) rectangle ++ (1.15, 0.63);
\node[] at (-0.25,1) {{$\mathcal{L}_{u^+_{ik}}(\cdot)$}};
\draw [-stealth] (1.4,0.85) -- (0.35,0.85);
\draw [-stealth] (0.35,1.15) -- (1.4,1.15);
\node[] at (-0.2,0.4) {\scriptsize{\textrm{LLR values}}};
\node[] at (-0.2,0.15) {\scriptsize{\textrm{of $\,\bm{U}^+$}}};

\node[] at (6.1,-0.5) {\small{\textrm{prior module}}};
\node[] at (6.1,-0.9) {\small{\textrm{of $\bm{X}^+$}}};
\draw[fill=carnelian!15] (5.5, 0.75) rectangle ++ (1.1, 0.55);
\node[] at (6.05,1) {{$\mathcal{E}_{x^+}(\cdot)$}};

\draw [-stealth] (5.5,0.85) -- (4.55,0.85);

\draw [-stealth] (4.55,1.15) -- (5.5,1.15);
\node[] at (6.05,0.4) {\scriptsize{\textrm{MMSE of}}};
\node[] at (6.05,0.15) {\scriptsize{\textrm{i.i.d. $\bm{X}^+$}}};

\end{tikzpicture}

%% file: SE-algorithm-LLR.tex
\begin{algorithm}[h!]
\footnotesize
\caption{UCS State Evolution}\label{algo:u-bi-vamp-SE}
\begin{algorithmic}[1]
\Statex $\mathbf{Require:}$ Noise precision $\gamma_w$; sensing matrix $\bm{A}$, set $\beta_u=1$ and $\beta_x=\frac{R}{M}$; $\widetilde\gamma_{\boldsymbol{U}_{ \mathsf{e}}}$ from (\ref{eq:SE-expectation-lambda-u}); number of $\textrm{iterations}~ T_\textrm{max}$. 
\vspace{0.1cm}
\Statex $\mathbf{Initialization:}$
\Statex precision of $\bm{X}$: $\bar\gamma_{\boldsymbol{X}_{ \mathsf{e}}^{-},1}$ \vspace{0.07cm}
\Statex extrinsic LLR values: $\overbar{\textrm{LLR}}^{+}_{{\text{diag}},1}$, $\overbar{\textrm{LLR}}^{-}_{{\text{diag}},1}$, $\overbar{\textrm{LLR}}^{+}_{{\text{off-diag}},1}$ and $\overbar{\textrm{LLR}}^{+}_{{\text{off-diag}},1}$ \vspace{0.07cm}
\For {$t=1,\dots, T_{\textrm{max}}$}
\Statex \LeftComment{1}{\scriptsize update the analytical posterior and extrinsic precision of $\textrm{\textbf{{\rv{x}}}}^-$ using (\ref{eq:bi-LMMSE-X})}
\State $\bar\gamma_{\boldsymbol{X}_{ \mathsf{p}}^{-},t+1} = \frac{1}{\mathcal{E}_{x^-}\big(\bar\gamma_{\boldsymbol{X}_{ \mathsf{e}}^{-},t+1}\big)}$
\State $\bar\gamma_{\boldsymbol{X}_{ \mathsf{e}}^{+},t+1}~=~\bar\gamma_{\boldsymbol{X}_{ \mathsf{p}}^{-},t+1}~-~\bar\gamma_{\boldsymbol{X}_{ \mathsf{e}}^{-},t}$\vspace{0.2cm}
\State update the extrinsic LLR values $\overbar{\textrm{LLR}}_{{\textrm{diag}}, t+1}^+$ and $\overbar{\textrm{LLR}}_{{{\textrm{off-diag}}}, t+1}^+$
\Statex \quad ~of  $\textrm{\textbf{{\rv{u}}}}^+$ using (\ref{eq:LLR-plus-diag-convergent-integral-1}) and (\ref{eq:LLR-plus-diag-convergent-integral})

\State update the extrinsic LLR values $\overbar{\textrm{LLR}}_{{\textrm{diag}}, t+1}^-$ and $\overbar{\textrm{LLR}}_{{{\textrm{off-diag}}}, t+1}^-$
\Statex \quad ~of  $\textrm{\textbf{{\rv{u}}}}^-$ using (\ref{eq:LR-values-diag-off-diag})\vspace{0.2cm}

\Statex \LeftComment{1}{\scriptsize compute the analytical posterior precision of  $\textrm{\textbf{{\rv{u}}}}^-$ using (\ref{eq:SE-Eu--final})}
\State $\bar\gamma_{ \boldsymbol{U}_{ \mathsf{p}}^{-},t+1}=\frac{1}{\mathcal{E}_{u^-}\left(\widetilde\gamma_{\boldsymbol{U}_{ \mathsf{e}}},\,\overbar{\textrm{LLR}}_{\textrm{diag},t+1}^-,\,\overbar{\textrm{LLR}}_{{\textrm{off-diag}},t+1}^-\right)}$
\Statex \LeftComment{1}{\scriptsize compute the analytical posterior and extrinsic precision of  $\textrm{\textbf{{\rv{x}}}}^+$ using (\ref{eq:x+-sensitivity-function})}
\State $\bar\gamma_{ \boldsymbol{X}_{ \mathsf{p}}^{+},t+1}=\frac{1}{\mathcal{E}_{x^+}\left(\bar\gamma_{\boldsymbol{X}_{ \mathsf{e}}^{+},t+1}\right)}$
\State $\bar\gamma_{\boldsymbol{X}_{ \mathsf{e}}^{-},t+1}=\bar\gamma_{\boldsymbol{X}_{ \mathsf{p}}^{+},t+1}-\bar\gamma_{\boldsymbol{X}_{ \mathsf{e}}^{+},t+1}$\vspace{0.1cm}
\EndFor
\State \textbf{Return}
$\bar\gamma_{ \boldsymbol{U}_{ \mathsf{p}}^{-},T_{\textrm{max}}+1}, \bar\gamma_{\boldsymbol{X}_{ \mathsf{p}}^{-},T_{\textrm{max}}+1}$
\end{algorithmic}
\end{algorithm}